\documentclass{article}

\usepackage{epubtk}

\usepackage{amssymb}
\usepackage{booktabs}
\usepackage{color}
\usepackage{ifpdf}
\usepackage{graphicx}
\usepackage{textcomp}
\usepackage{nameref}

\newcommand{\eqref}[1]{(\ref{#1})}

\pdfoutput=1

\begin{document}

\title{Testing General Relativity with Low-Frequency, Space-Based
  Gravitational-Wave Detectors}

\author{%
\epubtkAuthorData{Jonathan Gair}{%
Institute of Astronomy, Madingley Road, \\
Cambridge, CB3 0HA, UK}{%
jgair@ast.cam.ac.uk}{%
http://www.ast.cam.ac.uk/~jgair}
\and
\epubtkAuthorData{Michele Vallisneri}{%
Jet Propulsion Laboratory, California Institute of Technology,\\
Pasadena, CA
91109, USA}{%
vallis@vallis.org}{%
http://www.vallis.org}%
\and
\epubtkAuthorData{Shane L.\ Larson}{%
Center for Interdisclipinary Research and Exploration in
Astrophysics\\
Department of Physics and Astronomy\\
Northwestern University, Evanston, IL 60208}{%
s.larson@northwestern.edu}{%
}%
\and
\epubtkAuthorData{John G.\ Baker}{%
Gravitational Astrophysics Lab, NASA Goddard Space Flight Center,\\ 8800
Greenbelt Rd., Greenbelt, MD 20771, USA}{%
John.G.Baker@nasa.gov}{}%
}

\date{}
\maketitle

\begin{abstract}
We review the tests of general relativity that will become possible with space-based gravita\-tional-wave detectors operating in the $\sim$~10\super{-5}\,--\,1\~Hz low-frequency band. The fundamental aspects of gravitation that can be tested include the presence of additional gravitational fields other than the metric; the number and tensorial nature of gravitational-wave polarization states; the velocity of propagation of gravitational waves; the binding energy and gravitational-wave radiation of binaries, and therefore the time evolution of binary inspirals; the strength and shape of the waves emitted from binary mergers and ringdowns; the true nature of astrophysical black holes; and much more. The strength of this science alone calls for the swift implementation of a space-based detector; the remarkable richness of astrophysics, astronomy, and cosmology in the low-frequency gravitational-wave band make the case even stronger.
\end{abstract}

\epubtkKeywords{general relativity, gravitational waves, LISA, eLISA, data analysis, black holes, gravitation}

\newpage


\section{Introduction}\label{sec.introduction}





The first direct detection of gravitational waves (GWs), widely expected in the mid 2010s
with advanced ground-based interferometers
\cite{2010CQGra..27h4006H,avirgo}, will represent the culmination of
a fifty-year experimental quest \cite{collins2004gravity}. Soon
thereafter, newly plentiful GW observations will begin to shed light
on the structure, populations, and astrophysics of mostly dark,
highly relativistic objects such as black holes and neutron stars. In
the low-frequency band that will be targeted by space-based
detectors (roughly 10\super{-5} to 1~Hz), GW observations will provide a census of the massive
black-hole binaries at the center of galaxies, and characterize their
merger histories; probe the galactic population of binaries that include
highly evolved, degenerate stars;
study the stellar-mass objects that spiral into the central black holes
in galactic nuclei;
and possibly detect stochastic GW backgrounds from the dynamical
evolution of the very early universe.

Thus, there are very strong astrophysical motivations to observe the
universe in GWs, especially because the systems and phenomena that
can be observed in this fashion are largely orthogonal to those
accessible to 
traditional electromagnetic (EM) and astroparticle astronomy.
The promise of GWs appears just as great for fundamental physics.
Einstein's theory of gravity, general relativity (GR), has been confirmed by extensive
experimental tests; but these have largely
been confined to the solar system, where gravity is well approximated by
Newtonian gravity with small corrections.
A few tests, based on observations of binary compact-object systems, 
have confirmed the weakest (leading-order) effects of GW generation.  
By contrast, observation of strong GWs will provide the first \emph{direct} 
observational probe of the \emph{dynamical, strong-field regime} of GR, 
where the nature and behavior of gravity can be
significantly different from the Newtonian picture.
GWs are \textit{prima facie} the perfect probe to investigate
gravitation, since they originate directly from the bulk motion
of gravitating matter, relieving the need to understand and model
the physics of other intermediate messengers, typically photons from
stellar surfaces or black-hole surroundings.

Already today we can rely on a very sophisticated understanding of
the analytical and numerical techniques required to model GW
sources and their GW emission, including the post-Newtonian expansion
\cite{lrr-2006-4,lrr-2007-2}, black-hole perturbation theory
\cite{Chandrasekhar1992}, numerical relativity for vacuum spacetimes~\cite{Pretorius2009}, 
spacetimes with gases or magnetized plasmas~\cite{lrr-2008-7}, and much more. 
That these techniques should have been developed so much in the
absence of a dialogue with experimental data (except for the binary
pulsar~\cite{2008LRR....11....8L}) is witness to the great perceived promise of GW astronomy.
For a bird's-eye view of the field, see the Living Review by
Sathyaprakash and Schutz~\cite{2009LRR....12....2S}, who cover
the physics of GWs, the principles of operation of GW detectors, the
nature of GW sources, the data analysis of GW signals, and the
science payoffs of GW observations for physics, astrophysics, and
cosmology.

This review focuses on the opportunities to challenge or
confirm our understanding of gravitational physics that will be
offered by forthcoming space-based missions to observe GWs in the
low-frequency band between 10\super{-5} and 1~Hz. Most of the literature on this subject has focused on one mission design, LISA (the Laser Interferometer Space Antenna
\cite{lisa-98,Jennrich:2009p1398,sciencecase}), which was studied
jointly by NASA and ESA between 2001 and 2011. In 2011, budgetary
and programmatic reasons led the two space agencies to end this
partnership, and to pursue space-based GW detection separately,
studying cheaper, rescoped LISA-like missions.

ESA's proposed eLISA/NGO~\cite{2012CQGra..29l4016A} would be smaller than LISA, fly on orbits closer to Earth, and operate interferometric links only along two arms. In 2012 eLISA was considered for implementation as ESA's first large mission (``L1'') in the Cosmic Vision program. A planetary mission was selected instead, but eLISA will be in the running for the next launch slot (``L2''), with a decision coming as soon as 2014.
NASA ran studies on a broader range of missions~\cite{pcosreport}, including several variants of LISA to be implemented by NASA alone, as well as options with a geocentric orbit (OMEGA~\cite{omegarfi}), and without drag-free control (LAGRANGE~\cite{lagrangerfi}). The final study report concludes that scientifically compelling missions can be carried out for less, but not substantially less, than the full LISA cost; that scientific performance decreases far more rapidly than cost; and that no design choice or technology can make a dramatic reduction in cost without much greater risks. 
The NASA study noted the possibility of participation in the ESA-led eLISA mission (if selected by ESA) as a minority partner.

Whatever specific design is eventually selected, it is likely that its architecture, technology, and scientific reach will bear a strong resemblance to LISA's (with the appropriate scalings in sensitivity, mission duration, and so on). Thus, the research reviewed in this article, which was targeted in large part to LISA, is still broadly relevant to future missions. 
Such LISA-like observatories are characterized by a few common elements: a set of three spacecraft in \emph{long-baseline (Mkm) orbits}, monitoring their relative displacements using \emph{laser interferometry}; \emph{drag-free} operation (except for LAGRANGE~\cite{lagrangerfi}), whereby displacement measurements are referenced to freely falling \emph{test masses} protected by the spacecraft, which hover around the masses using precise \emph{micro-Newton thrusters}; \emph{frequency correction} of laser noise using a variety of means, including onboard cavities and interferometers, arm locking, and a LISA-specific technique known as \emph{Time-Delay Interferometry}. 

The predictions of GR that can be tested by space-based GW observatories include the absence of gravitational fields other than the metric tensor; the number and character of GW polarization states; the speed of GW propagation; the detailed progress of binary inspiral, as driven by nonlinear gravitational dynamics and loss of energy to GWs; the strength and shape of the GWs from binary merger and ringdown; the true nature of astrophysical black holes; and more.

Some of these tests will also be performed with ground-based GW detectors and pulsar-timing observations~\cite{2013arXiv1304.3473Y}, but space-based tests will almost always have superior accuracy and significance, because low-frequency sources are intrinsically stronger, and will spend a larger time within the band of good detector sensitivity. For binary systems with very asymmetric mass ratios, such as extreme mass-ratio inspirals (EMRIs), LISA-like missions will measure hundreds of thousands of orbital cycles; because successful detections require matching the phase of signals throughout their evolution, it follows that these observations will be exquisitely sensitive to source parameters. The data-analysis detection problem will be correspondingly delicate, but has been tackled both theoretically~\cite{dastatus}, and in a practical program of mock data challenges for LISA~\cite{2008CQGra..25r4026B,mldc3,2009CQGra..26i4024V}.

The rest of this review is organized as follows.
Section~\ref{sec.gravityTheory} provides the briefest overview of Einstein's GR, of the theoretical framework in which it can be tested, and of a few leading alternative theories. It also introduces the ``black-hole paradigm,'' which augments Einstein's equations with a few assumptions of physicality that lead to the prediction that the end result of gravitational collapse are black holes described by the Kerr metric. 
Section~\ref{app.lisa} reviews the ``classic'' LISA architecture, as well as possible options for LISA-like variants.
Section~\ref{sec.sources} summarizes the main classes of GW sources that would be observed by LISA-like detectors, and that can be used to test GR.
Section~\ref{sec.tests} examines the tests of gravitational dynamics that can be performed with these sources, while Section~\ref{sec.BHstructure} discusses the tests of the black-hole nature and structure.
(A conspicuous omission are possible stochastic GW backgrounds of cosmological origin~\cite{2012JCAP...06..027B}; indeed, in this article we do not discuss the role of space-based detectors as probes of cosmology and early-universe physics.)
Last, Section~\ref{sec.discuss} presents our conclusions and speculations.

\newpage






\section{The Theory of Gravitation}
\label{sec.gravityTheory}



Newton's theory of gravitation provided a description of the \emph{effect} of gravity through the inverse square law without attempting to explain the \emph{origin} of gravity. The inverse square law provided an accurate description of all measured phenomena in the solar system for more than two hundred years, but the first hints that it was not the correct description of gravitation began to appear in the late 19th century, as a result of the improved precision in measuring phenomena such as the perihelion precession of Mercury. Einstein's contribution to our understanding of gravity was not only practical but also aesthetic, providing a beautiful explanation of gravity as the curvature of spacetime. In developing GR as a generally covariant theory based on a dynamical spacetime metric, Einstein sought to extend the principle of relativity to gravitating systems, and he built on the
crucial 1907 insight that the equality of inertial and gravitational
mass allowed the identification of inertial systems in homogeneous
gravitational fields with uniformly accelerated frames -- the principle
of equivalence~\cite{0034-4885-56-7-001}.  Einstein was also guided by
his appreciation of Ricci and Levi-Civita's absolute
differential calculus (later to become differential geometry),
arguably as much as by the requirement to reproduce Newtonian gravity
in the weak-field limit.  Indeed, one could say that GR was born ``of
almost pure thought''~\cite{willLR}.

Einstein's theory of GR is described by the action
\begin{equation}
S_{\mathrm{GR}} = \int \sqrt{-g}\,R \, \mathrm{d}^4 x \,,
\end{equation}
in which $g$ is the determinant of the spacetime metric and $R = g_{\mu\nu}R^{\mu\nu}$ is the Ricci scalar, where $R^{\mu\nu}=R^\alpha_{\mu\alpha\nu}$ is the Ricci Tensor, $R^\alpha_{\beta\gamma\delta}=\Gamma^\alpha_{\beta\delta,\gamma} - \Gamma^\alpha_{\beta\gamma,\delta}+\Gamma^\alpha_{\gamma\lambda}\Gamma^\lambda_{\beta\delta}-\Gamma^\alpha_{\delta\lambda}\Gamma^\lambda_{\beta\gamma}$ the Riemann curvature tensor, $\Gamma^\alpha_{\beta\gamma}=g^{\alpha\delta}(g_{\beta\delta,\gamma}+g_{\delta\gamma,\beta}-g_{\beta\gamma,\delta})$ the affine connection, and a comma denotes a partial derivative. When coupled to a matter distribution, this action yields the field equations 
\begin{equation}
G_{\mu\nu} \equiv R_{\mu\nu}-\frac{1}{2} g_{\mu\nu}R = \frac{8\pi G}{c^4} T_{\mu \nu} \,,
\end{equation}
where $T_{\mu\nu}$ denotes the stress-energy tensor of the matter.

Since the development of the theory, GR has withstood countless
experimental tests
\cite{willLR,2008ARNPS..58..207T,1063-7869-52-1-R01} based on
measurements as different as atomic-clock precision~\cite{2009SSRv..148..233R}, orbital dynamics (most
notably lunar laser ranging~\cite{2010LRR....13....7M}), astrometry
\cite{1989racm.book.....S}, and relativistic astrophysics (most
exquisitely the binary pulsar
\cite{2008LRR....11....8L,willLR}, but not only
\cite{psaltis08}). It is therefore the correct and natural benchmark against which to compare alternative theories using future observations and we will follow the same approach in this article.
Unlike in the case of Newtonian gravity at the time that GR was
developed, there are no current observations that GR cannot explain
that can be used to guide development of alternatives.%
\epubtkFootnote{Some have argued that measurement of cosmological dark
energy effects may be explained by long-range modifications of
Einstein's equations. While space-based GW observations may
potentially help to refine redshift-distance measurements, we
generally take the possibility of cosmological scale GR alternatives
as falling outside the scope of this review.}
Nonetheless, there are crucial aspects of Einstein's theory that have never been probed directly, such as its strong-field dynamics and the propagation of field perturbations (GWs). 
Furthermore, it is known that classical GR must ultimately fail at the
Planck scale, where quantum effects become important, and traces of
the quantum nature of gravity may be accessible at lower energies
\cite{2009SSRv..148...15S}.  As
emphasized by Will~\cite{willLR}, GR has no adjustable
constants, so every test is potentially deadly, and a probe that could reveal new physics.

\subsection{Will's ``standard model'' of gravitational theories}
\label{standardmodel}

Will's Living Review~\cite{willLR} and his older
monograph~\cite{willTEGP} are the fundamental references about the
experimental verification of GR. In this section, we give only a brief
overview of what may be called Will's ``standard model'' for
alternative theories of gravity, which proceeds through four steps: a)
strong evidence for the equivalence principle supports a \emph{metric}
formulation for gravity; b) metric theories are classified according
to what gravitational fields (scalar, vector, tensor) they prescribe;
c) slow-motion, weak-field conservative dynamics are described in a
unified \emph{parameterized post-Newtonian} (PPN) formalism, and
constrained by experiment and observations; d) finally, equations for
the slow-motion generation and weak-field propagation of gravitational
radiation are derived separately for each metric theory, and again
compared to observations.  Many of the tests of gravitational physics
envisaged for LISA belong in this last sector of Will's standard model,
and are discussed in Section~\ref{classictests} of this review. 
This scheme however leaves out two other important points of contact between gravitational
phenomenology and LISA's GW observations: the strong-field, nonlinear
dynamics of black holes and their structure and excitations,
especially as probed by small orbiting bodies.  We will deal with
these in Sections~\ref{sec.tests} and \ref{sec.BHstructure},
respectively; but let us first delve into Will's standard model.

\paragraph*{The equivalence principle and metric theories of
gravitation.}
Einstein's original intuition~\cite{1989ehgr.conf....5N} placed the
equivalence principle~\cite{2001LNP...562..195H} as a cornerstone for
the theories that describe gravity as curved spacetime.  As formulated
by Newton, the principle states simply that inertial and gravitational
mass are proportional, and therefore all ``test'' bodies fall with the
same acceleration (in modern usage, this is known as the \emph{weak
equivalence principle}, or WEP).  Dicke later recognized that in
developing GR Einstein had implicitly posited a broader principle
(\emph{Einstein's equivalence principle}, or EEP) that consists of WEP
plus \emph{local Lorentz invariance} and \emph{local position
invariance}: that is, of the postulates that the outcome of local
\emph{non-gravitational} experiments is independent of, respectively,
the velocity and position of the local freely-falling reference frames
in which the experiments are performed.


\begin{table}[htb]
\caption{Hierarchy of formulations of the equivalence principle.}
\centering
\begin{tabular}{|p{13cm}|}
\hline
\begin{enumerate}
\item \textbf{Weak Equivalence Principle (WEP):} Gravitational and inertial masses are equivalent (neglecting self gravity).
\item \textbf{Einstein Equivalence Principle (EEP):} Local position invariance and local Lorentz invariance apply in addition to the WEP.
\item \textbf{Strong Equivalence Principle (SEP):} EEP applies also for self-gravitating objects.
\end{enumerate}\\
\hline
\end{tabular}
\end{table}

Turyshev~\cite{2008ARNPS..58..207T} gives a current review of the
experimental verification of WEP (shown to hold to parts in 10\super{13}
by differential free-fall tests~\cite{2008PhRvL.100d1101S}), local
Lorentz invariance (verified to parts in 10\super{22} by clock-anisotropy
experiments~\cite{1986PhRvL..57.3125L}), and local position invariance
(verified to parts in 10\super{5} by gravitational-redshift experiments
\cite{2002PhRvD..65h1101B}, and to much greater precision when looking
for possible time variations of fundamental constants
\cite{2011LRR....14....2U}).  Although these three parts of EEP appear
distinct in their experimental consequences, their underlying physics
is necessarily related in any theory of gravity, so Schiff conjectured
(and others argued convincingly) that any complete and self-consistent
theory of gravity that embodies WEP must also realize EEP
\cite{willLR}.

EEP leads to \emph{metric} theories of gravity in which spacetime is
represented as a pseudo-Rieman\-nian manifold, freely-falling test
bodies move along the geodesics of its metric, and non-gravitational
physics is obtained by applying special-relativistic laws in local
freely-falling frames.  GR \emph{is}, of course, a
metric theory of gravity; so are scalar-vector-tensor theories such
as Brans--Dicke theory, which include other gravitational fields in
addition to the metric.  By contrast, theories with dynamically
varying fundamental constants and theories (such as superstring
theory) that introduce additional WEP-violating gravitational fields
\cite[Section~2.3]{willLR} are not metric.  Neither are most theories
that provide short-range and long-range modifications to Newton's
inverse-square law~\cite{2003ARNPS..53...77A}.

The scalar and vector fields in scalar-vector-tensor theories cannot
directly affect the motion of matter and other non-gravitational
fields (which would violate WEP), but they can intervene in the
generation of gravity and modify its dynamics.  These extra fields can
be \emph{dynamical} (i.e., determined only in the context of solving
for the evolution of everything else) or \emph{absolute} (i.e.,
assigned \textit{a priori} to fixed values).  The Minkowski metric of
special relativity is the classic example of absolute field; such
fields may be regarded as philosophically unpleasant by those who
dislike feigning hypotheses, but they have a right of citizenship in
modern physics as ``frozen in'' solutions from higher energy scales or
from earlier cosmological evolution.

The additional fields can potentially alter the outcome of local
\emph{gravitational} experiments: while the local gravitational
effects of different metrics far away can always be erased by
describing physics in a freely-falling reference frame (which is to
say, the local boundary conditions for the metric can be arranged to
be flat spacetime), the same is not true for scalar and vector fields,
which can then affect local gravitational dynamics by their
interaction with the metric.  This amounts to a violation not of EEP,
but of the \emph{strong equivalence principle} (SEP), which states
that EEP is \emph{also} valid for self-gravitating bodies and
gravitational experiments.  SEP is verified to parts in 10\super{4} by
combined lunar laser-ranging and laboratory experiments
\cite{2004PhRvL..93z1101W}.  So far, GR appears to be the only
viable metric theory that fully realizes SEP.

\paragraph*{The PPN formalism.}
Because the experimental consequences of different metric theories follow from the specific metric that is generated by matter (possibly with the help of the extra gravitational fields), and because all these theories must realize Newtonian dynamics in
appropriate limiting conditions, it is possible to parameterize them in terms of the coefficients of a slow-motion, weak-field
expansion of the metric.  These coefficients appear in front of
gravitational potentials similar to the Newtonian potential, but
involving also matter velocity, internal energy, and pressure.  This
scheme is the \emph{parameterized post-Newtonian formalism}, pioneered
by Nordtvedt and extended by Will (see~\cite{willTEGP} for details).

Of the ten PPN parameters in the current version of the formalism, two
are the celebrated $\gamma$ and $\beta$ (already introduced by
Eddington, Robertson, and Schiff for the ``classical'' tests of GR)
that rule, respectively, the amount of space curvature produced by
unit rest mass and the nonlinearity in the superposition of
gravitational fields.  In GR, $\gamma$ and $\beta$ each have the value 1.  The
other eight parameters, if not zero, give origin to violations of
position invariance ($\xi$), Lorentz invariance
($\alpha_{1\mbox{\,--\,}3}$), or even of the conservation of total
momentum ($\alpha_3$, $\zeta_{1\mbox{\,--\,}4}$) and total angular
momentum ($\alpha_{1\mbox{\,--\,}3}$, $\zeta_{1\mbox{\,--\,}4}$).

The PPN formalism is sufficiently accurate to describe the tests of
gravitation performed in the solar system, as well as many tests using
binary-pulsar observations.  The parameter $\gamma$ is currently
constrained to 1~\textpm\ a few 10\super{-5} by tests of light delay around
massive bodies using the Cassini spacecraft~\cite{bertotti03}; $\beta$
to 1~\textpm\ a few 10\super{-4} by Lunar laser ranging
\cite{2004PhRvL..93z1101W}.\epubtkFootnote{In this context, laser ranging
constrains the Nordtvedt effect (the dependence of free fall for
massive bodies on gravitational self-energy), which is also a
violation of SEP. The $\beta$ constraint given in
\cite{2004PhRvL..93z1101W} assumes the Cassini result
\cite{bertotti03} for $\gamma$.} The other PPN parameters have
comparable bounds around zero from solar-system and pulsar
measurements, except for $\alpha_3$, which is known exceedingly well
from pulsar observations~\cite{willLR}.


\subsection{Alternative theories}
\label{sec:alttheory}

Tests in the PPN framework have tightly constrained the field
of viable alternatives to GR, largely excluding theories with absolute
elements that give rise to preferred-frame effects
\cite{willLR}.  The (indirect) observation of GW emission from
the binary pulsar and the accurate prediction of its $\dot{P}$ by
Einstein's quadrupole formula have definitively excluded other
theories~\cite{willLR,lrr-2003-5}. Yet more GR alternatives were conceived to illuminate
points of principle, but they are not well motivated physically and
therefore are hardly candidates for experimental verification.  Some of the theories that are still ``alive''  are described in the following. More details can be found in~\cite{willTEGP}.

\subsubsection{Scalar-tensor theories}

The addition of a single scalar
field $\phi$ to GR produces a theory described by the
\emph{Einstein-frame} action (see, e.g.,~\cite{willLR}),
\begin{equation}
\tilde I=(16\pi G)^{-1} \int [\tilde R -2\tilde g^{\mu\nu}
\partial_\mu \varphi \partial_\nu \varphi -V(\varphi)] (-\tilde
g)^{1/2}\, \mathrm{d}^4x + I_{\mathrm{matter}}(\psi_{\mathrm{m}}, A^2(\varphi) \tilde
g_{\mu\nu}) \,,
\end{equation}
where $\tilde{g}_{\mu \nu}$ is the metric, the Ricci curvature scalar
$\tilde{R}$ yields the general-relativistic Einstein--Hilbert action,
and the two adjacent terms are kinetic and potential energies for the
scalar field.  Note that in the action $I_{\mathrm matter}$ for matter dynamics, the
metric couples to matter through the function $A(\varphi)$, so this
representation is not manifestly metric; it can however be made so by
a change of variables that yields the \emph{Jordan-frame} action,
\begin{equation}
I=(16\pi G)^{-1} \int [\phi R - \phi^{-1} \omega(\phi)
g^{\mu\nu}\partial_\mu \phi \partial_\nu \phi - \phi^2 V] (-g)^{1/2}\,
\mathrm{d}^4x + I_{\mathrm{matter}}(\psi_{\mathrm{m}}, g_{\mu\nu}) \,,
\end{equation}
where $\phi \equiv A(\varphi)^{-2}$ is the transformed scalar field,
$g_{\mu\nu} \equiv A^2(\varphi) \tilde g_{\mu\nu}$ is the
\emph{physical metric} underlying gravitational observations, and
$3+2\omega(\phi) \equiv [\mathrm{d}(\ln A(\varphi))/\mathrm{d}\varphi]^{-2}$.

The ``classic'' \emph{Brans--Dicke theory} corresponds to fixing
$\omega$ to a constant $\omega_\mathrm{BD}$, and it is
indistinguishable from GR in the limit $\omega_\mathrm{BD} \rightarrow
\infty$.  In the PPN framework, the only parameter that differs from GR is $\gamma =
(1 + \omega_\mathrm{BD})/(2 + \omega_\mathrm{BD})$.  Damour and
Esposito-Far\`ese~\cite{1993PhRvL..70.2220D} considered an expansion
of $\log A(\varphi)$ around a cosmological background value,
\begin{equation}
\log A(\varphi) = \alpha_0 (\varphi - \varphi_0) + \frac{1}{2} \beta_0
(\varphi - \varphi_0)^2 + \cdots \,,
\end{equation}
where $\beta_0$ (and further coefficients) $= 0$ reproduces
Brans--Dicke with $\alpha_0^2 = 1/(2 \omega_\mathrm{BD} + 3)$,
$\beta_0 > 0$ causes the evolution of the scalar field toward
$\varphi_0$ (and therefore toward GR); and $\beta_0 < 0$ may allow
a phase change inside objects like neutron stars,
leading to large SEP violations.  These parameters are bound by
solar-system, binary-pulsar, and GW observations
\cite{1998PhRvD..58d2001D,2012MNRAS.423.3328F}.

Scalar-tensor theories have found motivation in string theory and cosmological models, and have
attracted the most attention in terms of tests with GW observations.

\subsubsection{Vector-tensor theories}

These are obtained by including a dynamical
vector field $u^{\mu}$ coupled to the metric tensor. The most general second-order action in such a theory takes the form~\cite{willLR}
\begin{eqnarray}
S &=& \frac{1}{16\pi G} \int \mathrm{d}^4x \, \sqrt{-g} \left[(1+\omega u_\mu u^\nu) R - K^{\mu\nu}_{\alpha\beta} u^\alpha_{;\mu} u^\beta_{;\nu}+\lambda (u_\mu u^\mu + 1)\right], \nonumber \\
\mbox{where\ } K^{\mu\nu}_{\alpha\beta} &=& c_1g^{\mu\nu}g_{\alpha\beta}+c_2\delta^\mu_\alpha \delta^\nu_\beta+c_3 \delta_\beta^\mu \delta_\alpha^\nu - c_4 u^\mu u^\nu g_{\alpha\beta}\,,
\end{eqnarray}
in which a semicolon denotes covariant differentiation, and the coefficients $c_i$ are arbitrary constants. There are two types of vector-tensor theories: in unconstrained theories, $\lambda \equiv 0$ and the constant $\omega$ is arbitrary, while in Einstein-aether theories the vector field $u^\mu$ is constrained to have unit norm, so the Lagrange multiplier $\lambda$ is arbitrary and the constraint allows $\omega$ to be absorbed into a rescaling of $G$. For the unconstrained theory, only versions of the theory with $c_4=0$ have been studied and for these the field equations are~\cite{willTEGP}
\begin{eqnarray}
R_{\kappa\lambda}-\frac{1}{2}Rg_{\kappa\lambda} + \omega
\Theta^{(\omega)}_{\kappa\lambda} +\eta
\Theta^{(\eta)}_{\kappa\lambda} +\epsilon
\Theta^{(\epsilon)}_{\kappa\lambda} +\tau
\Theta^{(\tau)}_{\kappa\lambda} + \Lambda g_{\kappa\lambda} = 0 \,,
\nonumber \\
\epsilon F_{\mu\nu}^{;\nu} + \frac{1}{2}\tau u_{\mu;\nu}^{;\nu} -\frac{1}{2}\omega u_{\mu}R -\frac{1}{2} \eta u^{\alpha}R_{\mu\alpha} = 0, \nonumber \\
 \Theta^{(\omega)}_{\kappa\lambda} = u_{\kappa}u_{\lambda}R + u^2
R_{\kappa\lambda}-\frac{1}{2} g_{\kappa\lambda} u^2 R - (u^2)_{;\kappa\lambda} +g_{\kappa\lambda} \Box_g u^2 \,, \nonumber \\
  \Theta^{(\eta)}_{\kappa\lambda} =
2u^{\alpha}u_{[\kappa}R_{\lambda]\alpha} - \frac{1}{2}
g_{\kappa\lambda} u^{\alpha}u^{\beta}R_{\alpha\beta} -(u^\alpha u_{[\kappa})_{;\lambda]\alpha}+\frac{1}{2} \Box_g (u_\kappa u_\lambda) + \frac{1}{2} g_{\kappa\lambda} (u^\alpha u^\beta)_{;\alpha\beta} \,, \nonumber \\
 \Theta^{(\epsilon)}_{\kappa\lambda} = -2(F^\alpha_\kappa F_{\lambda\alpha} - \frac{1}{4} g_{\kappa\lambda}F_{\alpha\beta}F^{\alpha\beta}) \,, \nonumber \\
 \Theta^{(\tau)}_{\kappa\lambda} = u_{\kappa;\alpha} u_{\lambda}^{;\alpha} + u_{\alpha;\kappa} u^\alpha_{;\lambda} - \frac{1}{2} g_{\kappa\lambda} u_{\alpha;\beta}u^{\alpha;\beta} + (u^\alpha u_{[\kappa;\lambda]} - u^\alpha_{;[\kappa} u_{\lambda]} - u_{[\kappa}u^{;\alpha}_{\lambda]})_{;\alpha} \,,
\end{eqnarray}
where $F_{\mu\nu}=u_{\nu;\mu}-u_{\mu;\nu}$, $u^2 \equiv u_{\mu}u^{\mu}$, $\eta = -c_2$, $\epsilon=-(c_2+c_3)/2$, and $\tau=-(c_1+c_2+c_3)$. We use the usual subscript notation, such that ``${}_{(,)}$'' and ``${}_{[,]}$'' denote symmetric and antisymmetric sums.

In the constrained Einstein-aether theory~\cite{2008arXiv0801.1547J} the field equations are
\begin{eqnarray}
{J^\alpha}_{\mu;\alpha} - c_4 \dot{u}^\alpha u^\alpha_{;\mu} &=& \lambda u_\mu \,, \nonumber \\
\mbox{where } {J^\alpha}_\mu &\equiv& K^{\alpha\beta}_{\mu\nu} u^\nu_{;\beta}, \nonumber \\
G_{\alpha\beta} &=& T^{(u)}_{\alpha\beta} + \frac{8\pi G}{c^4}T^{\mathrm{matter}}_{\alpha\beta} \,, \nonumber \\
T^{(u)}_{\alpha\beta}&\equiv& \left({J_{(\alpha}}^\mu u_{\beta)} - {J^\mu}_{(\alpha} u_{\beta)} -J_{(\alpha\beta)} u^\mu\right)_{;\mu} + c_1 \left( u^\alpha_{;\mu} u_\beta^{;\mu} - u_{\mu;\alpha} u^\mu_{;\beta}\right) + c_4 \dot{u}_\alpha \dot{u}_\beta \nonumber \\ 
&& \hspace{1in} + \left[ u_\nu J^{\mu\nu}_{;\mu} - c_4 \dot{u}^2\right] u_\alpha u_\beta - \frac{1}{2} g_{\alpha\beta} L_u \,,
\end{eqnarray}
where $\dot{u}^\alpha \equiv u^\beta u^\alpha_{;\beta}$, $L_u=-K^{\alpha\beta}_{\mu\nu}u^\mu_{;\alpha}u^\nu_{;\beta}$ is the aether Lagrangian, and $T^{\mathrm{matter}}_{\alpha\beta}$ is the usual matter stress-energy tensor~\cite{2004gr.qc....10001E}. Via field redefinition this theory can be shown to be equivalent to GR if $c_1+c_4=0$, $c_1+c_2+c_3=0$, and $c_3 = \pm \sqrt{c_1 (c_1-2)}$~\cite{2003PhRvD..68h7501B}. Field redefinition can also be used to set $c_1+c_3=0$~\cite{2005PhRvD..72d4017F}; if this constraint is imposed then equivalence to GR is only achieved if the $c_i$ are all zero. This constraint is therefore appropriate to pose Einstein-aether theory as an alternative to test against GR, since then any non-zero values of the $c_i$ would represent genuine deviations from GR.

Unconstrained vector-tensor theories were introduced in the 1970s as a
straw-man alternative to GR~\cite{willTEGP}, but they have four
arbitrary parameters and leave the magnitude of the vector field
unconstrained, which is a serious defect. Interest in Einstein-aether
theories was prompted by the desire to construct a covariant theory
that violated Lorentz invariance under boosts, by having a preferred
reference frame -- the aether, represented by the vector $u^\mu$. The preferred reference frame also provides a universal notion of time~\cite{1987CQGra...4..485G}. Interest in theories that violate Lorentz symmetry has recently been revived as a possible window onto aspects of quantum gravity~\cite{2003gr.qc.....9054A}.

\subsubsection{Scalar-vector-tensor theories}

The natural extension of scalar-tensor and vector-tensor theories are scalar-vector-tensor theories in which the gravitational field is coupled to a vector field and one or more scalar fields. These theories are relativistic generalizations of Modified Newtonian Dynamics (MoND), which was proposed in order to reproduce observed rotation curves on galactic scales. The relativistic extensions were designed to also satisfy cosmological observations on larger scales. The action takes the form
\begin{equation}
S = \frac{1}{16\pi} \int \mathrm{d}^4x\,\, \sqrt{-g} \left( L_G + L_S + L_u + L_{\mathrm{matter}} \right),
\end{equation}
where $L_G = (R-2\Lambda)/G$ and $L_{\mathrm{matter}}$ are the usual gravitational and matter Lagrangians. There are two main versions of the theory, which differ in the choice of the scalar-field and vector-field Lagrangians $L_S$ and $L_u$.

In Tensor-Vector-Scalar gravity (TeVeS)~\cite{2004PhRvD..70h3509B} the dynamical vector field $u^\mu$ is coupled to a dynamical scalar field $\phi$. 
A second scalar field $\sigma$ is here considered non-dynamical. 
The Lagrangians are
\begin{eqnarray}
L_S&=&\frac{1}{2G} \left[ \sigma^2 \left(g^{\alpha\beta}-u^\alpha u^\beta\right) \phi_{,\alpha} \phi_{,\beta} +\frac{1}{2} \frac{G}{l^2} \sigma^4 F(G\sigma^2)\right], \nonumber \\
L_u&=&\frac{K}{2G}\left[ g^{\alpha\beta}g^{\mu\nu} B_{\alpha\mu} B_{\beta\nu} + 2\frac{\lambda}{K}\left(u^\mu u_\mu+1\right)\right],
\end{eqnarray}
where $B_{\alpha\beta}=u_{\beta,\alpha}-u_{\alpha,\beta}$, 
$F$ is an unspecified dimensionless function, $K$ is a dimensionless parameter, and 
$l$ is a constant length parameter.
The Lagrange multiplier $\lambda$ is spacetime dependent, set to enforce normalization of the vector field $u^\mu u_\mu=-1$. In TeVeS the physical metric that governs the gravitational dynamics of ordinary matter does not coincide with $g_{\mu\nu}$, but is determined by the scalar field through
\begin{equation}
\hat{g}_{\mu\nu} = e^{2\phi}g_{\mu\nu} - 2u_\mu u_\nu \sinh(2\phi).
\end{equation}
An alternative version of TeVeS, called Bi-Scalar-Tensor-Vector gravity (BSTV) has also been proposed~\cite{2005MNRAS.363..459S}, in which the scalar field $\sigma$ is allowed to be dynamical. TeVeS is able to explain galaxy rotation curves and satisfies constraints from cosmology and gravitational lensing, but stars are very unstable~\cite{2007PhRvD..76f4002S} and the Bullet cluster~\cite{2006ApJ...648L.109C} observations (which point to dark matter) cannot be explained.

In Scalar-Tensor-Vector Gravity (STVG)~\cite{2006JCAP...03..004M} the Lagrangian for the vector field is taken to be
\begin{equation}
L_u = \omega\left[B^{\mu\nu}B_{\mu\nu} - 2 \mu^2 u^\mu u_\mu + V_u(u)\right],
\end{equation}
with $B_{\mu\nu}$ defined as before. The three constants $\omega$, $\mu$, and G that enter this action and the gravitational action are then taken to be scalar fields governed by the Lagrangian
\begin{equation}
L_S = \frac{16 \pi}{G} \left[\frac{1}{2} g^{\nu\rho} \left( \frac{G_{,\nu} G_{,\rho}}{G^2} + \frac{\mu_{,\nu} \mu_{,\rho}}{\mu^2} -\omega_{,\nu}\omega_{,\rho}\right) + \frac{V_G(G)}{G^2} +  \frac{V_\mu(\mu)}{\mu^2} + V_\omega(\omega)\right] .
\end{equation}
It is claimed that STVG predicts no deviations from GR on the scale of the solar system or for small globular clusters~\cite{2008ApJ...680.1158M}, and that it can reproduce galactic rotation curves~\cite{2006ApJ...636..721B}, gravitational lensing in the Bullet cluster~\cite{2007MNRAS.382...29B}, and a range of cosmological observations~\cite{2007arXiv0710.0364M}. TeVeS-like theories are constrained by binary-pulsar observations~\cite{2012MNRAS.423.3328F}. It has been proposed that an extension of the ESA-led LISA Pathfinder technology-demonstration mission may allow additional constraints on this class of theories~\cite{2012PhRvD..85d3527M}.  To date the consequences of TeVeS or STVG for GW observations have not been investigated.

\subsubsection{Modified-action theories}

\paragraph*{\textit{f(R)} gravity.} This theory is derived by replacing $R$ with an arbitrary function $f(R)$ in the Einstein--Hilbert action. There are two versions of $f(R)$ gravity. In the \emph{metric formalism} the action is extremized with respect to the metric coefficients only, and the connection is taken to be the metric connection, depending on the metric components in the standard way. The resulting field equations are
\begin{equation}
(-R_{;\kappa}R_{;\lambda} + g_{\kappa\lambda} R_{;\mu}R^{;\mu})
f'''(R) + (-R_{;\kappa\lambda} + g_{\kappa\lambda} \Box R)f''(R) +
R_{\kappa\lambda}f'(R)-\frac{1}{2} g_{\kappa\lambda} f(R) = 0 \,.
\end{equation}
In the \emph{Palatini formalism}, the field equations are found by extremizing the action
over both the metric and the connection. For an $f(R)$ action the
resulting equations are
\begin{equation}
R_{\kappa\lambda} f'(R) - \frac{1}{2} g_{\kappa\lambda} f(R) = 0 \,,
\qquad \nabla_{\alpha} \left[ \sqrt{-g} f'(R) g^{\kappa\lambda}
\right] = 0 \,.
\end{equation}
If the second derivative $f''(R)\neq0$, metric $f(R)$ gravity can be shown to be equivalent to a Brans--Dicke theory with $\omega_\mathrm{BD} = 0$, while Palatini $f(R)$ gravity is equivalent to a Brans--Dicke theory with $\omega_\mathrm{BD}= -3/2$, with no constraint imposed on $f(R)$~\cite{2004CQGra..21..417F,2010RvMP...82..451S,2010LRR....13....3D}. In both cases, the Brans--Dicke potential depends on the exact functional form $f(R)$.

$f(R)$ theories have attracted a lot of interest in a cosmological context, since the flexibility in choosing the function $f(R)$ allows a wide range of cosmological phenomena to be described~\cite{Nojiri2007,Capozziello2007a}, including inflation~\cite{Starobinsky1980,Vilenkin1985} and late-time acceleration~\cite{2003astro.ph..3041C,2004PhRvD..70d3528C}, without violating constraints from Big-Bang Nucleosynthesis~\cite{2008PhRvD..77h3514E}. 
However, metric $f(R)$ theories are strongly constrained by solar-system and laboratory measurements if the scalar degree of freedom is assumed to be long-ranged, which modifies the form of the gravitational potential~\cite{2003PhLB..575....1C}.
This problem can be avoided by assuming a short-range scalar field, but then $f(R)$ theories can only explain the early expansion of the universe and not late-time acceleration. The Chameleon mechanism~\cite{2004PhRvL..93q1104K} has been invoked to circumvent this, as it allows the scalar-field mass to be a function of curvature, so that the field can be short ranged within the solar system but long ranged on cosmological scales.

There are also other issues with $f(R)$ theories. For example, in Palatini $f(R)$ gravity the post-Newtonian metric depends on the local matter density~\cite{2006GReGr..38.1407S}, while in metric $f(R)$ gravity with $f''(R)<0$ there is a Ricci-scalar instability~\cite{2003PhLB..573....1D} that arises because the effective gravitational constant increases with increasing curvature, leading to a runaway instability for small stars~\cite{2008CQGra..25f2001B,2008CQGra..25j5008B}. We refer the reader to~\cite{2010RvMP...82..451S,2010LRR....13....3D} for more complete reviews of the current understanding of $f(R)$ gravity.

\paragraph*{Chern--Simons gravity.}
Yunes and others
\cite{yunesCSpropa,2007PhRvL..99x1101A,2011PhRvD..83l4050A,AliHamCSBound,2012PhRvD..86d4010C,2012JPhCS.363a2019C,2009PhRvD..80f4008C,2012PhRvD..86l4031D,2010arXiv1005.1911F,2008PhRvD..77d4015G,2010CQGra..27j5010H,2009PThPh.122..561K,2012PhRvD..85d4054M,SYCSemri,2012PhRvD..86d4037Y,yunesCSpropb,YPCSBH,YSCSqnm,2009PhRvD..80d2004Y,2010PhRvD..81f4020Y,2012PhRvL.109w1101O} have recently developed an extensive analysis
of the observational consequences of Jackiw and Pi's
\textit{Chern--Simons gravity}~\cite{2003PhRvD..68j4012J}, which
extends the Hilbert action with an additional \emph{Pontryagin term} ${}^*\!RR$ that is
quadratic in the Riemann tensor~\cite{2008PhRvD..77d4015G}:
\begin{equation}
I=(16\pi G)^{-1} \int [\tilde R - \frac{1}{4} \theta \, {}^*\!RR]
(-g)^{1/2}\, \mathrm{d}^4x + I_{\mathrm{matter}}(\psi_{\mathrm{m}}, g_{\mu\nu}) \,;
\end{equation}
here ${}^*\!RR =
{}^*\!{{R^a}_b}^{cd}{R^b}_{acd}$ is built with the help of the dual
Riemann tensor ${}^*\!{{R^a}_b}^{cd} = \frac{1}{2} \epsilon^{cdef}
{R^a}_{bef}$, and it can be expressed as the divergence of the
gravitational Chern--Simons topological current;%
\epubtkFootnote{In field theory, topological currents are those whose
conservation follows not from the equations of motion, but from their
very geometric construction.}
the scalar field $\theta$ can be treated either as a
dynamical quantity, or an absolute field.  In both cases, ${}^*\!RR$
vanishes, either dynamically, or as a constraint on acceptable
solutions, needed to enforce coordinate-invariant matter dynamics,
which restricts the space of solutions available to GR.

Chern--Simons gravity is motivated by string theory and by the attempt
to develop a quantum theory of gravity satisfying a gauge
principle. The Pontryagin term arises in the standard model of
particle physics as a gauge anomaly: the classical gravitational
Noether current that comes from the symmetry of the gravitational
action is no longer conserved when the theory is quantized, but has a
divergence proportional to the Pontryagin term. This anomaly can be
canceled by modifying the action via the addition of the Chern--Simons
Pontryagin term. The same type of correction arises naturally in string theory through the Green--Schwarz anomaly-canceling mechanism, and in Loop Quantum Gravity to enforce parity and charge-parity conservation.

The presence of the Chern--Simons correction leads to parity violation, which has various observable consequences, with magnitude depending on the Chern--Simons coupling, which string theory predicts will be at the Planck scale. If so, these effects will never be observable, but various mechanisms have been proposed that could enhance the strength of the Chern--Simons coupling, such as non-perturbative instanton corrections~\cite{2006JHEP...06..051S}, fermion interactions~\cite{2008PhRvD..77l4040A}, large intrinsic curvatures~\cite{2007PhRvD..75l4022A} or small string couplings at late times~\cite{2005PhRvD..72f3513W}. For further details on all aspects of Chern--Simons gravity, we refer the reader to~\cite{2009PhR...480....1A}.

\paragraph*{General quadratic gravity.} This theory arises by adding to the action all
possible terms that are quadratic in the Ricci scalar, Ricci tensor, and
Riemann tensor. For the action
\begin{equation}
S\equiv\frac{1}{16\pi G} \int\sqrt{-g} \left(-2\Lambda+R+\alpha R^2 + \beta R_{\sigma\tau}R^{\sigma\tau}+\gamma R_{\alpha\beta\gamma\delta}R^{\alpha\beta\gamma\delta} \right) \mathrm{d}^4 x
\end{equation}
the field equations are~\cite{psaltis08}
\begin{eqnarray}
\label{quadraticFEs}
&R_{\kappa\lambda}& - \frac{1}{2} R g_{\kappa\lambda} + \alpha
K_{\kappa\lambda} + \beta L_{\kappa\lambda} + \Lambda
g_{\kappa\lambda} = 0, \nonumber \\
&K_{\kappa\lambda}& \equiv -2R_{;\kappa\lambda} + 2g_{\kappa\lambda}
\Box R - \frac{1}{2} R^2 g_{\kappa\lambda} + 2RR_{\kappa\lambda},
\nonumber \\
&L_{\kappa\lambda}& \equiv -2{R_{\kappa}^{\sigma}}_{;\sigma\lambda} +
\Box R_{\kappa\lambda} + \frac{1}{2} g_{\kappa\lambda} \Box R -
\frac{1}{2}g_{\kappa\lambda} R_{\sigma\tau}R^{\sigma\tau} + 2
R_{\kappa}^{\sigma}R_{\sigma\lambda} \,.
\end{eqnarray}
This class of theories is parameterized by the coefficients $\alpha$, $\beta$, and $\gamma$.
More recently, Stein and Yunes~\cite{SYBHinalt} considered a more
general form of quadratic gravity that includes the Pontryagin term from Chern--Simons gravity. Their action was
\begin{eqnarray}
S&\equiv&\int\sqrt{-g} \Bigg\{ \kappa R + \alpha_1 f_1(\theta) R^2  +
\alpha_2 f_2(\theta) R_{ab} R^{ab} +  \alpha_3 f_3(\theta) R_{abcd}
R^{abcd} \nonumber\\
&& +  \alpha_4 f_4(\theta) R_{abcd}^*R^{abcd} - \frac{\beta}{2}
\left[ \nabla_a\theta\nabla^a\theta + 2 V(\theta) \right] + {\cal
L}_{\mathrm{matter}}\Bigg\}, \label{genquadgrav}
\end{eqnarray}
in which the $\alpha_i$ and $\beta$ are coupling constants,
$\theta$ is a scalar field, and ${\cal L}_{\mathrm{matter}}$ is the matter Lagrangian
density as before. There are two versions of this theory: a \emph{non-dynamical} version in which the functions $f_i(\theta)$ are constants, and a \emph{dynamical} version in which they are not.

General quadratic theories are known to exhibit \emph{ghost fields} --
negative mass-norm states that violate unitarity (see, e.g., \cite{2010RvMP...82..451S} for a discussion and further references). These occur generically, although models with an action that is a function only of $R$ and $R^2-4R_{\mu\nu}R^{\mu\nu}+R_{\mu\nu\rho\sigma}R^{\mu\nu\rho\sigma}$ only are ghost-free~\cite{2005PhRvD..72f4018C}. Ghost fields are also present in Chern--Simons modified gravity~\cite{2012PhRvD..85d4054M,2012PhRvD..86l4031D}, which places strong constraints on the parameters of that model.

\newpage
\subsubsection{Massive-graviton theories}
\label{subsec:massgrav}

Massive-graviton theories were first considered by Pauli and Fierz~\cite{Pauli:1939xp,Fierz:1939zz,1939RSPSA.173..211F}, whose theory is generated by an action of the form
\begin{eqnarray}
\label{PFaction}
S_{\mathrm{PF}}&=&M_P^2 \int \mathrm{d}^4 x\left[ -\frac{1}{4} (\partial_\mu h_{\nu\rho})^2 + \frac{1}{4} (\partial_\mu h)^2 - \frac{1}{2} (\partial_\mu) (\partial^\nu h_\nu^\mu) + \frac{1}{2} (\partial_\mu h_{\nu\rho})(\partial^\nu h^{\mu\rho}) \right.\nonumber \\
&&\left. \hspace{1in} -\frac{1}{4} m^2 \left(h_{\mu\nu}h^{\mu\nu} - h^2\right) + M_P^{-2} T_{\mu\nu}h^{\mu\nu}\right],
\end{eqnarray}
in which $h_{\mu\nu}$ is a rank-two covariant tensor, $m$ and $M_P$ are mass parameters, $T_{\mu\nu}$ is the matter energy-momentum tensor, indices are raised and lowered with the Minkowski metric $\eta_{\mu\nu}$, and $h=h_{\mu\nu} \eta^{\mu\nu}$. The terms on the first line of this expression are generated by expanding the Einstein--Hilbert action to quadratic order in $h_{\mu\nu}$. The massive graviton term is $m^2\left(h_{\mu\nu}h^{\mu\nu} - h^2\right)$; it contains a spin-2 piece $h_{\mu\nu}$ and a spin-0 piece $h$. 

This model suffers from the van Dam--Velten--Zakharov discontinuity~\cite{1970NuPhB..22..397V,1970JETPL..12..312Z}: no matter how small the graviton mass, the Pauli--Fierz theory leads to different physical predictions from those of linearized GR, such as light bending. The theory also predicts that the energy lost into GWs from a binary is twice the GR prediction, which is ruled out by current binary-pulsar observations. It might be possible to circumvent these problems and recover GR in the weak-field limit by invoking the Vainshtein mechanism~\cite{1972PhLB...39..393V,2010PhRvD..82j4008B}, which relies on nonlinear effects to ``hide'' certain degrees of freedom for source distances smaller than the Vainshtein radius~\cite{2013arXiv1304.7240B}. The massive graviton can therefore  become effectively massless, recovering GR on the scale of the solar system and in binary-pulsar tests, while retaining a mass on larger scales. In such a scenario, the observational consequences for GWs would be a modification to the propagation time for cosmological sources, but no difference in the emission process itself.

There are also non-Pauli--Fierz massive graviton theories~\cite{2003IJMPD..12.1905B}. For these, the action is the same as that in Eq.~(\ref{PFaction}), but the first term on the second line (the massive graviton term) takes the more general form
\begin{equation}
- \left(k_1 h_{\mu\nu}h^{\mu\nu} + k_2 h^2\right),
\end{equation}
where $k_1$ and $k_2$ are new constants of the theory that represent the squared masses of the spin-2 and spin-0 gravitons respectively. This theory can recover GR in the weak field, since $k_1$ and $k_2$ can independently be taken to zero, with modifications to weak-field effects that are on the order of the graviton mass squared. These theories are generally thought to suffer from instabilities~\cite{Pauli:1939xp,Fierz:1939zz,1939RSPSA.173..211F}, which arise because the spin-0 graviton carries negative energy. However, it was shown in~\cite{2003IJMPD..12.1905B} that the spin-0 graviton cannot be emitted without spin-2 gravitons also being generated. The spin-2 graviton energy is positive and greater than that of spin-2 gravitons in GR, which compensates for the spin-0 graviton's negative energy. The total energy emitted is therefore always positive, and it converges to the GR value in the limit that the spin-2 graviton mass goes to zero.

These alternative massive-graviton theories are therefore perfectly compatible with current observational constraints, but make very different predictions for strong gravitational fields~\cite{2003IJMPD..12.1905B}, including the absence of horizons for black-hole spacetimes and oscillatory cosmological solutions. Despite these potential problems, the existence of a ``massive graviton'' can be used as a convenient strawman for GW constraints, since the speed of GW propagation can be readily inferred from GW observations and compared to the speed of light. These proposed tests generally make no reference to an underlying theory but require only that the graviton has an effective mass and hence GWs suffer dispersion. This will be discussed in more detail in Section~\ref{sub.propagation}.

\subsubsection{Bimetric theories of gravity}

As their name suggests, there are two metrics in bimetric theories of gravity~\cite{1973GReGr...4..435R,1975GReGr...6..259R}. One is dynamical and represents the tensor gravitational field;  the other is a metric of constant curvature, usually the Minkowski metric, which is non-dynamical and represents a prior geometry. There are various bimetric theories in the literature.

Rosen's theory has the action~\cite{1971PhRvD...3.2317R,1973GReGr...4..435R,1974AnPhy..84..455R,1975GReGr...6..259R}
\begin{equation}
S=\frac{1}{64\pi G}  \int \mathrm{d}^4
x\left[ \sqrt{-\det(\eta)} \eta^{\mu\nu} g^{\alpha \beta}
g^{\gamma\delta} \left(g_{\alpha\gamma|\mu} g_{\alpha\delta|\nu}
- \frac{1}{2} g_{\alpha\gamma|\mu} g_{\alpha\delta|\nu} \right)\right]
+S_{\mathrm{matter}} \,,
\end{equation}
in which $\eta_{\mu\nu}$ is the fixed flat, non-dynamical metric, $g_{\mu\nu}$ is the dynamical gravitational metric and the vertical line in subscripts denotes a covariant derivative with respect to $\eta_{\mu\nu}$. The final term, $S_{\mathrm{matt}}$, denotes the action for matter fields. The field equations may be written
\begin{equation}
\Box_\eta g_{\mu\nu} - g^{\alpha\beta}\eta^{\gamma\delta}g_{\mu\alpha|\gamma}g_{\nu\beta|\delta} = -16\pi G \sqrt{\det(g)/\det(\eta)} \left(T_{\mu\nu} - \frac{1}{2} g_{\mu\nu} T \right) .
\end{equation}

Lightman and Lee~\cite{1973PhRvD...8.3293L} developed a bimetric theory based on a non-metric theory of gravity due to Belinfante and Swihart~\cite{1957AnPhy...1..168B}. The action for this ``BSLL'' theory is
\begin{equation}
S=\frac{1}{64\pi G}  \int \mathrm{d}^4 x \, \sqrt{-\det(\eta)} \left( \frac{1}{4} h^{\mu\nu|\alpha} h_{\mu\nu|\alpha} - \frac{5}{64} h_{,\alpha} h^{,\alpha} \right) +S_{\mathrm{matter}}\,,
\end{equation}
in which $\eta$ is the non-dynamical flat background metric and $h_{\mu\nu}$ is a dynamical gravitational tensor related to the gravitational metric $g_{\mu\nu}$ via
\begin{eqnarray}
g_{\mu\nu} &=& \left( 1- \frac{1}{16} h\right)^2 \Delta_\mu^\alpha \Delta_{\alpha\nu}, \nonumber \\
\delta^\mu_\nu &=& \Delta^\alpha_\nu \left( \delta_\alpha^\mu - \frac{1}{2} h_\mu^\alpha \right),
\end{eqnarray}
in which $\delta^\mu_\nu$ is the Kronecker delta and
$\Delta^\alpha_\nu$ is defined by the second equation. Indices on
$\Delta_{\alpha\beta}$ and $h_{\alpha\beta}$ are raised and lowered
with $\eta_{\mu\nu}$, but on all other tensors indices are raised and
lowered by $g_{\mu\nu}$.  Both the Rosen and BSLL bimetric theories
give rise to alternative GW polarization states, and have been used to
motivate the construction of the parameterized post-Einsteinian (ppE)
waveform families discussed in Section~\ref{sec:ppE}.

There is also a bimetric theory due to Rastall~\cite{1979CaJPh..57..944R}, in which the metric is an algebraic function of the Minkowski metric and of a vector field $K^\mu$. The action is
\begin{equation}
S=\frac{1}{64\pi G} \int \mathrm{d}^4 x \, \left[ \sqrt{-\det(g)} F(N) K^{\mu;\nu} K_{\mu;\nu} \right] +S_{\mathrm{matter}} \,,
\end{equation}
in which $F(N)=-N/(N+2)$, $N=g^{\mu\nu} K_\mu K_\nu$ and a semicolon denotes a derivative with respect to the gravitational metric $g_{\mu\nu}$. The metric follows from $K^\mu$ by way of
\begin{equation}
g_{\mu\nu}=\sqrt{1 + \eta^{\alpha\beta} K_\alpha K_\beta} \left( \eta_{\mu\nu} + K_\mu K_\nu \right),
\end{equation}
where $\eta_{\mu\nu}$ is again the non-dynamical flat metric. This theory has not been considered in a GW context and we will not mention it further; more details, including the field equations, can be found in~\cite{willTEGP}.


\subsection{The black-hole paradigm}
\label{gravtheory:bhstruct}

The present consensus is that all of the compact objects observed to reside in galactic centers are supermassive black holes, described by the Kerr metric of GR~\cite{1984ARAA..22..471R}. This explanation follows naturally in GR from the black-hole \emph{uniqueness theorems} and from a set of additional \emph{assumptions of physicality}, briefly discussed below. If a deviation from Kerr is inferred from GW observations, it would imply that the assumptions are violated, or possibly that GR is not the correct theory of gravity.
Space-based GW detectors can test black-hole ``Kerr-ness'' by measuring the GWs emitted by smaller compact bodies that move through the gravitational potentials of the central objects (see Section~\ref{sec:emritests}). Kerr-ness is also tested by characterizing multiple \emph{ringdown modes} in the final black hole resulting from the coalescence of two precursors (see Section~\ref{sec:ringdowntests}).

The current belief that Kerr black holes are ubiquitous follows from work on mathematical aspects of GR in the middle of the 20th century. Oppenheimer and Snyder demonstrated that a spherically-symmetric, pressure-free distribution would collapse indefinitely to form a black
hole~\cite{1939PhRv...56..455O}. This result was assumed to be a curiosity due to
spherical symmetry, until it was demonstrated by
Penrose~\cite{penrose65} and by Hawking and Penrose~\cite{hawkpen70}
that singularities arise inevitably after the formation of a \emph{trapped
surface} during gravitational collapse. Around the same time, it was
proven that the black-hole solutions of Schwarschild~\cite{schwarz16}
and Kerr~\cite{kerr63} are the only static and axisymmetric black-hole solutions in GR~\cite{israel67,carter71,rob75}. These results together indicated the inevitability of black-hole formation in gravitational collapse.

The assumptions that underlie the proof of the uniqueness theorem are that the spacetime is a stationary vacuum solution, that it is asymptotically flat, and that it contains an event horizon but no closed timelike curves (CTCs) exterior to the horizon~\cite{1973lsss.book.....H}. The lack of CTCs is needed to ensure causality, while the requirement of a horizon is a consequence of the cosmic-censorship hypothesis (CCH)~\cite{penrose69}.
The CCH embodies this belief by stating that any singularity that forms in nature must be hidden behind a horizon (i.e., cannot be naked), and therefore cannot affect the rest of the universe, which would be undesirable because GR can make no prediction of what happens in its vicinity. However, the CCH and the non-existence of CTCs are not required by Einstein's equations, and so they could in principle be violated.

Besides the Kerr metric, we know of many other ``black-hole--like'' solutions to Einstein's equations: these are vacuum solutions with a very compact central object enclosed by a high-redshift surface. In fact, \emph{any} metric can become a solution to Einstein's equation: it is sufficient to insert it in the Einstein tensor, and postulate the resulting ``matter'' stress-energy tensor as an input to the equations.
However, such matter distributions will not in general satisfy the \emph{energy conditions} (see, e.g.,~\cite{2004rtmb.book.....P}):
\begin{itemize}
\item \textbf{The weak energy condition} is the statement that all timelike observers in a spacetime measure a non-negative energy density, $T_{\mu\nu}v^\mu v^\nu \geq 0$, for all future-directed timelike vectors $v^\mu$. The \emph{null energy condition} modifies this condition to null observers by replacing $v^\mu$ by an arbitrary future-directed null vector $k^\mu$.
\item \textbf{The strong energy condition} requires the Ricci curvature measured by any timelike observer to be non-negative, $(T_{\mu \nu} - T^\alpha_\alpha g_{\mu\nu}/2) v^\mu v^\nu \geq 0$, for all timelike $v^\mu$.
\item \textbf{The dominant energy condition} is the requirement that matter flow along timelike or null world lines: that is, that $-T_\nu^\mu v^\nu$ be a future-directed timelike or null vector field for any future-directed timelike vector $v^\mu$.
\end{itemize}
These conditions make sense on broad physical grounds; but even after imposing them, there remain several black-hole--like solutions~\cite{2003esef.book.....S} besides Kerr.
Thus, space-based GW detectors offer an important test of the ``black-hole paradigm'' that follows from GR plus CCH, CTC non-existence, and the energy conditions. This paradigm is especially important: putative black holes are observed to be ubiquitous in the universe, so their true nature has significant implications for our understanding of astrophysics.

If one or many non-Kerr metrics are found, the hope is that observations will allow us to tease apart the various possible explanations:
\begin{itemize}
\item Does the spacetime contain matter, such as an accretion disk,
exterior to the black hole?
\item Are the CCH, the no-CTC assumption, or the energy conditions violated?
\item Is the central object an exotic object, such as a boson star~\cite{ryanBS,kesden05}?
\item Is gravity coupled to other fields?  This can lead to different black-hole solutions~\cite{1998PhRvD..58h4006K,2004GReGr..36.1361S,2005CQGra..22.3561S}, although some such solutions are known~\cite{straumann90}) or suspected~\cite{droz91} to be unstable to generic perturbations. 
\item Is the theory of gravity just different from GR? For instance, in Chern--Simons gravity black holes (to linear order in spin) differ from Kerr in their octupole moment~\cite{YPCSBH}, and this correction may produce the most significant observational signature in GW observations~\cite{SYCSemri}.
\end{itemize}
While these questions are challenging, we can learn a lot by testing black-hole structure with space-based GW detectors. These tests are discussed in detail in Section~\ref{sec.BHstructure}.

\newpage

\newpage

\section{Space-Based Missions to Detect Gravitational Waves}
\label{app.lisa}

The experimental search for
GWs began in the 1960s with Joseph Weber's resonant
bars (and resonant claims~\cite{collins2004gravity}); it has since
grown into an extensive international endeavor that has produced a
network of km-scale GW interferometers (LIGO~\cite{ligoweb}, EGO/VIRGO~\cite{virgoweb}, GEO600~\cite{geoweb}, and Tama/KAGRA~\cite{tamaweb}),
the proposals for space-based observatories such as LISA, and the
effort to detect GWs by using an array of pulsars as reference clocks
\cite{2010CQGra..27h4013H}.

In this section we briefly describe the architecture of LISA-like space-based GW observatories, beginning with the ``classic'' LISA design, and then discussing the variations studied in the 2011\,--\,2012 ESA and NASA studies~\cite{2012CQGra..29l4016A,pcosreport}. We also discuss proposals for detectors that would operate in a higher frequency band (between 0.1 and 10~Hz) that would bridge the gap in sensitivity between LISA-like and ground-based observatories.

\subsection{The classic LISA architecture}
\label{sec:lisa}

The most cited LISA reference is perhaps the 1998
pre-phase A mission study~\cite{lisa-98}; a more up-to-date review of
the LISA technical architecture is given by
Jennrich~\cite{Jennrich:2009p1398}, while the LISA science-case
document~\cite{sciencecase} describes the state of LISA science at the
end of the 2000s.  Here we give only a quick review of the elements of the
mission, referring the reader to those references for in-depth
discussions.

\paragraph*{LISA principles.} LISA consists of three identical
cylindrical spacecraft, approximately 3~m wide and 1~m high, that are launched together and, after a 14-month cruise, settle into an
Earth-like heliocentric orbit, 20\textdegree\ behind the Earth.  The orbit
of each spacecraft is tuned slightly differently, resulting in an
equilateral-triangle configuration with $5 \times 10^6\mathrm{\ km}$ arms
(commensurate with the frequencies of the LISA sensitivity band), inclined by 60\textdegree\
with respect to the ecliptic.  This configuration is maintained
to 1\% by orbital dynamics alone for the lifetime of the mission,
nominally 5~years, with a goal of 10.  In the course of a year, the
center of the LISA triangle completes a full revolution on the
ecliptic, while the triangle itself, as well as its normal vector, rotate
through 360\textdegree.

LISA detects GWs by monitoring the fluctuations in the distances
between freely-falling reference bodies -- in LISA, platinum-gold
\emph{test masses} housed and protected by the spacecraft.  LISA
uses three pairs of laser interferometric links to measure
inter-spacecraft distances with errors near $10\pm\mathrm{\
Hz}^{-1/2}$. The phase shifts accumulated along different
interferometer paths are proportional to an integral of GW strain
along those trajectories. The test masses suffer from residual
acceleration noise at a level of $3 \times 10^{-15}\mathrm{\ m\
s}^{-2}$, which is dominant below 3~mHz. Together, the position and
acceleration noises (divided by a multiple of the armlength) determine
the LISA sensitivity to GW strain, which reaches $10^{-20}\mathrm{\
Hz}^{-1/2}$ at frequencies of a few mHz.

Alternatively (but equivalently), we may describe LISA as measuring
the GW-induced relative Doppler shifts between the local lasers on
each spacecraft and the remote lasers.  These Doppler shifts are
directly proportional to a difference of instantaneous GW strains (as
experienced by the local spacecraft at the time of measurement, and by
the distant spacecraft at a time retarded by the LISA armlength
divided by the speed of light).  In either description, the 1\%
``breathing'' of the LISA constellation occurs at frequencies that are
safely below the LISA measurement band.

\paragraph*{LISA technology.} The distance measurement between the test
masses along each arm is in fact split in three: an inter-spacecraft
measurement between the two optical benches, and two local
measurements between each test mass and its optical bench.  To achieve
the inter-spacecraft measurement, 2~W of 1064-nm light are sent and
received through 40-cm telescopes.  The diffracted beams deliver only
100~pW of light to the distant spacecrafts, so they cannot be
reflected directly; instead, the laser phase is measured and
transponded back by modulating the frequency of the local lasers.
Each of the two optical assemblies on each spacecraft include the
telescope and a Zerodur optical bench with bonded fused-silica optical
components, which implements the inter-spacecraft and local
interferometers, as well as a few auxiliary interferometric
measurements, which are needed to monitor the stability of the
telescope structure, to control \emph{point-ahead} corrections, and to
compare the two lasers on each spacecraft.  The LISA phasemeters
digitize the signals from the optical-bench photodetectors at 50 MHz,
and multiply them with the output of local oscillators, computing
phase differences and driving the oscillators to track the frequency
of the measured signal.  This heterodyne scheme is needed to handle
Doppler shifts (as well as laser frequency offsets) as large as 15
MHz.

Because intrinsic fluctuations in the laser frequencies are
indistinguishable from GWs in the LISA output, laser frequency noise
needs to be suppressed by several orders of magnitude, using a
hierarchy of techniques~\cite{Jennrich:2009p1398}: the lasers are
prestabilized to local frequency references; \emph{arm locking} may be
used to further stabilize the lasers using the LISA arms (or their
differences) as stable references; finally, \emph{Time-Delay
Interferometry}~\cite{2005LRR.....8....4D} is applied in post
processing to remove residual laser frequency noise by algebraically
combining appropriately-delayed single-link measurements, in such a
way that all laser-frequency--noise terms appear as canceling pairs in
the combination.  With this final step, the LISA measurements combine
into \emph{synthesized-interferometer} observables
\cite{2005PhRvD..72d2003V} analogous to the readings of ground-based
interferometers such as LIGO.

The LISA \emph{disturbance reduction system} (DRS) minimizes the
deviations of the test masses from free-fall trajectories, by
shielding them from solar radiation pressure and interplanetary
magnetic fields.  On each spacecraft, the DRS includes \emph{two
gravitational reference sensors} (GRSs): a GRS consists of a 2-kg test
mass, enclosed in an electrode housing with capacitive reading and
control of test-mass position and orientation, and accompanied by
additional components to cage and reposition the test mass, to
maintain vacuum, and to control the accumulation of charge.  The other
crucial part of the DRS are the micro-Newton thrusters (on each
spacecraft, three clusters of four colloid or field-emission--electric
propulsion systems), which are controlled in response to the GRS
readings to maintain the nominal position of the test mass with
respect to the spacecraft.  This is known as \emph{drag-free} control.
The thrusters need to provide up to $100\,\mu\mathrm{N}$ force with $< 0.1\,\mu\mathrm{N}$ noise.  In addition to this active correction, test-mass
acceleration noise is minimized by the accurate knowledge and
correction of spacecraft self-gravity, by enforcing magnetic
cleanliness, and by controlling thermal fluctuations.  A version of
the LISA DRS with slightly lower performance will be flown and tested
in \emph{LISA Pathfinder}~\cite{2008CQGra..25k4034M,2009CQGra..26i4001A}, a
single-spacecraft technology precursor mission.

\paragraph*{The LISA response to GWs.} Compared to
ground-based interferometers, the LISA response to GWs is both richer
and more complex
\cite{springerlink:10.1007/BF00762449,springerlink:10.1007/BF00759146,2005PhRvD..71b2001V,2003PhRvD..67b2001C}.
First, the revolution and rotation of the LISA constellation imprint a
sky-position--dependent signature on long-lasting GW signals (which for
LISA include all binary signals): the revolution causes a
time-dependent Doppler shift with a period of a year, and a fractional
amplitude of $10^{-4}$; the rotation introduces a 1-year periodicity
in the LISA equivalent of the ground-based--interferometer
\emph{antenna patterns}~\cite{Maggiore:1900zz}, endowing an incoming monochromatic GW
signal with eight sidebands, separated by $\mathrm{yr}^{-1} = 3.17
\times 10^{-8}\mathrm{\ Hz}$.  These effects are sometimes referred to as the
LISA AM and FM modulations.  Thus, GW signals as measured by LISA
carry information regarding the position of the source; on the other
hand, data analysis must account for these effects, possibly including
sky position among the parameters of matched-filtering search
\emph{templates}.

Second, the \emph{long-wavelength approximation}, whereby the entire
``interferometer'' shrinks and expands as one, cannot be used
throughout the LISA band; indeed, the wavelength of GWs reaches the LISA
armlength at a frequency of 60~mHz.  As a consequence, the LISA
response to a few-second impulsive GW is not a single pulse, but a
collection of pulses with amplitudes and separations dependent on the
sky position and polarization of the source; the effect on high-frequency chirping
signals is more subtle, but still present.  This further complicates
data analysis, and introduces an independent mechanism (a
triangulation of sorts) to localize GW sources of short duration and high frequency.

\epubtkImage{first-gen-b-2.png}{%
\begin{figure}
\centerline{\includegraphics[width=0.85\textwidth]{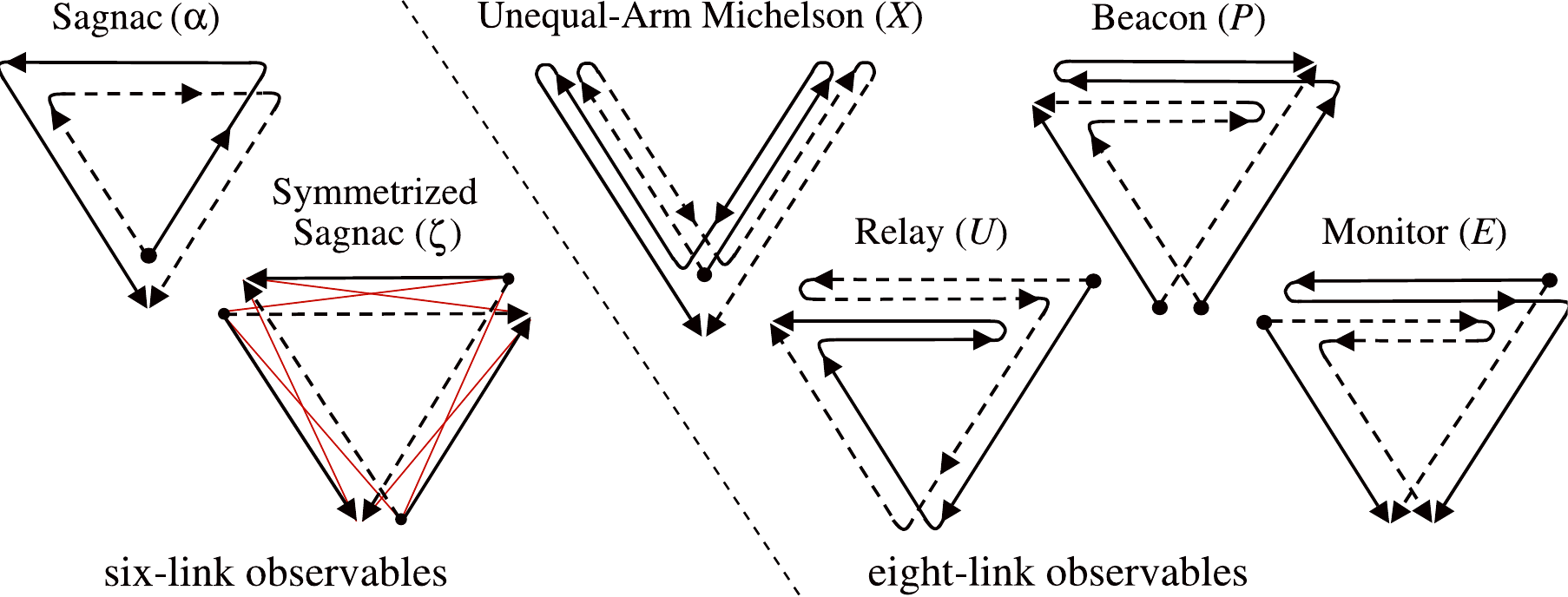}}
\caption{Time-Delay Interferometry (TDI). LISA-like detectors measure
GWs by transmitting laser light between three spacecraft in triangular configuration, and comparing the optical phase of the incident lasers against reference lasers on each spacecraft.  To avoid
extreme requirements on laser-frequency stability over the course of
the many seconds required for transmission around the triangle, data analysts will generate time-delayed linear combinations of the phase comparisons; the combinations simulate nearly
equal-delay optical paths around the sides of the triangle, and (much like an equal-arm Michelson interferometer) they suppress laser frequency noise.  Many such combinations, including 
those depicted here, are possible, but altogether they comprise at most three independent gravitational-wave observables. Image reproduced by permission from~\cite{2005PhRvD..72d2003V}, copyright by APS.}
\label{fig:TDI}
\end{figure}}

Third, LISA is in effect \emph{three detectors in one}: this can be
understood most easily by considering that subsets of the three LISA
arms form three separate Michelson-like interferometers (known in LISA
lingo as $X$, $Y$, and $Z$) at 120\textdegree\ angles.  More formally, the
LISA interferometric measurements can be combined into many different
TDI observables (see Figure~\ref{fig:TDI}), some resembling actual optical setups, others quite
exotic~\cite{2005PhRvD..72d2003V}, although at most three observables
are independent in the sense that any other observable can be
reconstructed by time-delaying and summing a generic basis of three
observables~\cite{0264-9381-25-6-065005}.  Furthermore, such a basis
can be chosen so that its components have uncorrelated noises, much
like widely separated ground-based detectors~\cite{2002PhRvD..66l2002P}.  
One of these must correspond, in effect, to $X + Y + Z$, and by symmetry it
must be relatively insensitive to GWs in the long-wavelength
limit, providing for an independent measurement of a combination of
instrument noises.

\subsection{LISA-like observatories}

The mission-concept studies ran in 2011\,--\,2012 by ESA~\cite{2012CQGra..29l4016A} and NASA~\cite{pcosreport} embrace several approaches to limiting cost.

Reducing mass is a broadly useful strategy, because it allows for launch on smaller, cheaper rockets, and because mass has been shown to be a good proxy for mission complexity, and therefore implementation cost. ESA's NGO design envisages interferometric links along two arms rather than LISA's three, resulting in an asymmetric configuration with one full and two ``half'' spacecraft. As a consequence, only one TDI observable can be formed, eliminating the capability of measuring two combinations of GW polarizations simultaneously.

Propellant may be saved by placing the spacecraft closer together (with arms $\sim$~1\,--\,3~Mkm rather than 5~Mkm) and closer to Earth. At low frequencies, the effect of shorter armlengths is to reduce the response to GWs proportionally; for the same test-mass noise, sensitivity then decreases by the same ratio. At high frequencies, the laser power available for position measurement \emph{increases} as $L^{-2}$, since beams are broadly defocused at millions of kms, improving shot noise, but not other optical noises, by a factor $L^{-1}$ (in rms). As a consequence, the sweet spot of the LISA sensitivity shifts to higher frequencies, although one may instead plan for a less powerful laser, for further cost savings. Spacecraft orbits that are different than LISA's, such as geocentric options, or flat configurations that lie in the ecliptic plane would also alter GW-signal modulations, and therefore parameter-estimation performance. The NASA report examines such effects briefly~\cite{pcosreport}. 

Reducing mission duration also saves money, because it reduces the cost of supporting the mission from the ground, and it allows for shorter ``warranties'' on the various subsystems. Conversely, missions are made cheaper by accepting more risk of failure or underperformance, since risk is ``retired'' by extensive testing and by introducing component redundancy, both of which are expensive.
The NASA study also explored replacing LISA's custom subsystems with variants already flown on other missions (as in OMEGA~\cite{omegarfi}), or eliminating some of them altogether (as in LAGRANGE~\cite{lagrangerfi}). However, the significant performance hit and additional risk incurred by such steps is not matched by correspondingly major savings, because the main cost driver for LISA-like missions is the necessity of launching and flying three (or more) independent spacecraft. Switching to atom interferometry would make for very different mission architectures, but the NASA study finds that an atom-interferometer mission would face many of the same cost-driving constraints as a laser-interferometer mission~\cite{pcosreport,2012PhRvL.108u1101B}.

Indeed, the overarching conclusion of the NASA study is that no technology can provide dramatic cost reductions, and that scientific performance decreases far more rapidly than cost. Thus, ``staying the course'' and pursuing a modestly descoped LISA-like design, whenever programs and budgets will allow it, may yet be the most promising strategy for GW detection in space.

\subsection{Mid-frequency space-based observatories}

The DECi-hertz Interferometer Gravitational wave Observatory (DECIGO~\cite{2001PhRvL..87v1103S,2006CQGra..23S.125K,2011CQGra..28i4011K}) is a proposed Japanese mission that would observe GWs at frequencies between 1~mHz and 100~Hz, reaching its best ($h \sim 10^{-23}$) sensitivity between 0.1 and 10~Hz, and thus bridging the gap between LISA-like and ground-based detectors. 
Prior to DECIGO, the possibility of observing GWs in the decihertz band had been studied in the context of a possible follow-up to LISA, the Big Bang Observer (BBO~\cite{phinney2003,2005PhRvD..72h3005C,2006PhRvD..73d2001C}).

The final DECIGO configuration (2024+) envisages four clusters in an Earth-like solar orbit, each cluster consisting of three drag-free spacecraft in a triangle with 1000-km arms. GWs are measured by operating the arms as a Fabry--P\'erot interferometer, which requires keeping the armlengths constant, in analogy to ground-based interferometers and in contrast to LISA's transponding scheme. DECIGO's test masses are 100~kg mirrors, and its lasers have 10~W power.
The roadmap toward DECIGO includes two pathfinders: the single-spacecraft DECIGO Pathfinder~\cite{2009CQGra..26i4019A} consists of a 30~cm Fabry--P\'erot cavity, and it could detect binaries of $10^{3}\mbox{\,--\,}10^{5}\,M_{\odot}$ black holes if they exist near the galaxy~\cite{2012CQGra..29g5005Y}; next, pre-DECIGO~\cite{2011CQGra..28i4011K} would demonstrate the DECIGO measurement with three spacecraft and modest optical parameters, resulting in a sensitivity 10\,--\,100 times worse than one of the final DECIGO clusters.

The DECIGO science objectives~\cite{2011CQGra..28i4011K} include measuring the GW stochastic background from ``standard'' inflation (with sensitivity down to $\Omega_\mathrm{GR} \simeq 2 \times 10^{-16}$), and determining the thermal history of the universe between the end of inflation and nucleosynthesis~\cite{2008PhRvD..77l4001N,2011PhRvD..84l3513K};
searching for hypothesized primordial black holes~\cite{2009PhRvL.102p1101S};
characterizing dark energy by using neutron-star binaries as standard candles (either with host redshifts~\cite{2011PhRvD..83h4045N}, or by the effect of cosmic expansion on the inspiral phasing~\cite{2001PhRvL..87v1103S,2012JPhCS.363a2052N});
illuminating the formation of massive galactic black holes by observing the coalescences of intermediate-mass ($10^3\mbox{\,--\,}10^{4}\,M_{\odot}$) systems;
constraining the structure of neutron stars by measuring their masses (in upwards of 100\,000 detections per year);
and even searching for planets orbiting neutron-star binaries.

DECIGO can also test alternative theories of gravity, as reviewed in~\cite{2013IJMPD..2241013Y}: 
by observing neutron-star--intermediate-mass-black-hole systems, it can constrain the dipolar radiation predicted in Brans--Dicke scalar-tensor theory (see Section~\ref{sub.quadrupole} in this review) better than LISA, and four orders of magnitude better than solar-system experiments~\cite{2010PThPh.123.1069Y};
it can constrain the speed of GWs, parametrized as the mass of the graviton (Section~\ref{sub.propagation}) three orders of magnitude better than solar-system experiments, although not as well as LISA~\cite{2010PThPh.123.1069Y}; by observing binaries of neutron stars and stellar-mass black holes, it can constrain the Einstein-Dilaton-Gauss--Bonnet~\cite{2012PhRvD..86h1504Y} and dynamical Chern--Simons~\cite{2012PhRvL.109y1105Y} theories;
it can measure the bulk AdS curvature scale that modifies the evolution of black-hole binaries in the Randall--Sundrum II braneworld model~\cite{2011PhRvD..83h4036Y};
and it can even look for extra polarization modes (Section~\ref{sub.polarization}) in the stochastic GW background~\cite{2010PhRvD..81j4043N}.

In the rest of this review we concentrate on the tests of GR that will be possible with low-, rather than mid-frequency GW observatories. Nevertheless, many of the studies performed for LISA-like detectors are easily extended to the ambitious DECIGO program, which would probe GW sources of similar nature, but of different masses or in different phases of their evolution.

\newpage


\section{Summary of Low-Frequency Gravitational-Wave Sources}\label{sec.sources}


The low-frequency GW band, which we define as $10^{-5}\mathrm{\ Hz} 
\lesssim f \lesssim 10^{-1}\mathrm{\ Hz}$, is astrophysically appealing 
because it is populated by a large number of sources, which are 
relevant to understanding stellar evolution and stellar populations, 
to probing the growth and evolution massive black holes (MBH) and their links to 
galaxies, and possibly to observing the primordial universe.  
Space-based interferometric detectors such as LISA and eLISA are 
sensitive over this entire band.  Figure~\ref{fig.discoverySpace} 
illustrates the GW strength, as a function of 
GW frequency, for a variety of interesting sources 
classes.  The LISA and eLISA baseline sensitivities are overlaid, 
illustrating the potential reach of interferometric detectors.  The area above 
the baseline curve is generically called ``discovery 
space;'' the height of a source above the curve provides a rough estimate 
of the expected signal-to-noise ratio (SNR) with which the source could be detected using matched-filtering techniques. SNR and matched filtering are important concepts that underlie much of what will be discussed elsewhere in this review, so we explain them briefly in this section.

In a particular GW search based on the evaluation of a detection ``statistic'' $\rho$,
the detection SNR of a GW signal is defined as the ratio of the expectation value $E[\rho]$ when the signal is present to the root-mean-square average of $\rho$ when the signal is absent.
Typically the output $\mathbf{s}$ of the detector is modeled as the sum of a signal $\mathbf{h}(\vec{\theta})$, depending on parameters $\vec{\theta}$, and of instrumental noise $\mathbf{n}$. It is normally assumed that the noise is stationary and Gaussian, and that it has uncorrelated frequency components $\tilde{n}(f)$. Under these assumptions, the statistical fluctuations of noise are completely determined by the one-sided power spectral density (PSD) $S_h(f)$, which is defined by the equation
\begin{equation}
\langle \tilde{n}(f) \tilde{n}(f')^* \rangle = \frac{1}{2} S_h(f) \delta(f-f'),
\end{equation}
where $\langle \rangle$ denotes the expectation value over realizations of the noise (also known as the ensemble average), and the asterisk denotes complex conjugation. This gives rise to a natural definition of an inner product on the space of possible waveform templates,
\begin{equation}\label{innerprod}
\left(\mathbf{h}_1 | \mathbf{h}_2 \right) = 2 \int_0^\infty \left[ \frac{\tilde{h}_1^* \tilde{h}_2 + \tilde{h}_1 \tilde{h}_2^*}{S_h(f)} \right]\, \mathrm{d}f.
\end{equation}
In particular, the (sampling) probability of a given realization of noise $\mathbf{n}_0$ is just
\begin{equation}\label{noiseprob}
p(\mathbf{n}=\mathbf{n}_0 ) \propto \exp\left[-\frac{1}{2} \left(\mathbf{n}_0 | \mathbf{n}_0 \right)\right] .
\end{equation}
Searching the data for GW signals usually involves applying a linear filter $K(t)$ to compute the statistic
\begin{equation}
W = \int_{-\infty}^\infty K(t) s(t)\, \mathrm{d}t = \int_{-\infty}^\infty \tilde{K}^*(f) \tilde{s}(f)\, \mathrm{d}f.
\end{equation}
Redefining the filter by setting $\tilde{F}(f) = \tilde{K}(f) S_h(f)$ yields the \emph{overlap} $(F | s)$. The corresponding SNR is 
\begin{equation}
\frac{S}{N}= \frac{(F | h)}{\sqrt{\langle (F|n) (F|n) \rangle}} = \frac{(F|h)}{\sqrt{(F|F)}},
\end{equation}
and from the Cauchy--Schwarz inequality we have
\begin{equation}
\left(\frac{S}{N}\right)^2 = \frac{(F|h)^2}{(F|F)} \leq \frac{(F|F) (h|h)}{(F|F)} = (h|h),
\end{equation}
with equality when $F = h$. This shows that the \emph{matched filter} obtained when $F = h$ is the \emph{optimal} linear statistic to search for the signal $h$ in noise characterized by the PSD $S_h(f)$. Furthermore, the \emph{optimal matched-filtering SNR} is just $\sqrt{(h|h)}$. Future references to SNR in this review will always refer to this mathematical object.

\epubtkImage{discoverySpaceeLISANew.png}{%
\begin{figure}[htb]
  \includegraphics[width=\textwidth]{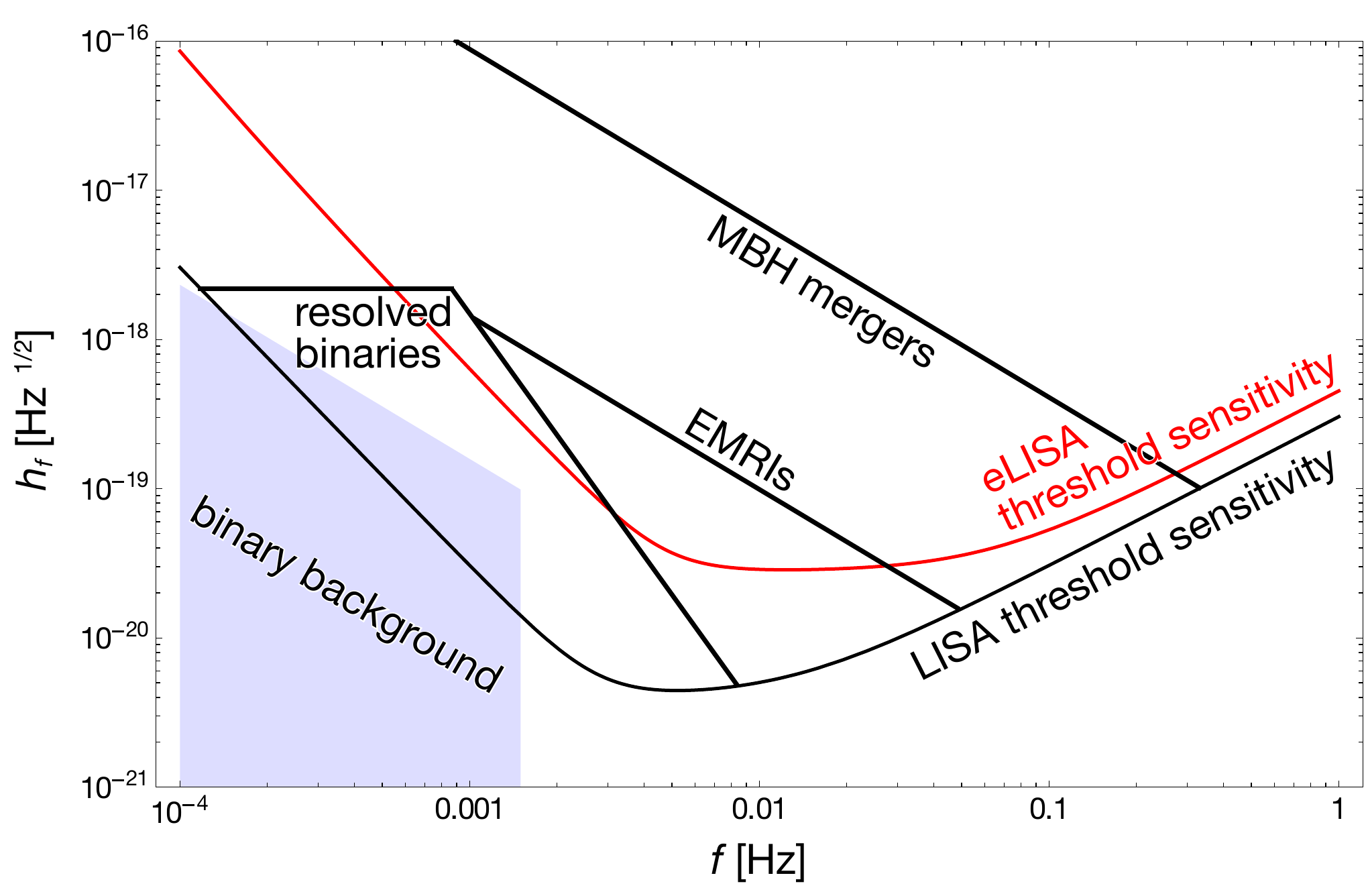}
   \caption{\label{fig.discoverySpace} The discovery space for space-based GW detectors, covering the low-frequency region of the GW spectrum, $10^{-5}\mathrm{\ Hz} \lesssim f \lesssim 0.1\mathrm{\ Hz}$. The discovery space is delineated by the LISA threshold sensitivity curve \cite{SCG} in black, and by the eLISA sensitivity curve in red~\cite{2013GWN.....6....4A} (the curves were produced using the online sensitivity curve and source plotting website~\cite{IoASensCurves}).
   This region is populated by a wealth of strong sources, often in large numbers, including mergers of MBHs, EMRIs of stellar-scale compact objects into MBHs, and millions of close-orbiting binary systems in the galaxy.  Thousands of the strongest signals from these galactic binary systems should be individually resolvable, while the combined signals of millions of them produce a stochastic background at low frequencies.  These systems provide ample opportunities for astrophysical tests of GR for gravitational-field strengths that are not well characterized and studied in conventional astronomy.}
\end{figure}}

The three main types of GW sources in the low-frequency band are
massive black hole (MBH) binaries, extreme mass-ratio inspirals
(EMRIs), and galactic binaries.

MBH binaries comprise two supermassive and nearly equal-mass black holes. These systems typically form following the merger of two galaxies, when the MBHs originally in the centers of the two pre-merger galaxies reach the center of the merged galaxy and form a binary. These binaries generate very strong GW emission and can be detected by space-based GW detectors out to cosmological distances. The signals will be observed from the time they enter the detector band, at $\sim 10^{-4}$ Hz, until the signal cuts off after the objects have merged. The systems evolve through an inspiral, as the objects orbit one another on a nearly circular orbit of gradually shrinking radius, followed by the merger, and then the ringdown. This last phase of the signal is generated as the black-hole merger product, which is initially perturbed in a highly asymmetric state, settles down to a stationary and axisymmetric configuration. During the inspiral phase the field strength is moderate and the velocity is low compared to the speed of light, so the system can be modeled using post-Newtonian theory~\cite{1990PhRvD..42.1123L,1992PhRvD..46.1517W,1993PhRvD..47.3281K, 1993PhRvD..48.4757W,lrr-2006-4}. During the merger the system is highly dynamical and requires full nonlinear modeling using numerical relativity~\cite{2005PhRvL..95l1101P,2006PhRvL..96k1101C,2006PhRvL..96k1102B,2009PhRvD..79b4003S}. The final ringdown emission can then be computed using black-hole perturbation theory, since the perturbations of the object away from stationarity and axisymmetry are small.

There have been several attempts to develop models that include all three phases of the emission in a single framework. The effective one-body (EOB) model was initially developed analytically, and is based on the idea of modeling the dynamics of a binary by describing the relative motion of binary components as the motion of a test particle in an external spacetime metric (the metric of the ``effective'' single body)~\cite{1999PhRvD..59h4006B,2000PhRvD..62f4015B,2001PhRvD..64l4013D}. The EOB model includes a smooth transition to plunge and merger, and then a sharp transition to ringdown. The model incorporates a number of free parameters that have now been estimated by comparison to numerical-relativity simulations~\cite{2009PhRvD..79l4028B,2013PhRvD..87h4035D}. The initial ``effective-one-body numerical-relativity'' (EOBNR) model described non-spinning black holes, but the formalism has now been extended to include the effects of non-precessing spins~\cite{2010PhRvD..81h4041P,2011PhRvD..84j4027B,2012PhRvD..86b4011T,2013PhRvD..87l4036B}. The other model that includes all phases of the waveform is the phenomenological inspiral-merger-ringdown (pIMR) developed by Ajith et al.~\cite{2008PhRvD..77j4017A,2011PhRvL.106x1101A}. This model was constructed by directly fitting an ansatz for the frequency-domain waveform, which was motivated by analytic and numerical results, to the output of numerical relativity simulations. This model has also now been extended to include the effect of non-precessing spins~\cite{2010PhRvD..82f4016S}.

EMRIs consist of stellar-mass ($\sim0.5\mbox{\,--\,}50\,M_{\odot}$) compact objects, either white dwarfs, neutron stars, or black holes, that orbit MBHs. These are expected to occur in the centers of quiescent galaxies, in which a central MBH is surrounded by a cluster of stars. Interactions between these stars can put compact objects onto orbits that come very close to the MBH, leading to gravitational capture. EMRIs are not as strong GW emitters as MBH binaries, but are expected to be observable (if sufficiently close to us) for the final few years before the smaller object merges with the central MBH, when the emitted GWs are in the most sensitive part of the frequency range of space-based detectors. EMRIs will also undergo inspiral, merger, and ringdown, but the signals from the latter two phases are likely to be too weak to be detected. During the inspiral, EMRIs are also expected to be on eccentric and inclined orbits, which colors the emitted GWs with multiple frequency components. During an EMRI, the smaller object spends many thousands of orbits in the strong-field regime, where its velocity is a significant fraction of the speed of light, so post-Newtonian waveforms are inapplicable. However, the extreme mass ratio means that black-hole perturbation theory can be used to compute waveforms, using the mass ratio as an expansion parameter~\cite{poissonLR,2003LRR.....6....6S}.

Stellar binaries in our own galaxy with two compact object components will also be sources for space-based detectors if the binary period is appropriate ($\sim 10^2\mbox{\,--\,}10^4\mathrm{\ s}$). The binary components must be compact to ensure that the binary can reach such periods without undergoing mass transfer. Theoretical models and observational evidence suggest that there may be many millions of such binaries that are potential sources. These systems are not expected to evolve significantly over the typical lifetime of a space mission, and will therefore produce continuous and mostly monochromatic GW sources in the band of space-based detectors. For a small number of systems it will be possible to observe a linear drift in frequency over the observation time, due to either GW-driven inspiral or mass transfer. These systems remain in the weak-field regime through their observation, and their emission can be represented accurately using the quadrupole formula \cite{PM,PM63}.

In addition to what they will teach us about astrophysics, all these sources are prospective laboratories for testing gravitation. The different character of sources in each class provides different opportunities. MBH
binaries are strong-field systems that yield GW signals with large SNRs, making signal
detection and characterization less ambiguous than for weaker sources.
This will allow detailed explorations of waveform deviations from the
predictions of GR.  EMRIs provide detailed probes of MBH spacetimes, thanks to the large number of strong-field waveform cycles that will be observable from these systems.  Individually-resolvable compact galactic binaries can also be used to test GR, because their waveforms and evolution are well understood and easily described using model templates. In addition, many such binaries may be detectable with both GW and EM observatories, providing opportunities to test the propagation of GWs relative to EM signals.

Many approaches to constraining alternative theories are
proposed as \textit{null experiments}, where the assumption is that
GW observations will validate the predictions of
GR to the level of the detector's instrumental noise.
The size of residual deviations from GR is then constrained by the 
size of the errors in the GW measurement. However, in order to perform these null experiments, it is necessary to have accurate models for the GWs predicted in GR. The appropriate modeling scheme, assuming GR and a system in vacuum, depends on the system under consideration, as described above. Additional modeling complications arise from astrophysical phenomena, since tidal coupling, extended body effects, and mass transfer can all
leave an impact on source evolution, complicating the interpretation of the observed GWs.

The very capabilities of space-based detectors also introduce complications: in contrast to the rare appearance of GW signals in the output of ground-based interferometers, the data from space-based missions will contain the superposition of millions of individual continuous signals. We expect that we will be able to
resolve individually thousands of the strongest sources. Thus, we
must deal with problems of \emph{confusion} (where the presence of many interfering sources 
complicates their individual detection),
\emph{subtraction} (where strong signals must be carefully modeled
and removed from the data before weaker sources, overlapping in
frequency, become visible), and \emph{global fit} (where the
parameters of overlapping sources have correlated errors that must be
varied together in searches and parameter-estimation studies). These
issues have not escaped the attention of LISA scientists, and have
been tackled both theoretically \cite{dastatus}, and in a practical
program of mock data challenges \cite{2008CQGra..25r4026B,mldc3,2009CQGra..26i4024V}. All these issues in data analysis and modeling will impact the prospects for tests of GR using the observations. Little work has been done to date to estimate these impacts, but we will discuss relevant studies where appropriate.

In the remainder of this section we will briefly describe the three principal source classes expected in the low-frequency band. For each source type, we will discuss the astrophysics of the systems, the estimated event rates for (e)LISA observations, their potential astrophysical implications, and their applications to testing GR, with pointers to later sections that describe the tests in more detail. We focus on the three types of source that we introduced above and that are most likely, from an astrophysical point of view, to be observed by LISA-like observatories: MBH coalescences, EMRIs, and galactic binaries.

It is possible that a space-based detector like LISA or eLISA could detect other sources that could be used for tests of relativity. For instance, if intermediate-mass black holes with mass between $100M_\odot$ and $10^4M_\odot$ do exist, they could be observed as intermediate-mass-ratio inspirals (IMRIs) when they inspiral into the MBHs in the centers of galaxies \cite{ASReview}. These systems would have the potential for the same kind of tests of fundamental physics as EMRIs, but with considerably larger SNR at the same distance and hence would be observable to much greater distances. Cosmological GW backgrounds might also be observed, generated by processes occurring at the TeV scale in the early universe, which would provide constraints on the physics of the early universe and inflation. We refer the reader to \cite{2013GWN.....6....4A,2012CQGra..29l4016A} for discussions of these sources. We regard them as somewhat more speculative than the other three source types and we do not consider them further here.

   \subsection{Massive black-hole coalescences}\label{sub.mbh}


Most (if not all) galactic nuclei come to harbor a MBH during their evolution~\cite{1971MNRAS.152..461L,1984ARAA..22..471R}, 
and individual galaxies are expected to undergo multiple mergers over their lifetime. It follows that the formation of MBH binaries following galaxy mergers is an expected outcome. The mergers of such binaries will be among the strongest sources of low-frequency GWs.  The rate at which MBH
binaries merge in the universe is uncertain at best, but these events will be
detectable by LISA-like detectors to extremely large distances, probing an enormous
volume of the visible universe.  The detection of \textit{any} MBH
mergers, even at a low rate, would produce interesting astrophysical
results.

For a circular binary, the mass of a system that merges at a frequency $f_{\mathrm{m}}$ is roughly
\begin{equation}
    M_{t} \approx 10^{5}\,M_{\odot}\left(\frac{40\,\mathrm{\ mHz}}{f_{m}}\right);
    \label{LISAmass}
\end{equation}
the frequency range accessible to space-based GW detectors extends from a few $10^{-4}\mathrm{\ Hz}$ to a few $10^{-1}\mathrm{\ Hz}$, which sets the sensitive mass range to $\sim 10^4\mbox{\,--\,}10^7\,M_{\odot}$.

\paragraph*{MBH mergers as low-frequency GW sources.} Predictions for the observable population of MBH mergers are
based on merger-tree structure-formation models. The overall merger rate depends on the detailed
mechanisms of evolutionary growth of the MBH
population.  The energy budgets of active galactic
nuclei suggest that MBHs could grow by efficient accretion
processes~\cite{YuTremaine2002}, while other considerations suggest that mergers could
contribute significantly to their early growth
\cite{2003MNRAS.340..647I}.  Studies of early cosmic structure
\cite{MadauRees2001} indicate that MBHs form from the
coalescence of many smaller \emph{seed} black holes.  Many models based on
this idea have been developed and simulated numerically
\cite{HaehneltKauffman2000,VMH2003,TanakaHaiman2009,BVR2006}. The number of events
detectable by LISA was estimated to be in the range 3\,--\,300 per year~\cite{SVH09}, with a spectrum of masses in the range $10^3\mbox{\,--\,}10^7\,M_{\odot}$. The predicted event rate is not much different for eLISA~\cite{2012CQGra..29l4016A}, although some of the marginally detectable events involving lighter black holes at high redshift will no longer be observable.

Figure~\ref{fig:eLISA-SNRctr} shows contours of constant SNR (as seen by eLISA) in the redshift--total-mass plane for equal-mass, nonspinning MBH mergers  (left panel) and in the total-mass--mass-ratio plane for MBH mergers at a fixed redshift of $z=4$ (right panel).  For typical events at $z\sim4$, the SNRs of eLISA
observations are expected to exceed 100 for total masses $\sim 5\times10^5\,M_{\odot}$.
For comparable-mass mergers at lower redshifts, eLISA observations could exceed
SNRs of 1000. SNRs for LISA are typically factors of 2\,--\,3 higher.
For total mass $>10^5\,M_{\odot}$, a significant fraction of this SNR comes from the final merger and ringdown phase.

\epubtkImage{NGOMBHSNR_Mz_q1-NGOMBHSNR_Mq_z4.png}{%
\begin{figure}[htb]
\centerline{
\includegraphics[width=0.45\textwidth]{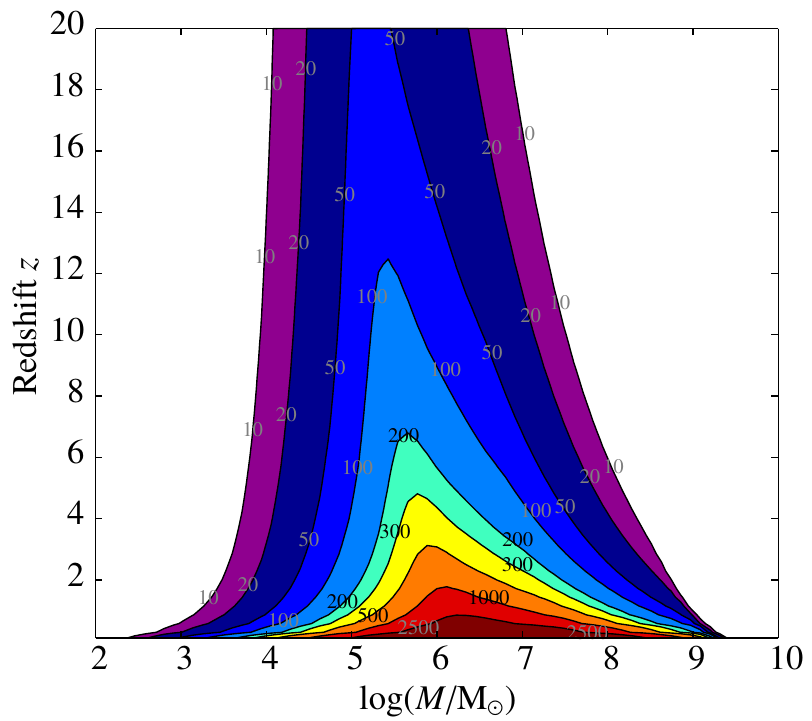}
\includegraphics[width=0.45\textwidth]{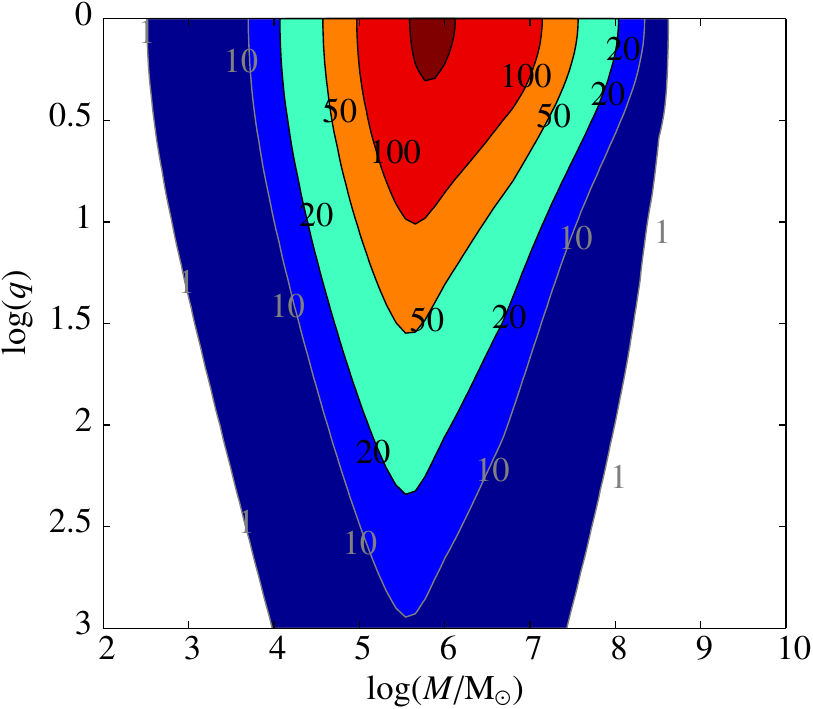}
}
\caption{Contours of constant SNR for MBH binaries observed with eLISA. The left-hand panel shows contours in the total-mass--redshift plane for equal-mass binaries, while the right-hand panel shows contours in the total-mass--mass-ratio plane for sources at redshift $z=4$. Image reproduced by permission from~\cite{2013GWN.....6....4A}.}
\label{fig:eLISA-SNRctr}
\end{figure}}

The practical detection of MBH systems will have to rely on careful data analysis. Several detection algorithms for MBH binaries have been studied by a number of different groups, encouraged in part by the Mock LISA Data Challenges~\cite{mldc3}. These include Markov Chain Monte Carlo techniques~\cite{2007CQGra..24.5729C}, time-frequency analysis~\cite{2007CQGra..24S.595B}, particle-swarm optimization~\cite{2009CQGra..26v5004G}, nested sampling~\cite{2009CQGra..26u5003F}, and genetic algorithms~\cite{Petiteau2010}. The most recent mock-data challenge included multiple spinning MBH binary signals in the same dataset, and multiple groups demonstrated their ability to recover these binaries from the LISA data stream~\cite{mldc3}.

\paragraph*{Astrophysics using MBH mergers.} GW observations will 
directly probe the character of the black holes on scales
comparable to the size of the event horizons. How the information
about the black holes (such as their spin and mass) is encoded in the waveform
is a topic of much focused research (see, e.g.,~\cite{PabloReview,ShapiroReview,2010RvMP...82.3069C,2013CRPhy..14..306S} for reviews). 

According to GR, any astrophysical black hole is fully described by just ten numbers, encoding its mass, position, momentum, and spin.  Correspondingly, a binary is fully described by 20 parameters, three of which, related to its center-of-mass momentum, cannot be measured from the GW signal.  Einstein's theory of gravity thus predicts a 17-dimensional parameter space of possible signals.  
The precision with which GW observations will be able to extract the
parameters characterizing a MBH binary system has been the subject of extensive investigation. The first 
comprehensive study~\cite{Cutler98} considered the LISA determination of
the sky location, luminosity distance, and mass of MBH systems,
using a simple model of LISA's orbital motion around the sun.
Later studies~\cite{MooreHellings} considered the angular
resolution of detectors in \textit{precessing-plane} configurations
like LISA's, as well as \textit{ecliptic-plane} interferometers that
lie flat in the ecliptic as they orbit the sun.  More recent
studies have considered more complete MBH-binary waveforms with spin effects, higher harmonics, and with merger and ringdown signals, and have looked at the effect of these corrections on LISA's parameter-estimation ability~\cite{Vecchio2004,2007PhRvD..76j4016A,2008PhRvD..77b4030T,2008PhRvD..78f4005P,2009CQGra..26i4027A,2009PhRvD..80f4027K,2011PhRvD..84f4003M}. These papers indicate that observations with a LISA-like detector will be able to determine the masses of the two binary components to 0.01\,--\,0.1\%, the spins to 0.001\,--\,0.1\%, the sky position of the binary to 1\,--\,10~deg\super{2}, and the luminosity distance to 0.1\,--\,10\%. For eLISA, the corresponding results are 0.1\,--\,1\%, 1\,--\,10\%, 10\,--\,100~deg\super{2} and 10\,--\,100\%. Such precise measurements will provide unprecedented information about MBHs in the universe.

These parameter-estimation studies witness the
quality of the information that will be available to astrophysicists for
the phenomenological exploration of the \textit{astrophysical
character} of MBH systems.  Of particular interest is the
evolution of black-hole spins and the final spin of merger remnants~\cite{2008ApJ...684..822B,HughesBlandford}, which GW observations should
be able to shed some light on.
In addition, the set of detected MBH merger systems will provide constraints on the formation history of MBHs~\cite{PJHLT2010,2011MNRAS.415..333P,hughesMerger,Menou2001,2011ApJ...732...82P}. In~\cite{MBHmodelA,MBHmodelB} it was shown that a detector like LISA would be able to tell with high confidence the difference between ten different models for the growth of structure in the universe. For many of the models, the difference would already be apparent after as little as three months of data collection. A similar differentiation among models would also be possible with eLISA~\cite{2013GWN.....6....4A}, and both detectors will also be able to constrain ``mixed models'' combining different MBH population models. MBH binaries can also be used as standard candles to probe cosmological evolution~\cite{hughesHolz}. 


\paragraph*{Testing relativity using MBH mergers.} MBH mergers are good laboratories for testing relativity because of the high SNR expected for the signals and the correspondingly large volume within which they can be observed. This makes them good systems for constraining GW polarizations (see Section~\ref{sub.polarization}) since relatively weak alternative-polarization signals could in principle be detected. They are also excellent systems for testing GW-propagation effects, such as subluminal propagation speeds and the presence of parity violations, since these effects accumulate with distance and MBH binaries can be seen to very high redshifts (see Section~\ref{sub.propagation}). MBH binaries are also well suited for generic tests of GR based on measuring the evolution of inspiral phasing and checking it for consistency with general-relativistic predictions, since the high SNR allows a large number of phasing parameters to be measured (see Section~\ref{frameworks}).

MBH binaries are also the only systems for which the quasinormal ringdown radiation, generated as the highly-distorted post-merger black hole settles down into a quiescent state, will be detectable. Although the amount of energy deposited into different ringdown modes is not well understood, the frequencies and decay rates of the ringdown modes are predicted precisely in GR from perturbation theory. The mass and spin of the final black hole can be measured from a single ringdown mode; if multiple modes are detected, the system can be used to check for consistency with the Kerr metric, and therefore provides a probe of black-hole structure. For a comprehensive review of black-hole ringdowns, we refer the reader to~\cite{QNMKokkotas,QNMNollert,2009CQGra..26p3001B}. A more detailed discussion of the use of ringdown radiation to probe black-hole structure may be found in Section~\ref{sec:ringdowntests}.

Last, the space-based GW observation of MBH binaries offers an unprecedented opportunity to probe the fully-dynamic strong-field regime of GR by comparison of the observed merger signals to the predictions of numerical relativity. This is an area that has not yet been explored in any detail and so we will not discuss it further in this review. However, it is an important subject that should be carefully explored before a space-based GW detector finally becomes a reality.



   \subsection{Extreme-mass-ratio inspirals}\label{sub.emri}

An EMRI is the
GW-emitting inspiral of a stellar-mass compact object, either a black
hole, neutron star or white dwarf, into a
MBH in the center of a galaxy. A main-sequence
star with mean density $\bar{\rho} = 10\mathrm{\ g\ cm}^{-3}$ will be tidally disrupted by a black hole of mass $10^{6}\,M_{\odot}$ at a distance of $\sim$~25 Schwarzschild radii~\cite{freitag03}. At such separations an extra-galactic EMRI system will not be generating detectable amounts of gravitational radiation in the low-frequency band. The tidal-disruption radius increases with decreasing stellar density and decreasing central--black-hole mass, and the reference values used above are at the upper end of suitable values for main-sequence stars and mHz GW sources. Therefore, it is not expected that the inspiral of a main-sequence star will be a candidate for an EMRI, with the possible exception of such sources in the galactic center, which is sufficiently nearby that radiation from an object orbiting at several tens of Schwarzschild radii might be detectable~\cite{freitag03}.

EMRIs occur in the dense stellar clusters
that are found surrounding the black holes in the cores of galaxies, and
are triggered by a variety of processes:
\begin{itemize}
\item \textbf{Direct capture:} This is the ``standard''
scenario, in which two-body encounters in the stellar cluster gradually perturb
the compact-object orbits, changing their angular momentum.  This
can put a compact object onto an orbit that passes very close to the central
black hole.  The orbit loses energy and angular momentum in bursts of
gravitational radiation emitted near periapse, and if sufficient
energy is radiated, the object can be left on an orbit that is bound to
the central black hole.  It will then gradually inspiral into
the central black hole as it loses orbital energy and angular momentum
to GW emission.
\item \textbf{Tidal splitting of binaries:} Galactic-center stellar clusters will also contain binaries, which can similarly be displaced onto orbits that pass close to the central MBH. If this happens, the binary will typically be disrupted, with one object remaining bound to the MBH and the other becoming unbound and being flung out with high velocity. If this happens to a binary containing one or two compact objects, the captured compact object will be left on a fairly tightly bound orbit a few hundred astronomical units from the MBH, and will become an EMRI on a nearly circular orbit~\cite{AS228}.
\item \textbf{Tidal stripping of giant stars:} Giant stars typically have a compact core surrounded by a diffuse hydrogen envelope. If such a star passes close to a MBH, the envelope can be partially or completely stripped by the tidal interaction with the MBH. This will deposit the dense core, which is essentially a white dwarf, on a close orbit about the MBH; the system will eventually become an EMRI~\cite{AS74,AS77},
\item \textbf{In-situ formation:} If a MBH has a massive accretion disc, the disc can become unstable to star formation. Stars formed in such a way are biased toward higher masses and will therefore tend to form black-hole remnants. These remnants will be in circular, equatorial orbits around the central MBH, and will inspiral as EMRIs under GW emission.
\end{itemize}
More details on all of these processes and further references can be found in~\cite{ASReview,PauLRR}.

\paragraph*{EMRIs as low-frequency GW sources.} The intrinsic rate at which EMRIs occur
depends on a lot of astrophysics that is rather poorly
understood. The rate is strongly influenced by several processes, including:
\begin{itemize}
\item \textbf{Mass segregation:} Stellar encounters tend on average to lead to equipartition of energy. This means that after a two-body interaction the heavier object tends to be moving slower, and the lighter object tends to be moving faster. If this occurs for two objects around a MBH, the heavier object sinks in the potential of the MBH and the lighter object moves further out. Components of the stellar distribution thus become segregated according to their mass. The more massive objects, which tend to be the black holes, end up closer to the MBH and are captured preferentially~\cite{ASReview,AS108,AS233}.
\item \textbf{Triaxiality:} The centers of galaxies tend to be approximately spherically symmetric, but on larger scales galactic nuclei can be triaxial. Orbits in triaxial potentials tend to be centrophilic: they have a tendency to pass close to the center. This process can funnel stars that are further away from the MBH toward the center and hence increase the inspiral rate~\cite{AS160,AS258}.
\item \textbf{Resonant relaxation:} In the standard relaxation picture, each encounter is random and uncorrelated, so stars undergo a random walk. The process is driven by the diffusion of energy which then leads to angular-momentum transfer. However, in a stellar cluster around a MBH, each star will be on a Keplerian orbit, which is a fixed ellipse in space. The orbits of two nearby stars will thus exert correlated torques on one another, which can lead to a direct resonant evolution of the angular momentum~\cite{AS271,AS272}. This process can lead to a much more rapid diffusion of orbits into the regime where they can become EMRIs.
\item \textbf{The ``Schwarzschild barrier'':} Resonant relaxation relies on the orbits having commensurate radial and azimuthal frequencies, so they remain in fixed planes over multiple orbits. In the strong-field potential of a massive object, orbits are no longer Keplerian but undergo significant perihelion precession. Resonant relaxation is only efficient in the regime where precession is negligible. The ``Schwarzschild barrier'' refers to the boundary between orbits with and without significant precession. Inside this point resonant relaxation is strongly quenched, potentially reducing inspiral rates~\cite{SchBarr}.
\end{itemize}
A full description of these processes and their relative importance can be
found in~\cite{ASReview,PauLRR}. The range of EMRI rates per galaxy reported
in the literature is $\sim1\mbox{\,--\,}1000\mathrm{\ Gyr}^{-1}$ for black holes, and
$\sim10\mbox{\,--\,}5000\mathrm{\ Gyr}^{-1}$ for white dwarfs, with most recent
best-guess estimates of $\sim 400\mathrm{\ Gyr}^{-1}$ / $20\mathrm{\
Gyr}^{-1}$ / $7\mathrm{\ Gyr}^{-1}$ for
black holes / white dwarfs / neutron stars~\cite{HopmanLISA7,2011CQGra..28i4017A}. 

Uncertainties in the event rate come not only from the intrinsic
rate of capture, but also from uncertainties in the number density of black holes of suitable mass.  At present, only three objects in the range
$10^4\mbox{\,--\,}10^7\,M_{\odot}$ are known from kinematic
measurements~\cite{AS90}.  Using galaxy luminosity functions, the
luminosity--velocity-dispersion relation and the MBH-mass--velocity dispersion relation, one infers the number density of
MBHs in the LISA mass range to be roughly constant per logarithmic
mass unit, and equal to a few $10^{-2}\mathrm{\ Mpc}^{-3}$.  However, when black holes merge, the asymmetric radiation of linear momentum over the final few orbits imparts linear momentum (a ``merger kick'') to the remnant black hole. These kicks can be large enough to eject the black hole from its host galaxy. If these black-hole merger kicks are significant for black holes merging in the centers of galaxies, the number of black holes with $M < 2 \times10^6\,M_{\odot}$ may therefore be significantly depleted~\cite{AS76}.

EMRI waveforms are complex, since the orbits are expected to be both
eccentric and inclined with respect to the equatorial plane of the central black hole. Various algorithms have
been suggested for EMRI detection, including
hierarchical matched filtering starting with short data
segments~\cite{emrirate}, time-frequency
analysis~\cite{tfWenGair,tfGairWen,tfGairJones}, searches with phenomenological templates~\cite{2012PhRvD..86j4050W}, and Markov Chain
Monte Carlo searches~\cite{mcmcstroeer,mcmcBGP,mcmcCornish}. Indeed, EMRIs have proven
to be the most difficult source to detect in the Mock LISA Data Challenges, but in the third challenge round two groups correctly
identified the three EMRIs with the highest-mass central black
holes~\cite{mldc3}. Several challenging aspects remain unsolved, such as dealing
with source confusion and detecting the EMRIs with lower-mass central
black holes, which are expected to dominate the data; recent progress encourages confidence that these problems will be solved 
by the time a space-based detector is launched.

Assuming that a matched-filtering SNR of 30 is required for EMRI detection, and using the previously-quoted EMRI rates per black hole~\cite{HopmanLISA7}, it was
estimated that LISA could see several hundred events~\cite{2009CQGra..26i4034G}
out to a redshift $z \lesssim 1$. The intrinsic rate uncertainties
mean that the actual number could be anywhere from a few tens to a few thousands. Most
of the events will be black-hole inspirals into MBHs of mass
$10^5\mbox{\,--\,}10^6\,M_{\odot}$. The estimate for eLISA is a few tens of events~\cite{eLISAEMRI,2012CQGra..29l4016A,2013GWN.....6....4A}, although this is based on a lower SNR detection threshold of 20. (The lower threshold is justified by the results of the Mock LISA Data Challenges, which successfully demonstrated detection of EMRI events with SNR of $\sim$~15, albeit with less-than-realistic signal confusion.)

\paragraph*{Astrophysics using EMRIs.}
EMRI observations will tell us about the properties of MBHs at low redshift and about the physics of nuclear stellar
clusters. Each EMRI observation can determine many of the parameters of the system, including the mass and spin of the central black hole, to accuracies of a fraction of a percent~\cite{AK,2009PhRvD..79h4021H}. With as few as ten EMRI observations, LISA would be able to
constrain the slope of the MBH mass function at low redshift
to $\sim\pm 0.3$, which is the level of the current observational
uncertainty~\cite{GTVemri}. eLISA would have the same capability given the same number of detections. If LISA observed as many EMRIs as current
estimates predict, the constraints on the slope would improve by an order of magnitude, although the corresponding improvement for eLISA would be only a factor of two. EMRIs
will also provide information on the distribution of black-hole
spins. The masses of the compact objects observed in EMRIs will
provide information about mass-segregation processes, while the
overall number of events encodes information about the complicated
physics that determines the event rate. Finally, the eccentricities
and inclinations of the observed EMRIs will provide information on
the relative efficiencies of the various channels through which EMRIs
could be formed~\cite{ASReview,PauLRR}.

\paragraph*{Testing relativity using EMRIs.}
EMRIs are well suited for testing relativity for several reasons:
the systems are expected to be clean since the small object is likely
to be a black hole with no internal structure, and the influence of
any gaseous material in the spacetime is expected to be negligible
(but see Section~\ref{sec.BHstructure}); the waveforms can be computed very
accurately using black-hole perturbation
theory~\cite{poissonLR,2007PhRvD..75f4021B}, allowing very small deviations
from the model to be identified; the inspiral is slow, meaning that
many waveform cycles are generated in the strong-field regime, and can
be used to test the spacetime structure~\cite{FinnThorne}; and the
waveform structure is complex, due to the expected eccentricity and
inclination of the orbits, and to the intrinsic richness of dynamics in Kerr (or Kerr-like) spacetimes~\cite{ASReview,DrascoGenEMRI}. 

The primary fundamental-physics application of EMRI observations is
probing black-hole structure, since the EMRI waveforms encode a
map of the spacetime geometry close to the central black hole. This
can be used to test whether the central object is indeed described by
the Kerr metric, as we expect, or is some other kind of object.
A discrepancy could indicate an astrophysical perturbation in the system, a violation of the cosmic-censorship
hypothesis~\cite{penrose69} or of the energy conditions of GR, or even evidence for the existence of an exotic massive compact object such as a boson star. All of this will be discussed
in detail in Section~\ref{sec.BHstructure}. EMRIs can also be used for tests
of GW polarization (Section~\ref{sub.polarization}), and to constrain alternative
theories of gravity through detection of modifications to the
inspiral rate (see Section~\ref{sub.quadrupole}).



   \subsection{Galactic binaries}\label{sub.binaries}



The most numerous source class in the low-frequency GW band are the ultra-compact binaries
of two stellar-mass compact objects: white dwarfs, neutron
stars, or black holes.  Most of these will be in the Milky Way.
Numerical population synthesis~\cite{nypz2001a,nypz2001b} and
theoretical studies of the highly-evolved stellar
population in the galaxy~\cite{IbenTutukov1984,IbenTutukov1986} predict that there
will be $\sim 10^{7}$ ultra-compact binaries generating GWs in the mHz band visible to space-based detectors like LISA or eLISA.

\paragraph*{Ultra-compact binaries as low-frequency GW sources.} For an instrument with sensitivity similar to the ``classic'' LISA design, the complete population of compact galactic binaries would separate into two
distinct observational classes: \textit{confused binaries} and
\textit{resolvable binaries}.  Confused binaries are those with
orbital frequencies that are close enough to other
binaries in the population that they cannot be readily distinguished
from one another in the data~\cite{HilsBender1997,
HilsBender2000,CornishLarson2003}; their signals are overlapping and cannot be disentangled from one another, unless one source happens to be much closer and stronger than the others. For LISA, the confused binaries make up a stochastic foreground signal from the galaxy, which would be
present in the data for all frequencies below the confusion limit $f_{c} \simeq 3\mathrm{\ mHz}$. Various attempts have been made to estimate the spectrum of this foreground~\cite{nypz2001c,trc, EdlundPRD,EdlundCQG,RBBL,2012ApJ...758..131N}, but the final profile would also depend on details of the algorithms used to analyze the LISA data~\cite{LarsonFinn2006,CornishLarson2003,stroeer2007,
CrowderCornish2007,2010PhRvD..81f3008B,2011PhRvD..84f3009L}. For an instrument with a sensitivity comparable to eLISA, which is approximately a factor of two worse in the relevant range, this confusion foreground would be much less prominent, and perhaps not detectable~\cite{2012CQGra..29l4016A,2013GWN.....6....4A}.

Resolvable binaries are found both above and below the $f_c$ confusion limit. Sources below $f_c$ are binaries that happen to be nearby, and so dominate the emission near their central frequency. For sources above $f_{c}$, the characteristic orbital frequencies are separated by a few frequency bin widths $\Delta f \sim 1/T_{\mathrm{obs}}$, where
$T_{\mathrm{obs}}$ is the length of observation. An instrument like LISA would be expected to resolve as many as twenty thousand individual binaries~\cite{nypz2001c,2012ApJ...758..131N}, while for eLISA the number is expected to be a few thousand~\cite{2012CQGra..29l4016A,2013GWN.....6....4A,2012ApJ...758..131N}.

Among the resolvable binaries there are several \textit{verification binaries} (see~\cite{sv2006}; for a current list see~\cite{NelemansWiki}).
These are binaries that are expected to be resolvable in the LISA/eLISA data, but have already been detected and characterized using electromagnetic telescopes. These known binaries will provide early verification that the mission is operating correctly at the nominal sensitivity.


\paragraph*{Astrophysics using ultra-compact binaries.} The large
number of compact binaries detectable throughout the entire
galaxy makes these sources an exceptional resource to address
questions about the stellar evolutionary history of the galaxy and the
astrophysical dynamics of strong-field interactions in compact systems.
If the confusion foreground is observed, it will be modulated by detector motion,
a fact that can be used to estimate the multipole moments of the
spatial distribution of binaries in the galaxy
\cite{SetoCooray2004,EdlundPRD}.  A detector like LISA might also be able to detect
the extra-galactic population of ultra-compact binaries, providing a valuable probe of its characteristics~\cite{FarmerPhinney2003,Cooray2004,CooraySeto2005}. However, an extra-galactic population is unlikely to be observed by eLISA. 

Detections of galactic binaries will be dominated by double--white-dwarf binaries, which are scarce in the galaxy, so (e)LISA will be in a unique position to illuminate their physics. This will shed light on their formation and evolution, including the poorly-understood phase of common-envelope evolution. This is the phase during the evolution of a binary that follows unstable mass transfer. In the frame that co-rotates with the binary, the region of space around a star where orbiting material is bound to that star is known as the Roche lobe.
As the orbit of a binary shrinks during inspiral, the Roche lobes also
shrink, and one of them can eventually become smaller than the
corresponding star. At that point the outer layers of the star are no
longer gravitationally bound to the star so they are lost, and this
mass is transferred to the other star in the binary. The process can
be stable if, when the material is transferred, the orbital radius
increases, the Roche lobe increases, and the star therefore no longer
overflows the Roche lobe. Typically as the star loses mass it also
becomes bigger, so stable mass transfer requires the Roche lobe to
increase in size more quickly than the star. Mass transfer can also
occur unstably if, when the material is lost, the orbit shrinks,
leading to further shrinking of the Roche lobe and more mass transfer
(or if the rate at which the star increases in size due to mass loss
is faster than the rate at which the Roche lobe increases). After an
unstable mass-transfer event, the companion star and the core of the
donor star will end up as a binary contained entirely within the
material that was transferred from the donor -- the ``common envelope.'' Subsequently, energy is transferred from the orbit to the envelope, leading to shrinking of the orbit until the objects merge or, in the case of interest for ultra-compact binaries, the envelope has been completely expelled. The physics of this phase of the evolution is very poorly understood, but it is normally characterized in terms of the amount of energy transferred between the binary and the envelope. The distribution of orbital periods and masses for ultra-compact galactic binaries will encode important clues about this process.

(e)LISA is also expected to observe interacting binaries where mass transfer is ongoing. GW observations can shed light on the fraction of mass-transferring systems; these systems may allow coincident electromagnetic observations, making them even more precise probes of astrophysical processes such as mass transfer and tidal coupling~\cite{Sepinsky2007,Willems2008,Sepinsky2009,Willems2010}.

\paragraph*{Testing relativity using ultra-compact binaries.} Resolvable compact binaries offer a good laboratory for testing GR because their waveforms and evolution are well understood and easily described using model
templates, and because a reasonable fraction of them will be visible both to (e)LISA and to electromagnetic telescopes.

Ultra-compact binaries will provide constraints on massive-graviton theories, either by comparison of
GW-propagation delays to the observable
electromagnetic signals from the binary, or by the dispersion of the
GW as the binary slowly evolves. This is discussed in Section~\ref{sub.propagation}. The ultra-compact binaries also provide laboratories for testing waveform 
evolution, and for GW polarization and dispersion studies, precisely because they
evolve slowly and are easily described by post-Newtonian analysis. This is discussed in Sections~\ref{sub.polarization} and \ref{frameworks}.



\newpage


\section{Gravitational-Wave Tests of Gravitational Physics}\label{sec.tests}


Almost since its inception, GR was understood to
possess propagating, undulatory solutions -- GWs,
described at leading order by the celebrated quadrupole formula~\cite{kennefick2007traveling}.
It took several decades to establish firmly that these waves
were real physical phenomena and not merely artifacts of gauge
freedom. 

How would GW observations test the GR description of strong
gravitational interactions, and possibly distinguish between GR and alternative theories?  To answer this question we need to take a quick detour through GW data analysis.  At least for foreseeable detectors,
individual GW signals will typically be immersed in overwhelming
noise, and therefore will need to be dug out with techniques akin to
\emph{matched filtering}~\cite{16155}, which by definition can only
recover signals of shapes known in advance (the \emph{templates}), or
very similar signals.  A matched-filtering search is set up by first
selecting a parameterized template family (where the parameters are the
source properties relevant to GW emission), and then filtering the
detector data through discrete samplings of the family that cover the
expected ranges of source parameters.  The best-fitting templates
correspond to the most likely parameter values, and by studying the
quality of fits across parameter space it is possible to derive
posterior probability densities for the parameters.

After a detection, the first-order question that we may ask is whether the best-fitting GR template is a satisfactory explanation for the measured data, or whether a large \emph{residual} is left that cannot be explained as instrument noise, at least within our understanding of noise statistics and systematics. (Slightly more involved tests are also possible: for instance, we may divide measured signals in sections, estimate source parameters separately for each, and verify that they agree.)
If a large residual is found, many hypotheses would be \emph{a priori} more likely than a violation of GR: the fitting algorithm may have failed; another GW signal, possibly of unexpected origin, may be present in the data; the data may reflect a rare or poorly understood instrumental glitch; the GW source may be subject to astrophysical effects from nearby astrophysical objects, or even from intervening gravitational lenses.

Having ruled out such non-fundamental explanations, the only way to quantify the evidence for or against GR is to consider it alongside an alternative model to describe the data. This alternative model could be a phenomenological one (discussed below) or a self-consistent calculation within an alternative theory of gravity. 
If the alternative theories under consideration include one or more adjustable parameters that connect them to GR (such as $\omega_\mathrm{BD}$ for Brans--Dicke theory, see Section~\ref{standardmodel}), and if those parameters can be propagated through the mathematics of source modeling and GW generation, then GR template families can be enlarged to include them, and the extra parameters can be estimated from GW observations.
These extra parameters may have a more phenomenological character, as would, for instance, a putative graviton mass that would affect GW propagation, without finding direct justification in a specific theory.
Indeed, many of the ``classic tests''
discussed below (Section~\ref{classictests})
fall within this class.
To test GR against ``unconnected'' theories without adjustable parameters, we would instead filter the data through separate GR and alternative-theory template families, and decide which model and theory are favored by the data using \emph{Bayesian model comparison}, which we now describe briefly.

In complex data-analysis scenarios such as those encountered for GW detectors, the techniques of Bayesian inference~\cite{2005blda.book.....G,2006book..Sivia} are particularly useful for making assessments about the information content of data and for studying tests of gravitational theory, where the goal is to examine the hypothesis that the data might be described by some theory other than GR. 
In a traditional ``frequentist'' analysis of data, one computes the value of a statistic and then accepts or rejects a hypothesis about the data (e.g., that it contains a GW signal) based on whether or not the statistic exceeds a threshold. The threshold is set on the basis of a false-alarm rate, which is a statement about how the statistic would be distributed if the experiment was repeated many times. Evaluating the distribution of the statistic relies on a detailed and reliable understanding of the measurement process (noise, instrument response, astrophysical uncertainties, etc.). By contrast, Bayesian inference attempts to infer as much as possible about a \emph{particular} set of data that has been observed, instead of making a statement about what would happen if the experiment were repeated.

Bayesian inference relies on the application of Bayes' 1763 theorem: given the observed data $d$ and a parameterized model $M(\vec{\theta})$, the theorem relates the \emph{posterior} probability of the parameters given the data, $p(\vec{\theta}|d,M)$ to the \emph{likelihood} $p(d | \vec{\theta}, M)$ of observing the data $d$ given the parameters $\vec{\theta}$, and the \emph{prior} probability $p(\vec{\theta}|M)$ that the parameters would take that value:
\begin{equation}
\label{eq:bayestheorem}
p(\vec{\theta}|d,M) = \frac{p(d | \vec{\theta}, M)\,p(\vec{\theta}|M)}{p(d|M)}.
\end{equation}
The term $p(d|M) = \int p(d | \vec{\theta}, M)\,p(\vec{\theta}|M) d\vec{\theta}$ in the denominator is the \emph{evidence} for the model $M$.
While Eq.~\eqref{eq:bayestheorem} follows trivially from the definition of conditional probability, its power comes from the idea of \emph{updating} the prior knowledge of a system given the results of observations. However, its practical application is complicated by the necessity of attributing priors, and the correct evaluation of likelihoods relies on the same detailed understanding of the statistical properties of the measurement noise as in the frequentist case.

The evidence represents a measure of the consistency of the observed data with the model $M$, and can be used to compare two models (e.g., the GR and modified-gravity descriptions of a GW-emitting system) by evaluating the \emph{odds ratio} for model 1 over model 2,
\begin{equation}
{\cal O}_{12} = \frac{p(d|M_1) p(M_1)}{p(d|M_2) p(M_2)},
\end{equation}
where $p(M_1)$ and $p(M_2)$ are the prior probabilities assigned for model 1 and 2 respectively. Either model would be preferred if the odds ratio is sufficiently large/small, but the decision on which hypothesis is best supported by the data is influenced by the choice of the priors, which will reflect the analyst's assessment of the relative correctness of the alternatives (see~\cite{2012PhRvD..86h2001V} for a discussion of this point).

In the absence of well-defined alternative-theory foils, it may be desirable to proceed along the lines of the PPN formalism (Section~\ref{standardmodel}) and immerse the GR predictions in expanded waveform families, designed to isolate differences in the resulting GW phenomenology (Section~\ref{frameworks}). Proposals to do so include schemes where the waveform-phasing post-Newtonian coefficients, which are normally deterministic functions of a smaller number of source parameters, are estimated individually from the data~\cite{2006CQGra..23L..37A,2006PhRvD..74b4006A};
the ambitiously-named parameterized post-Einstein (ppE) framework~\cite{2009PhRvD..80l2003Y};
and the parameterization of Feynman diagrams for nonlinear graviton interactions~\cite{2009PhRvD..80l4035C}.
In Section~\ref{sec:mergerringdown} we discuss ideas (so far rather sparse) to use the GWs from binary mergers-ringdowns to test GR.

We close these introductory comments by discussing two methodological \emph{caveats}.
First, GW observations are often characterized as ``clean'' tests of
gravitational physics -- whereby the ``clean'' emission of GWs from the bulk motion of matter (already emphasized above) is contrasted to ``dirty'' processes such as mass transfer, dynamical equation-of-state effects, magnetic fields, and so on.
An even stronger notion of cleanness is important for the purpose of
testing GR: for the best sources, the waveform signatures of
alternative theories cannot be reproduced by changing the
astrophysical parameters of the system -- this orthogonality is quantified by the \emph{fitting factor} between the GR and alternative-theory waveform families~\cite{2012PhRvD..86h2001V}. The degeneracy of the alternative-theory and source parameters would also lead to a ``fundamental bias.'' Fundamental bias arises from the assumption that the underlying theory in the analysis, generally taken to be GR, is the correct fundamental description for the physics being observed, which will impact the estimation of astrophysical quantities~\cite{2009PhRvD..80l2003Y,2013PhRvD..87j2002V}.

Second, many of the results presented in this section rely on the \emph{Fisher-matrix formalism} for evaluating the expected parameter-estimation accuracy of GW observations~\cite{2008PhRvD..77d2001V}.  As described at the beginning of Section~\ref{sec.sources}, the output of a GW detector is normally modeled as a linear combination of a signal, $\mathbf{h}(\vec{\theta})$, and noise $\mathbf{n}$, $\mathbf{d} = \mathbf{h}(\vec{\theta}) + \mathbf{n}$. If the detector noise is assumed to be Gaussian and stationary, the probability $p(\mathbf{n}=\mathbf{n}_0)$ is given by Eq.~(\ref{noiseprob}). The likelihood $p(\mathbf{d} | \vec{\theta},M)$ is just the probability that the noise takes the value $\mathbf{n} = \mathbf{d} - \mathbf{h}(\vec{\theta})$, which is
\begin{equation}
p(\mathbf{d} | \vec{\theta},M) \propto \exp\left[-\frac{1}{2} \left( \mathbf{d} - \mathbf{h}(\vec{\theta}) \Big | \mathbf{d} - \mathbf{h}(\vec{\theta}) \right) \right],
\end{equation}
where $(\mathbf{a} | \mathbf{b})$ is the inner product defined in Eq.~(\ref{innerprod}). Writing $\mathbf{d} = h(\vec{\theta}_0) + \mathbf{n}$ and assuming that we are close to the true parameters $\vec{\theta}_0$, so that we can use the linear approximation $h(\vec{\theta})=h(\vec{\theta}_0) + \partial_j h \, \Delta \theta^j + \cdots$ (with $\Delta \theta^j = \vec{\theta} - \vec{\theta}_0$), we find that at quadratic order in $\Delta \theta_j$
\begin{equation}
p(\mathbf{d} | \vec{\theta},M) \propto \exp\left[ -\frac{1}{2} {\Gamma}_{jk} \left(\Delta \theta^j - ({\Gamma}^{-1})^{jl} (\mathbf{n}|\partial_l h)  \right) \left(\Delta \theta^k - ({\Gamma}^{-1})^{km} (\mathbf{n}|\partial_m h)  \right) \right],
\end{equation}
where
\begin{equation}
   \Gamma_{ij} = \left(\frac{\partial h}{\partial \theta^{i}} \Big | \frac{\partial h}{\partial \theta^{j}}\right)
   \label{eqn.FisherMtx}
\end{equation}
is the Fisher information matrix. Thus, to leading order the shape of the likelihood in the vicinity of its maximum is that of a multivariate Gaussian with covariance matrix $({\Gamma}^{-1})^{jk}$ (independent of $\mathbf{n}$),  
and the variance of the one-dimensional marginalized posterior probability density of parameter $\theta^i$ is approximately $(\Gamma^{-1})_{ii}$ (no sum over $i$). This will be achieved in the limit of ``high'' signal-to-noise ratio where the errors $\Delta \theta^j$ are small and the linear approximation is valid.
The Fisher matrix arises also as the Cram\'er--Rao lower bound on the variance of an unbiased estimator of the waveform parameters $\vec{\theta}$. A full discussion of the various routes to the Fisher-matrix formula and its applications may be found in~\cite{2008PhRvD..77d2001V}.

As emphasized by one of us~\cite{2008PhRvD..77d2001V}, because the Fisher matrix is built with the first derivatives of waveforms with respect to source parameters, it can only ``know'' about the close neighborhood of the true source parameters. If the estimated errors take the waveform outside that neighborhood, then the formalism is simply inconsistent and unreliable. Higher SNRs reduce expected errors and therefore would generally make the formalism ``safer,'' but the meaning of ``high'' is problem dependent, depending on the number of parameters that need to be estimated, on their correlation, and on the strength of their effects on the waveforms.

In practice, only by carrying out a full computation of the posterior probability using, for example, Monte Carlo methods will it be known if the Fisher matrix is providing a good guide to the shape of the posterior. However, the Fisher matrix is generally much easier to compute than the full posterior, so it is widely used as a guide to the precision with which parameters of the model can be determined. In the context of testing GR, the Fisher matrix can be evaluated for an expanded waveform model that includes non-GR-correction parameters, but at a set of parameters that correspond to GR. The estimated error in the correction parameter, $(\Gamma^{-1})_{ii}$, can then be interpreted as the minimal size of a correction that would be detectable with a GW observation.


\subsection{The ``classic tests'' of general relativity with gravitational waves}
\label{classictests}

As Will points out~\cite[ch.\ 10]{willTEGP}, virtually any Lorentz-invariant metric theory of gravity must predict gravitational radiation, but alternative theories will differ
in its properties. Will identifies three main properties
that can be measured with GW detectors. These are the \emph{polarization},
\emph{speed}, and emission \emph{multipolarity} (monopole, dipole, quadrupole, etc.) of GWs in GR. In this paper, we broaden the scope
of the third to include changes to the loss of energy
to GWs in inspiraling systems.

In analogy to the three classic tests of GR (the
perihelion of Mercury, deflection of light, and gravitational
redshift) we like to refer to the verification that these
properties have the predicted GR values,
rather than the values predicted by alternative theories, as the ``classic tests'' of
GR using GWs.  Just as PPN tests probe weak-field, slow-motion dynamics, these tests
can be seen as probing the weak-field far zone, where waves
have propagated far from their sources. However, the multipolarity of GWs at emission and the energy that they carry away can be influenced by strong-field properties
in the near zone where waves are generated.

\subsubsection{Tests of gravitational-wave polarization}
\label{sub.polarization}

GR predicts the existence of two transverse quadrupolar polarization modes for GWs (also described as ``spin-2'' and ``tensor'' using the language of group theory), usually labeled $h_+$ and $h_\times$.
Alternative metric theories of gravity predict as many as \emph{six} polarizations~\cite{willTEGP} (three transverse and three longitudinal), corresponding to the independent electric-type components of the Riemann curvature tensor, $R_{0i0j}$. Schematically, these components are measured by GW detectors by monitoring the geodesic deviation of nearby reference masses. The effect of different polarization modes is best illustrated by the induced motion of a ring of test particles, as in Figure~\ref{polfig}. The response of a standard right-angle interferometer to a scalar wave is maximal when the wave propagates along one arm; by contrast, tensor modes elicit maximal response when the wave propagates in a direction perpendicular to the plane of the detector.

\epubtkImage{GWPolStates.png}{%
\begin{figure}[htb]
\centerline{\includegraphics[width=4in]{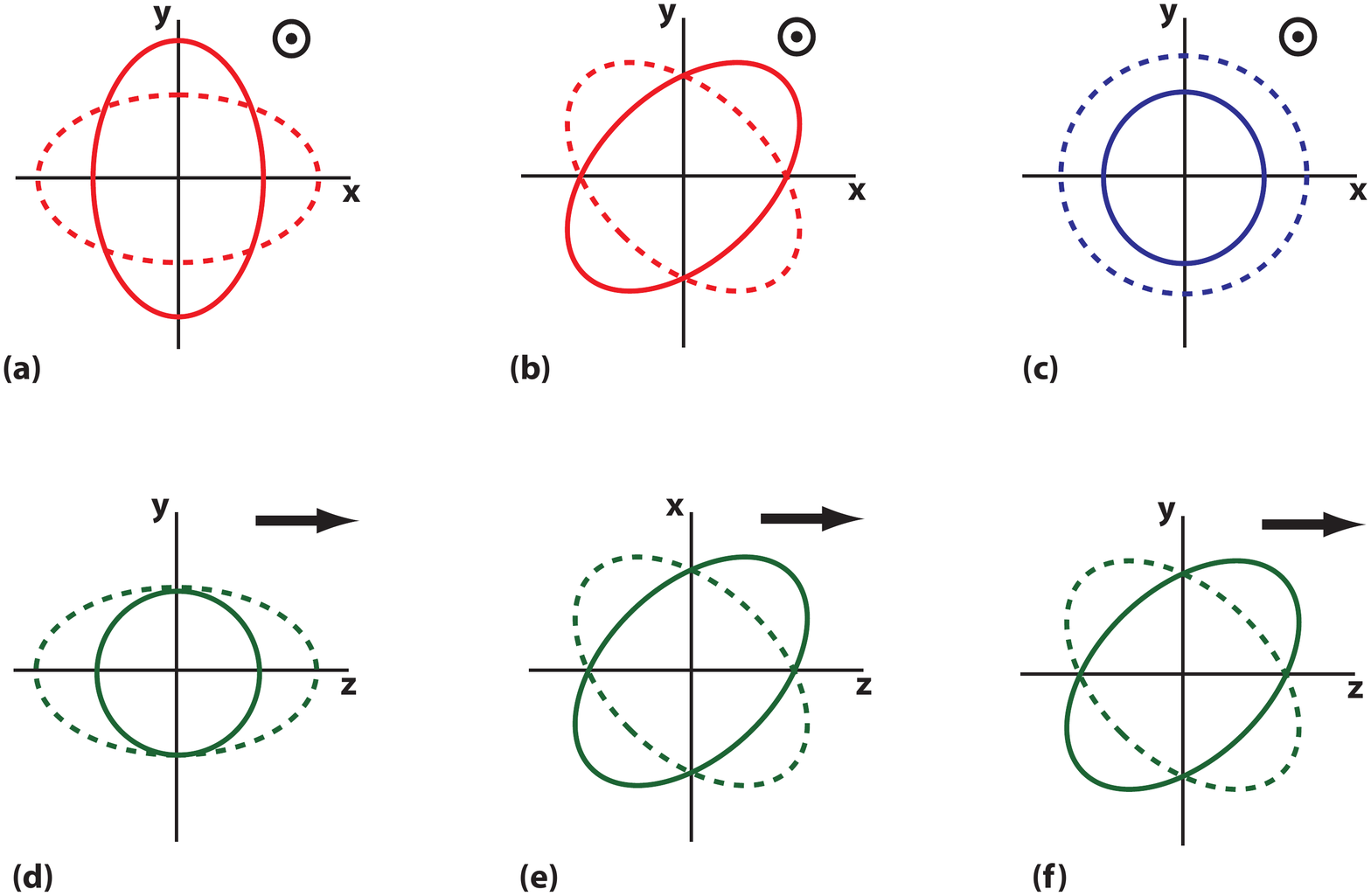}}
\caption{Effect of the six possible GW polarization modes on a ring of test particles. The GW propagates in the \textit{z}-direction for the upper three transverse modes, and in the \textit{x}-direction for the lower three longitudinal modes. Only modes (a) and (b) are possible in GR. Image reproduced by permission from~\cite{willLR}.}
\label{polfig}
\end{figure}}

\paragraph*{Direct detection.}
The use of GW polarization modes to test GR was first proposed in 1973~\cite{eardleyprl,eardleyprd}. The sensitivity of resonant and interferometric detectors, as well as Doppler-tracking and pulsar-timing measurements, to the extra modes was considered in several studies~\cite{1977PhRvD..15..409P,1978PhRvD..17.3158H,1994PhRvD..50.7304S,1997rggr.conf..419W,1998grwa.conf..168L,2000PhRvD..62b4004M,2001PhRvD..63h2001N,2008ApJ...685.1304L,2009PhRvD..79h2002N,2011PhRvD..83l3529A,2012PhRvD..85h2001C,2012PhRvD..85d3005B,
2013PhRvD..87f2003H}.
In the most general setting, the problem of disentangling the modes has eight unknowns -- the time series for the six polarizations, plus two direction angles that affect the projection of the modes on the detector -- but only six observables, corresponding to the $R_{0i0j}$ components. Thus, the problem is indeterminate, unless the source position is known from other observations (such as time-of-flight delays for a long-baseline network of detectors), or unless we restrict GWs to transverse modes on theoretical grounds.

Space-based observatories similar to LISA have either a single independent interferometric observable or three (if laser links are maintained across all three arms). Each observable measures a different admixture of polarizations. Thus, a detector with three active arms could in principle discriminate a non-GR polarization mode if the direction to the source is known, or if it can be determined from the measured signal (by means of the modulations produced by orbital motion, or by triangulation between the signals measured at the three spacecraft).

The LISA sensitivity to alternative polarization modes was assessed in~\cite{LISAaltpolTDI}, using the full TDI response (see Section~\ref{sec:lisa}). At frequencies larger than the inverse light-travel time along the arms, LISA would be ten times more sensitive to scalar-longitudinal and vector modes [(d) to (f) in Figure~\ref{polfig}] than to tensor and scalar-transverse modes [(a) to (c) in Figure~\ref{polfig}], because longitudinal effects can accumulate as the lasers travel between the spacecraft. At lower frequencies, the sensitivity to all modes is approximately the same. These results have not yet been used to work out the constraints that LISA could place on specific alternative theories using different types of sources.

In~\cite{2012CQGra..29g5011A}, a generic model for a system emitting dipole radiation in addition to quadrupole radiation was constructed. The model was similar in structure to the ppE models which will be discussed in Section~\ref{sec:ppE}. This model included both the dipolar component of the waveform, at the orbital frequency, and modifications to the gravitational wave phasing of both the quadrupole and dipole waveform components that arise from the additional energy lost into the dipole mode. In~\cite{2012CQGra..29g5011A}, the model was used to determine the constraints on dipole radiation emission that would be possible using ground-based GW detectors. Results for space-based detectors were included in a subsequent review~\cite{2013IJMPD..2241012A}. This demonstrated that eLISA would be able to place bounds on the parameter $\alpha$, that describes the observed amplitude of the dipole radiation relative to the quadrupole, of $10^{-3}$, and bounds on the parameter $\beta$, which describes the amount of binary orbital energy lost into the dipole radiation, of $10^{-6}$. The parameter $\beta$ affects the phase evolution and so stronger bounds would be obtained for less massive systems, for which more waveform cycles will be observed in band. These bounds are, in both cases, comparable to those from observations with the Einstein Telescope, one order of magnitude better than those possible with Advanced LIGO and one order of magnitude worse than what would be possible with LISA.

\paragraph*{Solar oscillations.} Finn~\cite{finn85} observed that solar oscillations with 5- to 10-minute periods produce gravitational strains $\sim 10^{-26}$ at Earth, possibly within reach of space-based detectors. The detectors would measure the sun's dynamical gravitational field in the \emph{transition} region where it is turning into radiation. Finn showed that the field develops a significant phase shift relative to solar oscillations, which depends on the GW polarizations, and which could distinguish between scalar, vector, tensor, and scalar--tensor theories of gravity. The limit placed by such observations on the Brans--Dicke parameter would be weaker than current bounds from solar-system tests; on the other hand, measuring incipient GWs in the transition zone makes this a novel and possibly unique test. However, we note that Finn's early exploration~\cite{finn85} predates our full understanding of the design and parameters of LISA-like missions, which are likely to be less sensitive to this signal. This problem was revisited in~\cite{1996PhRvD..54.1287C}, in which the authors assessed the sensitivity of LISA to the quadrupole ($l=2$) low-order normal modes ($p$, $f$ and $g$-modes) of the sun. They estimated that the energy in these modes would have to exceed $10^{30}\mathrm{\ ergs}$ to allow a LISA detection, and that the required mode energy would be even higher for eLISA.

\paragraph*{Galactic binaries.} Among the compact galactic binaries that would be detected by a LISA-like detector, several have orbital inclinations known from optical observations. For these systems we can compute the specific linear combination(s) of polarizations that would be appear in the data, which can then be checked for consistency. A single inconsistent binary may indicate an error in the determination of inclination or distance, but systematically inconsistent sources would hint at \emph{large} non-tensor GW components. However, from general arguments the measurement accuracy for polarization amplitudes is $\sim 1/\mathrm{SNR}$ (with SNR a few tens at most for galactic binaries), so only very large corrections to GR would be detectable in this way.

\subsubsection{Tests of gravitational-wave propagation}
\label{sub.propagation}


In GR, gravitational radiation propagates at the speed of light: $v_g = c$. The experimental validation of this prediction can be posed as a bound on the graviton mass $m_g$, which is exactly zero in GR (see~\cite{2009arXiv0910.4066B,2010RvMP...82..939G} for a broader context). However, it may be advisable to consider $m_g$ as a purely phenomenological parameter, since certain massive-graviton theories do not recover GR predictions such as light bending, as discussed in Section~\ref{subsec:massgrav}.

Weak-field measurements in the solar system already provide bounds on $m_g$ on the basis of the massive-graviton Yukawa correction to the Newtonian potential:
\begin{equation}
V(r) = -\frac{M}{r}\exp\left(-r/\lambda_{g}\right),
\label{eqn.Yukawa}
\end{equation}
where $\lambda_{g} = h/m_{g}$ is the Compton wavelength of the graviton. The corresponding GW propagation speed would be given by
\begin{equation}
\Bigl(\frac{v_g}{c}\Bigr)^{2} = 1 - 
\Bigl(\frac{c}{f_g \lambda_{g}}\Bigr)^{2} \,,
\label{eqn.gravitonSpeed}
\end{equation}
with $f_g$ the frequency of radiation. The best solar-system tests provide the bound $\lambda_{g} > 2.8 \times 10^{12}\mathrm{\ km}$ (or $m_{g} < 4.4 \times 10^{-22}\mathrm{\ eV}$)~\cite{1988PhRvL..61.1159T}. By contrast, binary-pulsar dynamics only provide a bound $m_g < 7.6 \times 10^{-20}\mathrm{\ eV}$~\cite{2002PhRvD..65d4022F}.
As we discuss in this section, observations of binary GWs with LISA-like detectors could provide bounds competitive with these results, with the advantage of examining a rather different sector of gravitational physics, wave propagation. Two distinct methods have been proposed for this.

\paragraph*{Comparing the phase of GW and EM signals.}
This technique offers a direct comparison of the speed of GWs with the speed of a radiation assumed to be null (light itself). For the technique to work, sources must be observable in both light and GWs, and the astrophysical delays (if any) between the two signals must be well understood and modeled.
The most prominent low-frequency sources for this purpose are compact galactic binaries. 

Let the difference between the arrival times of GWs and EM signals be
\begin{equation}
\Delta t = \Delta t_{a} - (1 + z)\Delta t_{e} \,,
\label{eqn.timeLag}
\end{equation}
where $\Delta t_a$ arises from propagation (the very effect we wish to measure), and $\Delta t_e$ from different emission mechanisms or geometries. Here $z \simeq D_{L} H_{0}/c$ is the redshift of the source, with $D_L$ the luminosity distance and $H_{0}$ the value of the Hubble parameter. In terms of $v_g$,
\begin{equation}
\epsilon_g \equiv 1 - \frac{v_{g}}{c} = 1 \times 10^{-11}\left(\frac{1\mathrm{\ kpc}}{D_{L}}\right)\left(\frac{\Delta t_{a}}{1\mathrm{\ s}}\right),
\label{eqn.speedLag}
\end{equation}
which we relate to $m_g$ using the total relativistic energy,
\begin{equation}
\epsilon_g \equiv 1 - \frac{v_{g}}{c} = \frac{1}{2} \Bigl(\frac{m_g c^2}{h f_g}\Bigr)^2 \,,
\label{eqn.speedDelay}
\end{equation}
where $f_g$ is the GW frequency.

The measurement of $\Delta t$ has been considered repeatedly in the literature~\cite{2000PhRvD..61j4008L,2003PhRvD..67b4015C,2004PhRvD..69j3502C}.
The main difficulty lies with modeling the emission delay $\Delta t_e$: 
consider for instance AM CVn binaries, where a low-mass helium donor has expanded to fill its Roche lobe and is spilling mass onto a white-dwarf primary. The EM signal from these systems is greatly affected by the light emitted from the overflow stream impacting the accretion disk, and the light curve oscillates as the system orbits, alternately flashing the impact point toward and away from the observer. The times of maximum emission can be taken as reference for the EM phase, but how are they related to GW emission? 

To evaluate this $\Delta t_e$, one may observe the compact binary at two epochs, ideally at opposite points across the Earth's orbit~\cite{2000PhRvD..61j4008L,2003PhRvD..67b4015C}. Under the assumption that $\Delta t_e$ is constant, differencing the total $\Delta t$ measured at the two epochs leaves a measure of $\Delta t_a$ alone. However, the subtraction reduces $\Delta t_a$ to what can be accumulated across the diameter of the Earth's orbit, rather than across the entire distance to the binary. As a consequence, the strongest bound from known LISA verification binary would be $\lambda_{g} > 3 \times 10^{13}\mathrm{\ km}$ ($m_{g} < 4.6 \times 10^{-23}\mathrm{\ eV}$).

Alternatively, one may concentrate on eclipsing compact binaries, where the light curve varies due to the mutual eclipses of the binary components, allowing the orientation geometry of the system to be precisely determined as a function of time, and yielding an accurate measure of $\Delta t_{e}$.
In this case the measured $\Delta t_a$ is accumulated over the entire distance to the source.
Only one eclipsing binary that would be observable with LISA-like detectors is currently known~\cite{2012ApJ...757L..21H}, but an analysis of their statistically-expected population suggests that LISA would obtain a bound $\lambda_{g} > 2 \times 10^{14}\mathrm{\ km}$ ($m_{g} < 6 \times 10^{-24}\mathrm{ eV}$)~\cite{2004PhRvD..69j3502C}.

The reader may question whether it is appropriate to compare gravitons to photons, when the current bound on the putative mass of the photon is as high as $m_{\gamma} < 2 \times 10^{-16}$ eV. However, the much higher frequency of optical photons compared to low-frequency gravitons leads to $\epsilon_\gamma < 3 \times 10^{-33}$, much smaller than $\epsilon_g < 10^{-8}$ (for solar-system tests)~\cite{2000PhRvD..61j4008L}, so a comparison based on speeds is indeed appropriate.

A related test using pulsar-timing observations would compare the GW-induced phase delays accumulated by photons traveling to Earth from different pulsars~\cite{2010ApJ...722.1589L}: the delays depend on the graviton speed through a geometric factor that alters the expected Hellings--Downs correlation~\cite{1983ApJ...265L..39H} that GWs will produce in the timing of pulsars located at different positions on the sky.

It might also be possible to observe simultaneous EM and GW signals from MBH mergers, using the approximate position of the source known from pre-merger GW observations to guide a follow-up campaign in the EM spectrum~\cite{2008ApJ...684..870K}. However, the nature of possible EM counterparts is extremely uncertain, so differences between the GW and EM phasing could be explained by uncertainties in the modeling of the EM signal. Therefore, it is unlikely that constraints from these systems will be competitive with galactic-binary constraints, or with the constraints from GW dispersion discussed in the following subsection.

\paragraph*{Measuring the dispersion of gravitational-wave chirps.}
\label{sec:GWdisp}
The chirping signals emitted by inspiraling binaries contain a range of frequency components: if the graviton has mass, the components propagate at different speeds, again given by Eq.~\eqref{eqn.speedDelay}. This effect can be modeled in the templates used to search for binary signals by including a $\lambda_{g}$ dependence in the waveform phasing~\cite{1998PhRvD..57.2061W}. In the frequency-domain representation, the propagation effect appears as a ``dephasing'' term $-\beta (\pi \mathcal{M}_c f)^{-1}$, where $\beta = \pi^2 D \mathcal{M}_c / \lambda_g^2 / (1 + z)$ with $\mathcal{M}_c$ the binary chirp mass, $f$ the GW frequency, $D$ the source distance, and $z$ the source redshift. By comparison, the leading-order term in the post-Newtonian expansion of the phasing is $3/128 (\pi \mathcal{M}_c f)^{-1}$, while the $\lambda_{g}$ correction contributes the same power of orbital frequency as the ``1PN'' term.

For space-based detectors, the best chirp-dispersion bounds will come from massive--black-hole systems; they improve slightly with the total binary mass and with better low-frequency ($10^{-5}\mbox{\,--\,}10^{-4}\mathrm{\ Hz}$) sensitivity. However, the expected bounds depend strongly on which other physical effects (such as spin-induced precessions, orbital eccentricity, higher waveform harmonics, the merger-ringdown phase) will be relevant in the detected systems. As a result, a variety of predictions have appeared in the literature~\cite{2004CQGra..21.4367W,2005PhRvD..71h4025B,2005CQGra..22S.943B,2009CQGra..26o5002A,2009PhRvD..80d4002S,2010PhRvD..81f4008Y,2010JPhCS.228a2049S,2010PhRvD..82l2001K,2012PhRvD..86h4028H}. Bounds as strong as $\lambda_g >$ a few $10^{16}\mathrm{\ km}$ seem possible, and would be strengthened by analyzing full catalogs of binary detections at once~\cite{2011PhRvD..84j1501B}.

Instead of the chirping signals from inspiraling binaries, Jones~\cite{2005ApJ...618L.115J} proposes a test of the GW dispersion relation using the waves from eccentric galactic binaries, which are emitted at multiple harmonics of the orbital frequencies; if at least one galactic binary has sufficient eccentricity, Jones claims sensitivity comparable to the chirp-dephasing measurements.
Mirshekari et al.~\cite{2012PhRvD..85b4041M} extend the graviton-mass formalism to more general modified-gravity theories that predict violations of Lorentz invariance and modified dispersion relations for GW modes, given by
\begin{equation}
E^2 = p^2 c^2 + m_g^2 c^4 + \mathbb{A} p^\alpha c^\alpha \,;
\end{equation}
both $m_g$ and $\mathbb{A}$ can be constrained together, given the $\alpha$ corresponding to specific theories, by inspiral-binary observations with ground and space-based detectors.

\paragraph*{Parity violations.}
In GR parity is a conserved quantity, so left and right-circular polarized gravitational radiation propagates alike.
Many attempts to formulate a quantum theory of gravity require the addition of a parity-violating Chern--Simons (CS) term to the Einstein--Hilbert action~\cite{AGWitt84,AlexGates06,Polchinski}:
\begin{equation}
S_{\mathrm{CS}} = \frac{1}{64\pi} \int \theta \, R^* R \, \mathrm{d}^4 x \,, \qquad
\mbox{where} \quad R^*R = \frac{1}{2}
R_{\alpha\beta\gamma\delta}\epsilon^{\alpha\beta\mu\nu}R^{\gamma\delta}_{\mu\nu} \,;
\end{equation}
here $R_{\alpha\beta\gamma\delta}$ is the Riemann tensor, $\epsilon^{\alpha\beta\mu\nu}$ is the Levi-Civita tensor density, and $\theta$ is a (possibly) position-dependent function that describes the coupling of the CS field to spacetime.
This correction creates a difference in the propagation equations for the left- and right-circular GW polarizations, resulting in their \emph{amplitude birefringence}: one circularly-polarized state is amplified through propagation, while the other is attenuated.

This effect is potentially observable with LISA-like detectors for MBH-binary inspirals at cosmological distances~\cite{yunesCSpropa} (see also~\cite{yunesCSpropb}), where the amplitude birefringence generates an apparent precession of the orbital plane of the binary. The CS correction accumulates with distance, and is larger for sources at higher redshifts.
%
%
%
%
Orbital-plane precession will also arise from general-relativistic spin--orbit coupling, but the scaling of the precession with frequency is different, so the two effects can be distinguished, at least in principle.

For an equal-mass binary with redshifted masses of $10^6\,M_{\odot}$ that is observed plane-on at a redshift $z=15$, LISA could constrain the integrated CS contribution at the level of $10^{-19}$~\cite{yunesCSpropa}. This is several orders of magnitude better than solar-system experiments, which furthermore can only provide local constraints. Thus, LISA-like detectors may provide some hints as to the very quantum nature of gravity.


\subsubsection{The quadrupole formula and loss of energy to gravitational waves}
\label{sub.quadrupole}

In theories that do not satisfy the strong equivalence principle, the
internal gravitational binding energies of bodies can create a
difference between the inertial dipole moment (i.e., the linear
momentum, which is conserved) and the GW-generating gravitational
dipole moment. Thus, alternative theories of gravity generally admit
dipole radiation, but it is forbidden in GR, where the two moments are
identical. Dipole radiation would be given at leading order by~\cite{willLR}
\begin{equation}
h_d \sim \frac{1}{D_L} \frac{\mathrm{d}}{\mathrm{d}t} \left(
m^G_1 \mathbf{x}_1 + m^G_2 \mathbf{x}_2
\right) \sim \frac{\mu^I \mathbf{v}}{D_L} \left(
\frac{m^G_1}{m^I_1} - \frac{m^G_2}{m^I_2}
\right),
\end{equation}
where $\mathbf{x}_{1,2}$ are the positions of the binary components,
$\mathbf{v}$ their relative velocity, $m^I_{1,2}$ and $m^G_{1,2}$ are
their inertial and gravitational masses respectively, $\mu^I$ is the
inertial reduced mass, and $D_L$ is the luminosity distance to the
observer.

For relativistic objects such as neutron stars (NS), the
gravitational binding energy can be considerable and so
can be the resulting loss of energy to dipolar GWs. Indeed, the
experimental result that the orbital decay of the binary pulsar PSR1913+16
\cite{2008LRR....11....8L} adhered closely to GR's quadrupole-formula prediction
was sufficient to definitely falsify GR alternatives such as bimetric
and ``stratified'' theories~\cite{willTEGP}. (Amusingly, certain theories
even predict that dipole radiation carries away \emph{negative} energy from
a binary~\cite{willTEGP}.)
Thus it is factually correct to state that the indirect detection of GWs has already provided a strong test of GR.

By contrast, the binary pulsar could not falsify scalar-tensor theories in this way, because these are ``close'' to GR. For instance, although dipole radiation is predicted by Brans--Dicke theory and changes the progression of orbital decay, the coupling parameter $\omega_\mathrm{BD}$ can be adjusted to approximate GR results to any desired accuracy. GR is reproduced for $\omega_\mathrm{BD} \rightarrow \infty$, so experimental bounds on Brans--Dicke are \emph{lower} bounds. The Hulse--Taylor binary pulsar does provide a bound on $\omega_\mathrm{BD}$, but one that is not competitive with solar-system tests, among which the best comes from the Doppler tracking of the Cassini spacecraft, which sets $\omega_\mathrm{BD} > 40\,000$~\cite{willLR,bertotti03}. However, other binary systems containing pulsars are known that provide constraints, which are competitive with solar-system constraints. The best constraints on scalar-tensor gravity (and also TeVeS gravity) come from the pulsar--white-dwarf binary J1738+0333~\cite{2012MNRAS.423.3328F}, which provides the limit $\omega_\mathrm{BD} > 25\,000$.

LISA-like detectors can constrain $\omega_\mathrm{BD}$ by looking for dipole-radiation--induced modifications in the GW phasing of binary inspirals (monopole radiation is also present, but suppressed relative to the dipole), as long as at least one of the binary components is not a black hole: because of the no-hair theorem, black holes cannot sustain the scalar field that would lead to a differing $m^G$ and $m^I$ (as was recently confirmed in full numerical-relativity simulations~\cite{2011arXiv1112.3928H}). This restriction can be circumvented by having non-asymptotically flat boundary conditions for the black hole~\cite{2012JCAP...05..010H}. If the scalar field is slowly varying far from the black hole (either as a function of time or space) then it can support a scalar field. This scenario was investigated numerically in~\cite{2013PhRvD..87l4020B}, which found that accelerated single black holes and black-hole binaries would emit scalar radiation, in the latter case at twice the orbital frequency. If the asymptotic scalar-field gradient that supports the black-hole scalar hair is cosmological in origin, this effect will be negligible, but the possibility does exist in general. Except for these considerations, the canonical source for detecting this effect is the inspiral of a neutron star into a relatively low-mass central black hole, although the number of detections of such systems is likely to be very low~\cite{2009CQGra..26i4034G}.

Early studies~\cite{ScharreWill2002,2004CQGra..21.4367W}, based on simplified models of the waveforms and of the LISA sensitivity, estimated that for a $1.4\,M_{\odot}$ neutron star inspiraling into a MBH, at fixed SNR~=~10, the $\omega_\mathrm{BD}$ bounds would scale as
\begin{equation}
\omega_\mathrm{BD} > 2 \times 10^{4} \left(\frac{\cal S}{0.3}\right)^{2}\left(\frac{100}{\Delta\Phi_{D}}\right)
\left(\frac{T}{1\mathrm{\ yr}}\right)^{7/8}
\left(\frac{10^{4}\,M_{\odot}}{M_\bullet}\right)^{3/4}\,,
\label{eqn.SWboundBD}
\end{equation}
where $\cal S$ (the ``sensitivity'') is a measure of the difference between the neutron-star and MBH self-gravitational binding energies per unit rest mass; $\Delta \Phi_{D}$ is the dipole contribution to the GW phasing; $T$ is the time of observation; and $M_\bullet$ is the MBH mass. However, this estimate is reduced by a factor of ten or more when more realistic waveforms are considered that include spin couplings~\cite{2005PhRvD..71h4025B,2005CQGra..22S.943B}, spin-induced orbital precession and eccentricity~\cite{2010PhRvD..81f4008Y}.
Bounds can also be derived for a massive-scalar variant of Brans--Dicke theory~\cite{2012PhRvD..85l2005B}, and are of order
\begin{equation}
\left( \frac{m_s}{\sqrt{\omega_{\mathrm{BD}}}} \right)
\left( \frac{\rho}{10} \right) \lesssim 10^{-19}\mathrm{\ eV},
\end{equation}
(where $m_s$ is the mass of the scalar and $\rho$ the detection SNR) for the intermediate--mass-ratio inspiral of a NS into a black hole with mass $\lesssim 10^3\,M_\odot$.

These results were obtained using only the leading order correction from the scalar radiation. In~\cite{2012PhRvD..85j2003Y} the authors extended this calculation to all post-Newtonian orders, but in the extreme-mass-ratio limit by using the Teukolsky formalism. The conclusion, that constraints on massless scalar-tensor theories from GW observations will, in general, be weaker than those from solar-system observations, was unchanged. The reason is that scalar-tensor theories are weak-field (infrared) corrections to GR and are therefore largest in the weak field, so the leading order correction captures the majority of the effect. Massive scalar-tensor theories were also considered in~\cite{2011PhRvL.107x1101C,2012PhRvD..85j2003Y}. In those theories, the primary observable consequence is the possible existence of ``floating orbits'' at which the scalar flux experiences a condition where GWs scatter off the central, massive body, emerging with more energy (extracted from the spin of the central body).  The waves transfer that energy to the small orbiting body, increasing its orbital energy.  This ``super-radiant resonance'' temporarily balances the GW flux. The transition of an EMRI through such a floating orbit is many orders of magnitude slower than the normal EMRI inspiral and can last more than a Hubble time. If an EMRI consistent with GR is observed it means that the EMRI not only did not pass through such a floating orbit during the timescale of the observation but could not have encountered one prior to the observation since it would not then have reached the millihertz band. Therefore, an observation of a single EMRI can constrain the massive scalar-tensor parameter space to many orders of magnitude greater precision than current solar-system observations.

\paragraph*{Other modifications to the inspiral phasing.}
A number of other suggestions have been made for low-frequency GW tests of GR that do not quite fit a ``modified energy-loss'' description. For instance, dynamical Chern--Simons theory introduces nonlinear modifications in the binary binding energy and dissipative corrections at the same PN order~\cite{SYTinalt,2012PhRvD..85f4022Y} that could be observed in the late inspiral, constraining the characteristic Chern--Simons length scale $\xi^{1/4}$ to $\lesssim 10^5\mbox{\,--\,}10^6\mathrm{\ km}$~\cite{2012PhRvL.109y1105Y}, 
comparable to current solar-system constraints~\cite{AliHamCSBound} (advanced ground-based detectors could do even better, placing bounds of $\lesssim 10\mbox{\,--\,}100\mathrm{\ km}$).

Corrections to the inspiral phasing will also arise if the spacetime outside the central object is not described by the Kerr metric or if additional energy is lost into scalar or other forms of radiation. This has been considered for various alternative theories of gravity; we discuss these results in detail in Section~\ref{sec:BHinalt}.

GW \emph{tails}, which are due to the propagation of gravitational radiation on the curved background of the emitting binary, appear at a relative 1.5PN order ($c^{-3}$) beyond the leading-order quadrupole radiation, and their observation would test the nonlinear nature of GR~\cite{1995PhRvL..74.1067B}. (This would be a \emph{null} test of GR, since tails are included in the ``standard'' post-Newtonian inspiral phasing; see also the PN-coefficient tests discussed in section~\ref{sec:PNphasemod}.

Promoting Newton's constant, $G$, to a function of time modifies both
a binary's binding energy and GW luminosity, and therefore its
phasing. A three-year observation of a
$10^4\mbox{\,--\,}10^5\,M_{\odot}$ inspiral would constrain
$\dot{G}/G$ to $\sim 10^{-11}\mathrm{\ yr}^{-1}$~\cite{2010PhRvD..81f4018Y}. The infinite Randall--Sundrum braneworld model~\cite{1999PhRvL..83.4690R} may predict an enormous increase in the Hawking radiation emitted by black holes~\cite{2002JHEP...08..043E,2002PThPS.148..307T}. The resulting progressive mass loss may be observed as an outspiral effect in the quasi-monochromatic radiation of galactic black-hole binaries, as a correction to the inspiral phasing of a black-hole binary~\cite{2011PhRvD..83h4036Y} and it would also affect the rate of EMRI events~\cite{2010PhRvL.104n1601M,2011PhRvD..83h4036Y}. The constraints on the size of extra dimensions coming from observations with LISA will, in general, be worse than those derivable from tabletop experiments. However, DECIGO observations of BH--NS binary mergers would be able to place a constraint about ten times better than tabletop experiments, assuming a detection rate of $\sim 10^5$ binaries per year~\cite{2011PhRvD..83h4036Y}.



\subsection{Tests of general relativity with phenomenological inspiral template families}
\label{frameworks}

As discussed above, quantitative tests of GR against modified theories of gravity evaluate how well the measured signals are fit by alternative waveform families, or (more commonly) by waveform families that extend GR predictions by including one or more modified-gravity parameters, such as $\omega_{\mathrm{BD}}$ for Brans--Dicke theory. To set up these tests we need to work within the alternative theory to derive sufficiently accurate descriptions of source dynamics, GW emission, and GW propagation.
An alternative approach is to operate directly at the level of the waveforms by introducing phenomenological corrections to GR predictions: for instance, by modifying specific coefficients, or by adding extra terms.

This section discusses the first attempts to do so. So far these have concentrated on post-Newtonian waveforms~\cite{lrr-2006-4} for circular, adiabatic inspirals, as described by the stationary-phase approximation in the frequency domain:
\begin{equation}
\label{eq:ArunFull}
  \tilde h(f) = A f^{-7/6}\exp\left[i\Psi(f)+i\frac\pi4\right]\,,
\end{equation}
where $f$ is the GW frequency; $A$ is the GW amplitude, given by geometrical projection factors $\times$ $\mathcal{M}_c^{5/6} / D_L$ (with $\mathcal{M}_c = (m_1 m_2)^{3/5} / (m_1 + m_2)^{1/5}$ the \emph{chirp mass} and $D_L$ the luminosity distance); and for simplicity we omit the nontrivial response of space-based detectors, as well as the PN amplitude corrections.
The phasing $\Psi(f)$ is expanded as
\begin{equation}
\label{eq:ArunPhase}
  \Psi(f)= 2\pi f t_c+ \Phi_c+\sum_{k \in  \mathbb{Z} }\left[\psi_k+\psi_{k}^\mathrm{log} \log f\right]f^{(k-5)/3}.
\end{equation}
For binaries with negligible component spins, the post-Newtonian phasing coefficients $\psi_k$ are currently known up to $k = 7$ (3.5 PN order), and in GR they are all functions of the two masses $m_1$ and $m_2$ alone (although $\psi_1 = \psi^{\log}_{0\mbox{\,--\,}4} = \psi^{\log}_{7} = 0$, and $\psi_5$ is completely degenerate with $\Phi_c$, so it is usually omitted)~\cite{PhysRevLett.74.3515,PhysRevD.65.061501,PhysRevLett.93.091101,PhysRevD.71.084008}.

\subsubsection{Modifying the PN phasing coefficients}
\label{sec:PNphasemod}

Arun et al.~\cite{2006CQGra..23L..37A} propose a test of GR based on estimating all the $\psi_k$ simultaneously from the measured waveform as if they were free parameters, in analogy to the post-Keplerian formalism~\cite[Section~4.5]{2008LRR....11....8L}.
The value and error estimated for each%
\epubtkFootnote{Because $\log f$ varies weakly in the relevant
  frequency ranges, it is considered constant
  in~\cite{2006CQGra..23L..37A}, and the $\psi_k^{\log} \log f$
  combinations are absorbed in the $\psi_k$.}
$\psi_k$, together with its PN functional form as a function of $m_1$ and $m_2$, determines a region in the $m_1$--$m_2$ plane. If GR is correct, all the regions must intersect near the true masses, as shown in Figure~\ref{fig:pheno}. The extent of the intersection provides a measure of how precisely GR is verified by a GW observation. A Fisher-matrix analysis~\cite{2006CQGra..23L..37A} suggests that, for systems at the optimistic distance of 3~Gpc, LISA could measure $\psi_0$ to $\sim$~0.1\% and $\psi_2$ and $\psi_3$ to 10\%, but that the fractional error on higher-order terms would be at best $\sim$~1.

\epubtkImage{m1m2.png}{%
\begin{figure}[htbp]
\centerline{\includegraphics[width=3in]{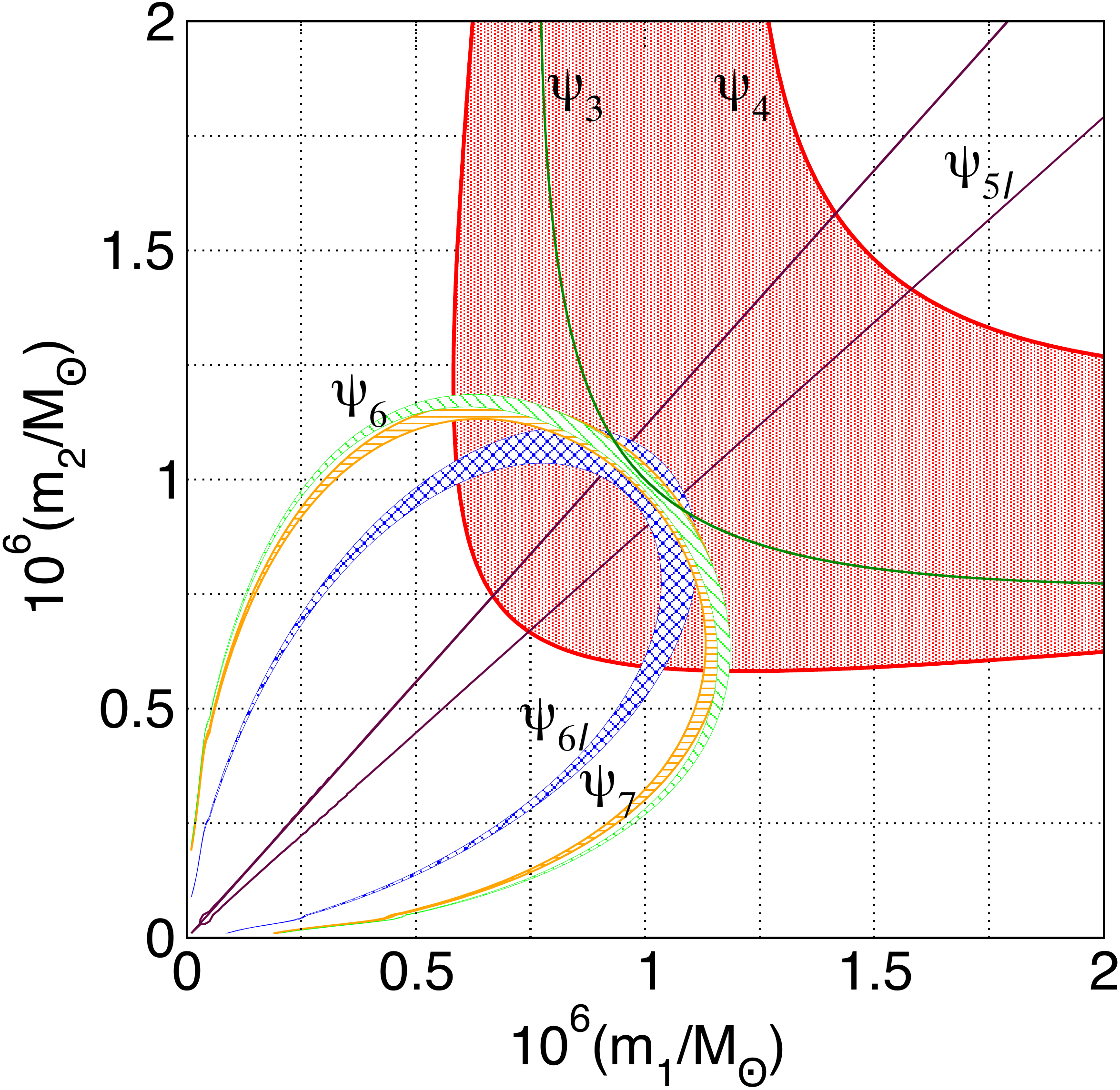}}
\caption{Estimating all the binary-inspiral phasing coefficients $\psi_k$ of Eq.~\eqref{eq:ArunPhase} yields differently shaped regions in the $m_1$--$m_2$ plane, which must intersect near true mass values if GR is correct. Image reproduced by permission from~\cite{2006PhRvD..74b4006A}, copyright by APS.}
\label{fig:pheno}
\end{figure}}

However, this setup may understate the power of this kind of test,
since most of the estimation uncertainty in the $\psi_k$ arises from
their mutual degeneracy -- that is, from the fact that it is possible
to vary the value of a subset of $\psi_k$ without appreciably
modifying the waveform. This degeneracy should not impact the degree
to which the data is deemed consistent with GR. In a follow-up
paper~\cite{2006PhRvD..74b4006A}, Arun et al.\ propose a revised test whereby the masses are determined from $\psi_0$ and $\psi_2$, while the other $\psi_k$ (as well as $\psi^\mathrm{log}_{5}$ and $\psi^\mathrm{log}_{6}$) are individually estimated and checked for consistency with GR.
In this case, even for sources at $z=1$ ($\sim$~7~Gpc), all the parameters can be constrained to 1\% (a few \% for $\psi_4$, 0.1\% for $\psi_3$), at least for favorable mass combinations.
Performing parameter estimation for the eigenvectors of the $\psi_k$ Fisher matrix~\cite{2013CQGra..30b5011P} indicates which combinations of coefficients can be tested more accurately for GR violations.

However, it is not clear what significance with regards to testing GR should be ascribed to the accuracy of measuring the $\psi_k$, since we do not know at what level we could expect deviations to appear. By contrast, if we were to find that, say, the $n$--$\sigma$ regions in the $m_1$--$m_2$ plane do \emph{not} intersect, we could make the statistically-meaningful statement that GR appears to be violated at the $n$--$\sigma$ level.

Del Pozzo et al.~\cite{2011PhRvD..83h2002D} and Li et
al.~\cite{2012PhRvD..85h2003L,2012JPhCS.363a2028L} propose a more
satisfying formulation for these tests, based on Bayesian model
selection~\cite{2005blda.book.....G}, which compares the
\emph{Bayesian evidence}, given the observed data, for the pure-GR
scenario against the alternative-gravity scenarios in which one or
more of the $\psi_k$ are modified. The issue of significance discussed
above reappears in this context as the inherent arbitrariness in
choosing prior probabilities for the $\psi_k$, but Del Pozzo et al.\ argue that this does not affect the efficacy of the model-comparison test in detecting GR violations.
(For a comprehensive discussion of model selection in the context of GW detection, rather than GR tests, see also~\cite{2008CQGra..25r4010V,2008PhRvD..78b2001V,2009PhRvD..80f3007L}. For more recent applications of this formalism to ground-based detectors, see~\cite{2010PhRvD..82f4010M}.)

\subsubsection{The parameterized post-Einstein framework}
\label{sec:ppE}

In~\cite{2009PhRvD..80l2003Y}, Yunes and Pretorius propose a similar but more general approach, labeling it the ``parameterized post-Einsteinian'' (ppE) framework. For adiabatic inspirals, they propose enhancing the stationary-phase inspiral signal with extra powers of GW frequency:
\begin{equation}\label{ppE}
\tilde h^{\mathrm{ppE}}(f)=\tilde h^\mathrm{GR}(f) \times \left( 1+\alpha(\pi{\cal
M}f)^a\right)\exp\left[i\beta(\pi{\cal M}f)^b\right],
\end{equation}
where $\tilde{h}^{\mathrm{GR}}(f)$ is given in Eqs.~\eqref{eq:ArunFull} and \eqref{eq:ArunPhase}.
While the initial suggestion in~\cite{2009PhRvD..80l2003Y} is to consider $a, b \in \mathbb{R}$, there are analytical arguments why $a$ and $b$ should be restricted to values $a = \bar{a}/3$ and $b = \bar{b}/3$, with $(\bar{a},\bar{b}) > (-10,-15)$~\cite{2012PhRvD..86b2004C}, which reproduces Arun's PN-coefficient scheme for $\bar{b} \geq -5$.
Nevertheless, this representation can reproduce the leading-order effects of several alternative theories of gravity (see Table~\ref{table:ppEtheories}).
\begin{table}
\caption[Leading-order effects of alternative theories of gravity, as represented in the ppE framework.]{Leading-order effects of alternative theories of gravity, as represented in the ppE framework [Eq.~\eqref{ppE}]. For GR $\alpha=\beta=0$. This table is copied from~\cite{2011PhRvD..84f2003C}, except for the two entries labeled with an asterisk. The quadratic curvature ppE exponent given in~\cite{2011PhRvD..84f2003C} was $b=-1/3$, coming from the conservative dynamics. However, it was shown in~\cite{2012PhRvD..85f4022Y} that the dissipative correction is larger, giving the value $b=-7/3$ quoted above. The dynamical Chern--Simons ppE exponent given in ~\cite{2011PhRvD..84f2003C} was $b=4/3$, which was derived using the slow-rotation metric accurate to linear order in the spin~\cite{YPCSBH}. At quadratic order in the spin~\cite{2012PhRvD..86d4037Y}, the corrections to both conservative and dissipative dynamics occur at lower post-Newtonian order, giving $b=1/3$~\cite{2012PhRvL.109y1105Y}.}
\label{table:ppEtheories}
\centering
\begin{tabular}{lcccc}
\toprule
~ & $a$ & $\alpha$ & $b$ & $\beta$\\
\midrule
Brans--Dicke            & -- & 0 & $-7/3$ &$\beta$\\
parity violating        & 1 & $\alpha$ & 0 & -- \\
variable $G(t)$         & $-8/3$ & $\alpha$ & $-13/3$ & $\beta$\\
massive graviton        & -- & 0 & $-1$ & $\beta$\\
quadratic curvature     & -- & 0 & $-7/3^*$ & $\beta$\\
extra dimensions        & -- & 0 & $-13/3$ & $\beta$\\
dynamical Chern--Simons & $+3$ & $\alpha$ & +1/3 & $\beta$\\
\bottomrule
\end{tabular}
\end{table}

In~\cite{2009PhRvD..80l2003Y}, Yunes and Pretorius are motivated by the possibility of detecting GR violations, but also by the ``fundamental bias'' that would be incurred in estimating GW-source parameters using GR waveforms when modified GR is instead correct.
In~\cite{2011PhRvD..84f2003C}, Cornish et al.\ reformulate the detection of GR violations described by ppE as a Bayesian model-selection problem, similar to the PN-coefficient tests discussed in section~\ref{sec:PNphasemod}. Figure~\ref{fig:CornishEA2011Figs5and7} shows the $\beta$ bounds, for various fixed $b$, that could be set with LISA observations of $m_{1,2} \sim 10^6\,M_{\odot}$ binary inspirals at $z = 1$ and 3.
For $b$ corresponding to modifications in higher-order PN terms (which require strong-field, nonlinear gravity conditions to become evident), the bounds provided by LISA-like detectors become more competitive with respect to solar-system and binary-pulsar results (where weak-field conditions prevail).

\epubtkImage{ppe-bounds.png}{%
\begin{figure}[htb]
\centerline{\includegraphics[width=0.8\textwidth]{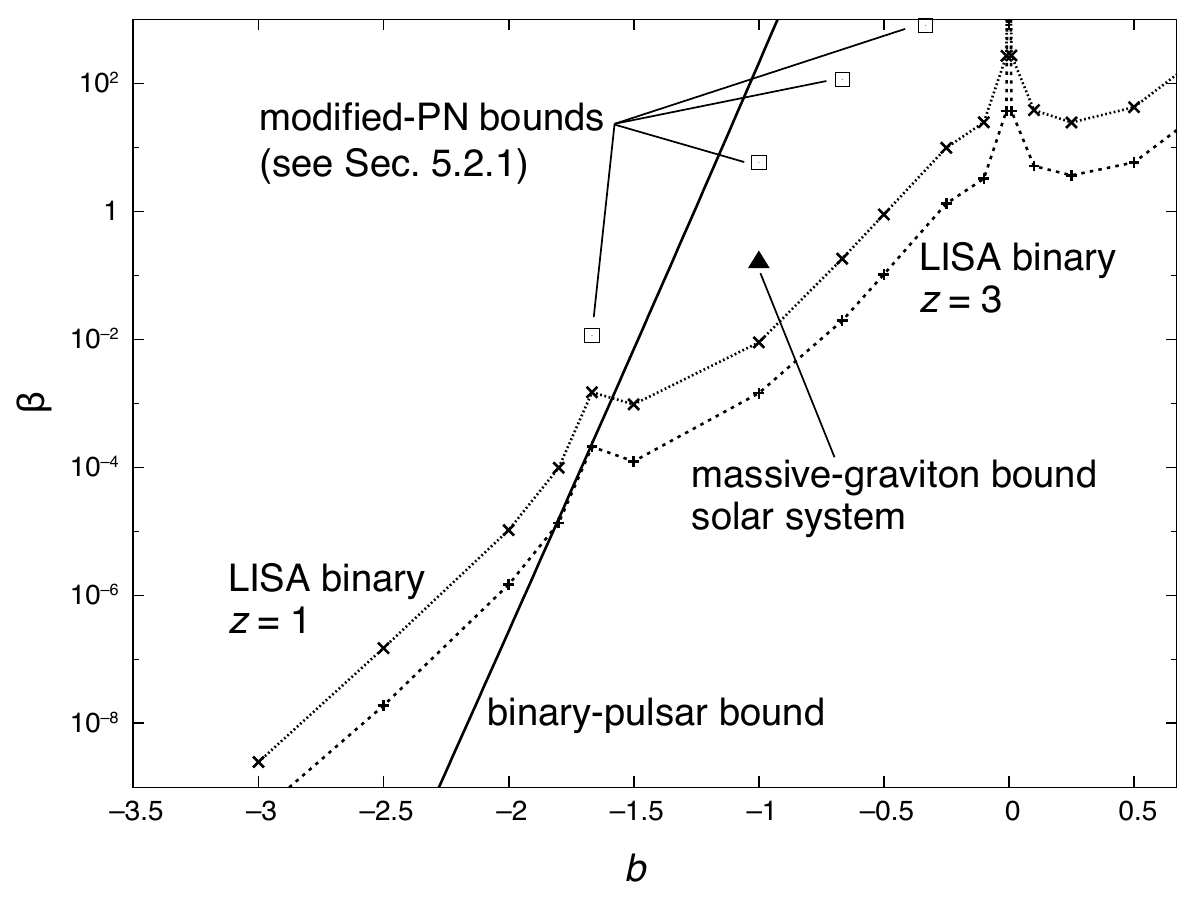}}
\caption{Constraints on phasing corrections in the ppE framework, as determined from LISA observations of $\sim 10^{6}\,M_{\odot}$ massive--black-hole inspirals at $z = 1$ and $z = 3$. 
The figure also includes the $\beta$ bounds derived from pulsar PSR~J0737--3039~\cite{2010PhRvD..82h2002Y}, the solar-system bound on the graviton mass~\cite{1988PhRvL..61.1159T}, and PN-coefficient bounds derived as described section~\ref{sec:PNphasemod}. The spike at $b = 0$ corresponds to the degeneracy between the ppE correction and the initial GW-phase parameter. (Adapted from~\cite{2011PhRvD..84f2003C}.)}
\label{fig:CornishEA2011Figs5and7}
\end{figure}}

A ppE-like model including dipole radiation in addition to quadrupole radiation but no other modifications to the waveform phasing was described in~\cite{2012CQGra..29g5011A} and was discussed in Section~\ref{sub.polarization} above. The full ppE framework was extended to include all additional polarization states and higher waveform harmonics in~\cite{2012PhRvD..86b2004C}. The final form was motivated by considering Brans--Dicke theory, Lightman--Lee theory and Rosen's theory. In the most general form, Eq.~\ref{ppE} is modified to
\begin{eqnarray}
\tilde h^\mathrm{ppE}(f)&=&\tilde h^\mathrm{GR}(f) \times \exp\left[i\beta(\pi{\cal M}f)^b\right] \nonumber \\
&& + \left(\alpha_+ F_+ + \alpha_\times F_\times + \alpha_b F_b + \alpha_L F_L + \alpha_{\mathrm{sn}} F_{\mathrm{sn}} + \alpha_{\mathrm{se}} F_{\mathrm{se}}\right) \nonumber \\
&& \hspace{2in} \times (\pi {\cal M}f)^a \exp\left[-i\Psi_{\mathrm{GR}}^{(2)} + i \beta (\pi{\cal M}f)^b \right] \nonumber \\
&& +\left(\alpha_+ F_+ + \alpha_\times F_\times + \alpha_b F_b + \alpha_L F_L + \alpha_{\mathrm{sn}} F_{\mathrm{sn}}+\alpha_{\mathrm{se}} F_{\mathrm{se}}\right) \nonumber \\
&& \hspace{2in} \times (2\pi {\cal M}f)^c \eta^{\frac{1}{5}}  \exp\left[-i\Psi_{\mathrm{GR}}^{(1)} + i \delta (2\pi{\cal M}f)^d \right] ,
\label{ppEPlusHarms} 
\end{eqnarray}
in which $\Psi_{\mathrm{GR}}^{(l)}$ is the GR phase of the $l$th waveform harmonic, $\eta=m_1 m_2/(m_1+m_2)^2$ is the symmetric mass ratio, and $F_A$ is the detector response to a GW in polarization mode $A$. The ppE parameters are $(\{\alpha_A\},\{\gamma_A\},a,\beta,b,c,\delta,d)$.

The authors of~\cite{2012PhRvD..86b2004C} considered two further variants of this scheme. One variant restricted the coefficients in the expansion so that they were not all independent, but were related to one another via energy conservation. The second variant included this interdependence of the parameters, and also accounted for modified propagation effects by introducing additional ``phase-difference'' parameters into the second and third terms. As yet, this fully extended ppE scheme has not been used to explore the constraints that will be possible with space-based detectors.

An analysis using a waveform model with higher harmonics and spin
precession, but not alternative polarization states, was carried out
in~\cite{2012PhRvD..86h4028H}. Its authors considered modifications to
a subset of the phase and amplitude parameters only, which
corresponded to certain post-Newtonian orders and could therefore also
be interpreted in terms of modifications to the pN phase coefficients
as discussed in Section~\ref{sec:PNphasemod}. The estimated bounds
derived using this more complete waveform model were typically one to
two orders of magnitude better than previous estimates for high-mass
systems, but basically the same for low-mass systems. This is
unsurprising, since the effects of spin-precession and higher
harmonics will only be important late in the inspiral. High-mass
systems generate lower frequency GWs and are therefore only observable
for the final stages of inspiral, merger and plunge. Therefore, late-time
corrections are proportionally more important for those
systems. For high-mass systems, the authors
of~\cite{2012PhRvD..86h4028H} estimated that LISA would be able to
measure deviations in the phasing parameters to a precision $\Delta
\Psi_n \sim 0.1,10,50,500,1000$ for $n = -1,0,1,3/2,2$ respectively,
where $n$ denotes the post-Newtonian order, with $\Delta \Psi_n$ the
coefficient of $f^{2n/3-5/3}$ in the waveform phase. Using the same
model, they also estimated that LISA could place a bound of $\lambda_g
> 1 \times 10^{16}\mathrm{\ km}$ on the graviton Compton wavelength
when allowing for correlations between the different
phase-modification parameters $\Delta \Psi_n$. This was discussed in
Section~\ref{sub.propagation}.

An extension of the ppE framework to EMRI systems requires a model in
which orbits can be both eccentric and inclined. To develop this,
Vigeland et al.~\cite{2011PhRvD..83j4027V} derive a set of near-Kerr spacetime metrics that satisfy a set of conditions, including the existence of a Carter-constant--like third integral of the motion, as well as asymptotic flatness. The solutions, which were previously found in~\cite{GeneralCarterConst}, are restricted to a physically interesting subset by setting to zero any metric coefficients not required to reproduce known black-hole solutions in modified gravity, and by applying the peeling theorem (i.e., by requiring that the mass and spin of the black hole not be renormalized by the perturbation).

The existence of a third integral is not a requirement for black-hole solutions, but in general its absence allows ergodic behavior in the orbits. This is discussed as a potential observable signature for deviations from GR in Section~\ref{intloss}. However, data-analysis pipelines designed for GR waveforms may be insensitive to such qualitatively different systems. Therefore the existence of a third integral is a practical assumption for interpretation once a GR-like EMRI has been observed.

In~\cite{GairYunesGenericEMRI}, Gair and Yunes construct gravitational waveforms for EMRIs occurring in the metrics of~\cite{2011PhRvD..83j4027V}, based on the analytic kludge model constructed for GR EMRIs~\cite{AK}. The waveforms provide a ppE-like model for EMRIs that can be used in the same way as the circular ppE framework. Parameter-estimation results with these ppE--EMRI models have not yet appeared in the literature.

\subsubsection{Other approaches}

In~\cite{2012PhRvD..86h2001V}, Vallisneri provides a unified model-comparison performance analysis of \emph{all} modified-GR tests that is valid for sufficiently-loud signals, and that yields the detection SNR required for a statistically-significant detection of GR violations as a simple function of the \emph{fitting factor} FF between the GR and modified-GR waveform families. The FF measures the extent to which one can reabsorb modified-GR effects by varying standard-GR parameters from their true values. Vallisneri's analysis is valid in the limit of large SNR, and may not be applicable to all realistic scenarios with finite SNRs.

An alternative to modifying frequency-domain inspiral waveforms is offered by Cannella et al.~\cite{2009PhRvD..80l4035C,2011arXiv1103.0983C}. They propose tests based on the effective-field-theory approach to binary dynamics~\cite{2006GReGr..38.1537G}, which expands the Hilbert+point-mass action as a set of Feynman diagrams. 
In this framework, GR corrections can be introduced by displacing the coefficients of interaction \emph{vertices} from their GR values. For instance, multiplying the three-graviton vertex by a factor $(1+\beta_3)$ affects the conservative dynamics of the theory in a manner similar to the PPN parameter $\beta$, but also has consequences on radiation. A similar modification to the four-graviton vertex (parameterized by $\beta_4$) yields effects at the second post-Newtonian order, so it has no analog in PPN. Cannella et al.\ argue that GR-violating values of $\beta_3$ and $\beta_4$ would not be detectable with GW signals, but they would instead generate small systematic errors in the estimation of standard binary parameters. However, a thorough analysis of the detectability of such deviations has not been carried out, so this conclusion may be modified in the future.

\subsection{Beyond the binary inspiral}
\label{sec:mergerringdown}

According to GR, black-hole mergers are the most energetically luminous events in the universe, with $L \sim 10^{23} \, L_\odot \sim 10^{56}$ erg/s, regardless of mass: at their climax, they outshine the combined power output of all the stars in the visible universe. 
Nevertheless, second-generation ground-based GW interferometers are expected to yield the first detections of black-hole mergers~\cite{2010CQGra..27q3001A}, but only with rather modest SNRs. By contrast, LISA-like GW detectors would observe the mergers of heavier black holes, with SNRs as high as hundreds or more throughout the universe, offering very accurate measurements of the merger waveforms. Massive--black-hole coalescences may feature significant spins and eccentricity, further enriching the merger phenomenology~\cite{2008ApJ...684..822B,2012JPhCS.363a2035R}.

The powerful merger events correspond to very relativistic velocities and very strong gravitational fields, so much that the PN expansion of the field equations cannot be applied, and we must resort to very complex and costly numerical simulations~\cite{2010RvMP...82.3069C}.
This makes it challenging to encode the effects of plausible GR modifications in the signal model. The first ppE paper~\cite{2009PhRvD..80l2003Y} makes such an attempt on the basis of a very crude model of merger-ringdown signals, which would probably be insufficient even to phase-match the GR signals themselves. Broad efforts are currently under way to build phenomenological full-waveform (inspiral-merger-ringdown) models 
\cite{2008PhRvD..77j4017A,2011PhRvD..84l4052P,2012PhRvD..86b4011T}; these involve tunable parameters that are adjusted to match the waveforms produced by numerical relativity. Such parameters could also be used to encode non-GR effects in the merger-ringdown. However, at this time designing such extensions in a principled way seems daunting.

A simpler approach, proposed by Hughes and Menou~\cite{2005ApJ...623..689H}, involves the \emph{golden binaries} for which system parameters can be estimated from \emph{both} inspiral and ringdown GWs. The former encode the parameters of the binary, while the latter encode the parameters of the final black hole formed in the merger. The functional relation of the two sets of parameters can then be compared with the predictions of numerical relativity, providing a null test of the strong-field regime of GR.

Hughes and Menou focus on measuring the remnant's \emph{mass deficit}, which equals the total energy carried away by GWs, so their definition of golden binaries selects those in which the mass deficit can be estimated to better than 5\%. For LISA, these systems tend to have component masses between a few $10^{5}\,M_{\odot}$ and a few $10^{6}\,M_{\odot}$, and to be found at $z \sim 2\mbox{\,--\,}3$, making up 1\,--\,10\% of the total merger rate depending on black-hole population models. The estimates of~\cite{2005ApJ...623..689H} are based on rather simple waveform models that omit a range of physical effects, so they could be seen as conservative, given that increased waveform complexity tends to improve parameter-estimation accuracy. A more complete analysis was carried out in~\cite{2006CQGra..23.3763L}, but in the context of ground-based GW detectors rather than space-based detectors.

\newpage


\section{Tests of the Nature and Structure of Black Holes}\label{sec.BHstructure}
It has become apparent over the last few decades that the centers of
most galaxies harbor massive, dark compact objects with masses in
the range $10^{6}\mbox{\,--\,}10^{9}\,M_{\odot}$. It is clear that these objects
play a very important role in the evolution of galaxies. This is
exemplified by the very tight measured correlation (the $M\mbox{--}\sigma$
relation) between the mass
of the central dark objects and the velocity dispersion of stars in
the central spheroid~\cite{Ferrarese00,tremaine02}. It is generally accepted that
the central dark objects are black holes described by the Kerr metric, but there is presently no
definitive proof of that assumption. The alternatives to the black-hole interpretation include dense star clusters, supermassive stars,
magnetoids, boson stars, and fermion balls. Support for the black-hole
interpretation has arisen as a result of both observational and
theoretical work. A short review of the evidence may be found
in~\cite{celotti99}, although we summarize some key details in Section~\ref{bhstruct:status}.

As described in Section~\ref{gravtheory:bhstruct}, the theoretical basis for the belief that these objects are Kerr black holes has arisen from proofs that singularities inevitably form during gravitational collapse~\cite{1939PhRv...56..455O,penrose65,hawkpen70} and that the Kerr solution is the unique stationary and axisymmetric black-hole solution in GR~\cite{israel67,carter71,rob75}. The uniqueness of the Kerr solution is sometimes referred to as the
``no-hair'' theorem, since the solution is characterized by just two
parameters, the black-hole mass $M$ and angular momentum (per unit mass) $a$.

The field of any vacuum, axisymmetric
spacetime in GR can be characterized by a sequence of
mass and current multipole moments, which we denote as $M_l$ and $S_l$
respectively~\cite{geroch70,hansen74}. For Kerr spacetimes these
multipole moments are all determined by the mass and spin via
\begin{equation}
M_l + \mathrm{i}S_l = M (\mathrm{i}a)^l \,,
\label{nohair}
\end{equation}
so these spacetimes require no additional independent parameters or ``hair''. The proof of the uniqueness theorem relies on various assumptions beyond the validity of the Einstein equations, so a demonstration of the non-Kerr nature of astrophysical black holes could reveal exotic physics within GR. It might also indicate the presence of material in the spacetime outside the black-hole horizon, or a deviation from GR in the true theory of gravity. In this section we discuss the potential of space-based low-frequency GW detectors to probe the structure of massive compact objects and the possible interpretation of these results. Short reviews of the prospects for testing relativity with measurements of black-hole ``hair'' can be found in~\cite{2006IJMPD..15.2209B,2008RSPTA.366.4365G}.

\subsection{Current observational status}
\label{bhstruct:status}

The observational evidence for the presence of black holes in the
centers of galaxies has come mainly from the studies of Active
Galactic Nuclei (AGN). These are known to be extremely energetic and
also compact -- typical luminosities of $10^{46}\mathrm{\ erg\ s}^{-1}$ are
produced in regions less than $10^{16}\mathrm{\ m}$ across~\cite{krolik99}. The
inferred AGN efficiency of $\sim$~10\% is much greater than the typical
efficiencies of nuclear fusion processes ($\sim$~1\%), implying the
need for a very deep relativistic potential. X-ray observations show
variability on timescales of less than an hour, while observations of
iron lines indicate the presence of gas moving at speeds of several
thousand km per second~\cite{krolik99}. Radio observations of
water maser discs are consistent with Keplerian motion around very
compact central objects. In the spiral galaxy NGC~4258 VLBA
observations have indicated a disc with an inner (detected) edge at
$\sim$~0.1~pc, around an object of mass
$3.6\times10^{7}\,M_{\odot}$~\cite{miyoshi95}. Such compactness cannot
be realized by a stellar cluster. In addition, about 10\% of AGNs
are associated with jets, which move at highly relativistic
velocities and persist for millions of years. This requires a
relativistic potential that has a preferred axis that is stable over
very long timescales. AGNs are also remarkably similar over several
decades of mass, which favors the black-hole explanation, again because Kerr
black holes are characterized by just $M$ and $a$.

In the Milky Way, evidence for the presence of a black hole
coincident with the Sgr~A* radio source has come from observations of
stellar motions. These are completely consistent with Keplerian
motion around a point source of mass
$M\sim4\times10^6\,M_{\odot}$~\cite{2009ApJ...692.1075G,ghez08}. One star, S2, has
been observed for one complete orbit and from this it has been
possible to put a limit of 0.066 on the extended fraction of the central
mass that could be contained between pericenter and apocenter of the orbit of S2. At perihelion S2
was $\sim$~100~AU from the central object, which provides a fairly
tight constraint on its compactness.

Electromagnetic observations can rule out stellar clusters as the explanation for the massive central objects, but some of the exotic alternatives remain.
X-ray emission comes from the very inner regions of accretion discs,
but uncertainties in the radius from which the emission is coming
and in the mass and spin of the central object limit their utility
for probing the structure of the central object~\cite{psaltis08}. It
is also possible to construct very compact boson star
spacetimes~\cite{kesden05} that could not be ruled out from
electromagnetic observations alone. The same applies to spacetimes
with a naked singularity. By contrast, GW observations will probe the spacetime structure as the object proceeds
through the inspiral and then passes the innermost stable orbit and
plunges into the horizon of the central object, if a horizon exists. We
discuss the prospects for using GW observations to
probe black-hole structure in the following
Sections~\ref{sec:emritests} and \ref{sec:ringdowntests}.

\subsection{Tests of black-hole structure using EMRIs}
\label{sec:emritests}

\subsubsection{Testing the ``no-hair'' property}
\label{testnohair}

Equation~(\ref{nohair}) tells us that a Kerr black hole is uniquely
characterized by two parameters. If we can measure three multipole
moments of the spacetime, we can check that they are consistent
with Eq.~(\ref{nohair}). If they are not, then the object cannot be a
Kerr black hole. Boson stars will typically have more independent multipole moments. In a certain class of models of rotating boson stars, the boson star can be uniquely characterized by three
multipole moments~\cite{ryanBS,2006IJMPD..15.2209B}, so a LISA measurement of four multipole moments could also exclude these models as an explanation of the data.

GWs from EMRIs are complicated superpositions of
different frequency components, found at harmonics of the three fundamental
frequencies of the orbit: the orbital frequency and the frequencies
of precession of the perihelion and of the orbital
plane~\cite{DrascoGenEMRI}. This complex structure encodes detailed
information about the spacetime in which the GWs are generated. The
details of this encoding were first worked out by Ryan~\cite{ryan95}. If the spacetime is assumed to be stationary and axisymmetric, it can be shown that the Einstein equations reduce to a single equation, the Ernst equation, for a complex scalar function, the Ernst potential~\cite{ernst68}. 
By using the Ernst potential and expressions due to
Fodor et al.~\cite{fodor89} that relate this potential to the
multipole moments of the spacetime, Ryan was able to study the
properties of orbits in vacuum and axisymmetric spacetimes that
possess an arbitrary set of mass and current multipole moments. 
Circular and equatorial orbits do not show perihelion or orbital plane
precession. However, if such an orbit is given a small radial or
vertical perturbation, it will undergo small oscillations at
frequencies (the ``epicyclic'' frequencies) that correspond to the
perihelion or orbital plane precession frequencies of nearly circular
and nearly equatorial orbits respectively. These frequencies can be
readily computed. For the arbitrary stationary axisymmetric spacetimes
considered in~\cite{ryan95} one finds
\begin{eqnarray}
\frac{\Omega_r}{\Omega} &=& 3(M\Omega)^{\frac{2}{3}} - 4
\frac{S_1}{M^2} (M\Omega) + \left(\frac{9}{2} - \frac{3}{2}
\frac{M_2}{M^3}\right) (M\Omega)^{\frac{4}{3}} - 10 \frac{S_1}{M^2}
(M\Omega)^{\frac{5}{3}} \nonumber\\ && +\left(\frac{27}{2}-2
\frac{S_1^2}{M^4}-\frac{21}{2}\frac{M_2}{M^3}\right) (M\Omega)^{2} +
\left(-48\frac{S_1}{M^2}-5 \frac{S_1
M_2}{M^5}+9\frac{S_3}{M^4}\right) (M\Omega)^{\frac{7}{3}} + \cdots ,
\label{omperi} \\
\frac{\Omega_z}{\Omega} &=& 2\frac{S_1}{M^2} (M\Omega) + \frac{3}{2}
\frac{M_2}{M^3} (M\Omega)^{\frac{4}{3}} + \left(7\frac{S_1^2}{M^4} +
3 \frac{M_2}{M^3}\right) (M\Omega)^2 \nonumber \\ && +
\left(11\frac{S_1M_2}{M^5} - 6 \frac{S_3}{M^4} \right)
(M\Omega)^{\frac{7}{3}} + \cdots \,,
\label{omplane}
\end{eqnarray}
where $\Omega$ is the angular ($\phi$) frequency of the circular orbit being perturbed, $\Omega_r$ and $\Omega_z$ are the perihelion and orbital plane
precession frequencies, and $M_l$/$S_l$ denote the mass/current
multipole moments of the spacetime metric, as in Eq.~(\ref{nohair}).
The primary conclusion from Eqs.~(\ref{omperi})\,--\,(\ref{omplane}) is
that the various multipole moments enter the different terms in this
expansion at different orders in $\Omega$. The precession frequencies
and orbital frequency could be extracted from
GW observations, so these expansions are, in
principle, observable. We can use this information to ``map'' the
spacetime structure near the central object and verify that the
multipole moments are consistent with the no-hair
property~(\ref{nohair}) that we expect for a Kerr black hole. This
technique is sometimes termed ``bothrodesy'' or
``holiodesy''\epubtkFootnote{The term \emph{bothrodesy} was coined by Sterl
Phinney in 2001, arising from the use of the Greek word $\beta
o\theta\rho o\varsigma$ (meaning ``sacrificial pit'') to describe
black holes. In modern Greek, $\beta o\theta\rho o\varsigma$ has come
to mean ``sewage pit'', so \textit{holiodesy} is a suitable replacement
that was first suggested by Marc Favata~\cite{collins2004}.} by
analogy with geodesy, in which observations of the motion of
satellites are used to probe the gravitational field of the Earth.

The multipole moments are also encoded in the total orbital energy
lost as the orbital frequency changes by a unit logarithmic interval
\begin{eqnarray}
\frac{\Delta E(f)}{\mu} &=& \frac{1}{3} (M\Omega)^{\frac{2}{3}} -
\frac{1}{2} (M\Omega)^{\frac{4}{3}} +  \frac{20}{9} \frac{S_1}{M^2}
(M\Omega)^{\frac{5}{3}} + \left(-\frac{27}{8} +
\frac{M_2}{M^3}\right) (M\Omega)^{2} \nonumber \\ && + \frac{28}{3}
\frac{S_1}{M^2} (M\Omega)^{\frac{7}{3}}  + \left(-\frac{225}{16} +
\frac{80}{27}\frac{S_1^2}{M^4} + \frac{70}{9}\frac{M_2}{M^3} \right)
(M\Omega)^{\frac{8}{3}}  \nonumber \\ &&\qquad+ \left(\frac{81}{2}
\frac{S_1}{M^2}+ 6 \frac{S_1 M_2}{M^5} -6 \frac{S_3}{M^4} \right)
(M\Omega)^{3} + \cdots
\end{eqnarray}
A more powerful observable than the three discussed so far is the
number of cycles that a trajectory spends near a particular frequency
\begin{equation}
\Delta {\cal N}(f) = \frac{f^2}{\mathrm{d}f/\mathrm{d}t} =
f^2\,\frac{\mathrm{d}E/\mathrm{d}f}{\mathrm{d}E/\mathrm{d}t},
\end{equation}
but this is not as clean an observable as the precession frequencies,
since it requires computing the rate of energy loss to
GWs in an arbitrary spacetime. Ryan used this
formalism in conjunction with a post-Newtonian waveform model to
estimate LISA's capability to measure the spacetime
multipoles~\cite{ryan97}. He considered nearly circular and nearly
equatorial inspirals, and found that LISA's ability to determine the
spacetime multipoles degraded as more multipoles were included in the
waveform model. The typical errors that Ryan found are in
Table~\ref{table:ryanacc}.
The conclusion was that LISA would be able to make moderately
accurate measurements of the lowest three multipole moments, but
probably no more.

\begin{table}[htb]
\caption[Accuracy with which a LISA observation could
determine the multipole moments of a spacetime decreases as more multipoles are
included in the model.]{Accuracy with which a LISA observation could
determine the multipole moments of a spacetime decreases as more multipoles are
included in the model, taken from Ryan~\cite{ryan97}. The third
column indicates the highest multipole included in the particular
model. Results are shown for two typical cases, a
$10\,M_{\odot}+10^{5}\,M_{\odot}$ inspiral and a
$10\,M_{\odot}+10^{6}\,M_{\odot}$ inspiral; in both cases the SNR of
the inspiral is 10.}
\label{table:ryanacc}
\centering
\begin{tabular}{cc|c|ccccc}
\toprule
$\mu/M_{\odot}$ & $M/M_{\odot}$ & $l_{\max}$ &
$\log_{10}(\delta{M}/M)$ & $\log_{10}(\delta{s_1})$ & $\log_{10}(\delta{M_2})$ & $\log_{10}(\delta{s_3})$ & $\log_{10}(\delta{M_4})$\\
\midrule
 ~ & ~ & 1 & --3.7 & --3.5 & ~ & ~ & ~ \\
10 & 10\super{5} & 2 & --3.0 & --2.9 & --1.8 & ~ & ~ \\
 ~ & ~ & 3 & --2.3 & --1.9 & --1.3 & --0.7 & ~ \\
 ~ & ~ & 4 & --1.5 & --1.3 & --1.1 & \phantom{--}0.1 & 1.0 \\
\midrule
 ~ & ~ & 1 & --3.3 & --2.8 & ~ & ~ & ~ \\
10 & 10\super{6} & 2 & --2.5 & --1.0 & --0.3 & ~ & ~ \\
 ~ & ~ & 3 & --1.2 & \phantom{--}0.1 & \phantom{--}0.8 & \phantom{--}0.9 & ~ \\
 ~ & ~ & 4 & --1.0 & \phantom{--}0.1 & \phantom{--}0.8 & \phantom{--}1.2 & 1.8 \\
\bottomrule
\end{tabular}
\end{table}

Ryan's analysis was restricted to circular and equatorial orbits, but
a counting argument suggests that spacetime mapping should still be
possible for generic orbits~\cite{LiLovelace}. One complication is
that the evolution of the orbital elements must also be inferred from
the observation, which spoils the nice form of the
expansions~(\ref{omperi})\,--\,(\ref{omplane})~\cite{gairbumpy}, since
all the multipole moments now enter at each order of the expansion.
However, this would also be true if the expansions for
circular-equatorial orbits were rewritten as an expansion in some
initial frequency, $\Omega_0$, which would more closely represent a
band-limited observation. In practice, the lowest-order multipole
dominates the lowest term in the expansion and so on, which makes
spacetime mapping possible in practice.

The Ryan formalism neatly illustrates why spacetime mapping with
EMRIs is possible, but it is not a very practical scheme. We expect
that the massive central objects are
indeed Kerr black holes and so we really want to consider what
imprint small deviations from Kerr will leave on the emitted
GWs. A multipole-moment expansion is not a very
practical way to do this, as the Kerr metric has an infinite number
of nonzero multipoles. Several authors have adopted the approach of
constructing spacetimes given by Schwarzschild--Kerr plus a small
deviation, and have examined the properties of geodesics in those
spacetimes.

Collins and Hughes~\cite{collins2004} considered a static deviation
from the Schwarzschild metric. This was constructed by writing the
metric in Weyl coordinates and adding a quadrupole perturbation to
the potential (in these coordinates, the potential equation reduces
to the flat-space Laplacian for the $(\rho, z)$ cylindrical
coordinates, which facilitates writing solutions). They
considered two types of quadrupole perturbation: a torus around
the black hole, and the addition of a point mass at each pole. In the
second case, the spacetime necessarily contains line singularities
running from the point masses either to infinity or to the black-hole horizon,
which are needed to support the point masses. The solutions
are perturbative, in that the authors kept only the terms that are
linear in the deviation from Schwarzschild. Collins and Hughes explored the
properties of orbits in these spacetimes by comparing the precession
and orbital frequencies of equatorial orbits in the spacetime to
orbits with the same orbital parameters in Kerr. They found that
there were measurable differences in the perihelion precession in
the strong field: for instance, at a radius of $\sim~20\,M$ for a 2\%
perturbation of the black hole, the trajectory would accumulate one radian
of dephasing in $\sim$~1000 orbits. Collins and Hughes coined the
term ``bumpy black hole'' to describe spacetimes of this type.

Glampedakis and Babak~\cite{GlamBab06} took a different approach to
studing deviations from Kerr. Starting from the Hartle--Thorne
metric~\cite{hartle67,hartlethorne}, which is an exact solution to Einstein's equations describing the spacetime outside of a slowly, rigidly rotating axisymmetric object, the authors constructed a spacetime with
metric of the form $g_{\mu\nu}=g^{\mathrm{Kerr}}_{\mu\nu} + \epsilon
h_{\mu\nu} + O(\delta M_{l\geq3}, \delta S_{l\geq2})$, working
perturbatively and keeping only the $\delta M_2$ perturbation in the
quadrupole mass moment. They termed the resulting spacetime a
``quasi-Kerr'' solution. A comparison of the frequencies of eccentric
equatorial geodesics in the quasi-Kerr spacetime to the same
geodesics in the Kerr spacetime indicated that it would take only
$\sim$~100 cycles to accumulate a $\pi/2$ phase shift for a $\sim$~1\%
deviation from Kerr. They also computed waveform overlaps and found
that, over the radiation-reaction timescale, the overlap of the
waveforms for an orbit with a semilatus rectum of ten geometric radii ($GM/c^2$) was $\sim$~20\%, 50\%, 90\%, 98\%
for inspirals with mass ratio $\mu/M = 10^{-6}$, $10^{-5}$, $10^{-4}$,
$10^{-3}$ respectively. 

A third approach to analyzing deviations from the Kerr spacetime was
considered by Gair et al.~\cite{gairbumpy}, who studied geodesic
motion in a family of exact spacetimes due to Manko and
Novikov~\cite{mankonov}, which include the Kerr spacetime for a specific choice of parameters. By using exact solutions of the
Einstein field equations, they obtained solutions that were valid
everywhere, in contrast to the perturbative solutions considered
in~\cite{collins2004,GlamBab06}, which break down near the central
object. However, this scheme offers less control over the multipole moments,
since it is possible to choose the lowest multipole that differs from
Kerr, but then the higher multipoles must also change. Gair et
al.~\cite{gairbumpy} studied observable properties of the orbits,
including the variation of the precession frequencies of nearly
circular and nearly equatorial orbits as a function of orbital
frequency, and the loss of the third integral of the motion (see
Section~\ref{intloss}).

These three papers outlined different ways to approach the problem of
spacetime mapping in practice. However, none of the analyses were
complete as they did not consider inspirals. Collins and Hughes and
Glampedakis and Babak also ignored waveform \emph{confusion} by assuming that the
orbital elements were the same between the orbits under
consideration. Glampedakis and Babak did discuss the importance of confusion and the role of the inspiral evolution in breaking such degeneracies, but no inspiral results were included in the published paper. Observationally, the correct orbits to compare will be
those with the same frequencies since we have no way to determine the
orbital radius or eccentricity directly. This was the approach
adopted in~\cite{gairbumpy}. Assessing the confusion problem is
relatively easy, but  including inspiral is very hard in general,
since the presence of excess multipole moments in the spacetime leads
to changes in the rates of energy and angular momentum loss, which
must also be included in the analysis. Progress can be made in the
presence of small deviations by including only the leading-order
contributions to the radiation reaction from the multipole moments.
This is an open area of research, although Barack and
Cutler~\cite{2007PhRvD..75d2003B} carried out a preliminary assessment using
post-Newtonian EMRI waveforms~\cite{AK} augmented with the leading-order contribution of an excess quadrupole moment to the precession
and inspiral rates. The resulting waveforms were an improvement in comparison to Ryan's
analysis~\cite{ryan97}, since they included orbital eccentricity and
inclination, and were filtered through an approximation of the LISA response. Barack and Cutler performed a
Fisher-matrix analysis of parameter-estimation uncertainties, and hence correctly accounted for the
confusion issue. For this simple model, they found that a single LISA
observation of the inspiral of a $10\,M_{\odot}$ black hole into a
$10^{5.5}\,M_{\odot}$, $10^6\,M_{\odot}$ or $10^{6.5}\,M_{\odot}$ black
hole could measure the deviation from Kerr of the quadrupole moment
to an accuracy of $\Delta(M_2/M^3) \sim 10^{-4}, 10^{-3}, 10^{-2}$,
while simultaneously measuring the mass and spin of the central
object to fractional accuracies of $10^{-4}$. This suggests that a LISA-like observatory
would be able to perform high-precision tests of the no-hair property of
massive compact objects in galactic centers. To put these numbers in
perspective, a boson star may have a quadrupole moment that is
$\sim100$ times that of a Kerr black hole with the same mass and
spin~\cite{ryanBS}, so it could easily be excluded.

\subsubsection{Probing the nature of the central object}

During an EMRI, the inspiraling object interacts gravitationally with
the horizon of the central black hole. This can be thought of as a
tidal interaction -- the gravitational field of the small body
raises a tide on the horizon that is dragged around through the orbit,
leading to dissipation of energy -- or as energy being lost by
GWs falling into the black hole. Fang and
Lovelace~\cite{fang05} explored the nature of the tidal-coupling
interaction by perturbing a Schwarzschild black hole with a distant
orbiting moon. They found that the time-dependent piece of the
perturbation affected the orbit in an unambiguous way: a
time-varying quadrupole moment is induced on the black-hole horizon
that is proportional to the time derivative of the moon's tidal
field. This quadrupole perturbation extracts energy and angular
momentum from the orbit at the same rate that energy and angular
momentum enter the horizon. However, the
effect of the time-independent piece of the perturbation remained
ambiguous. Working in the Regge--Wheeler gauge, Fang and Lovelace found that this piece
vanished, in contrast to a previous result by Suen~\cite{suen86}, who used
a different gauge. This ambiguity leads to an ambiguity in
the phase of the induced quadrupole moment as measured in a local
asymptotic rest frame (LARF), although the phase of the bulge
relative to the orbiting moon is well defined (using a spacelike
connection between the moon and the black hole, Fang and Lovelace
found that the horizon shear led the horizon tidal field by an angle
of $4M\Omega$, where $\Omega$ is the angular velocity of the moon).
The ambiguity of interpretation in the LARF makes it impossible to
define the polarizability of the horizon or the phase shift of the
tidal bulge in a body-independent way. Fang and Lovelace left open
the possibility of developing a body-independent language to describe
the response of the central object to tidal coupling, but as yet this
has not happened.

Although the nature of the response of the central object to tidal
coupling may be difficult to characterize from GW
observations, the total energy lost to tidal interactions is a good
observable. Ryan's original theorem~\cite{ryan95} ignored tidal coupling, but
it was later generalized by Li and Lovelace~\cite{LiLovelace}. They
found that the GWs propagating to infinity depended
only very weakly on the inner boundary conditions (i.e., on
the nature of the central object). This means that the spacetime's
multipole structure can be inferred from the outgoing radiation field
in the usual way, and hence the expected rate of orbital energy loss,
assuming no energy loss into the central body, can be calculated. The
rate of inspiral measures the actual rate of orbital energy loss, and
the difference then gives the rate at which energy is lost to
the central object, which is a direct measure of the tidal coupling.
Li and Lovelace estimated that the ratio of the change in energy
radiated to infinity due to the inner boundary condition to the
energy in tidal coupling scales with the orbital velocity as $v^{5}$.
Therefore, it should be possible to simultaneously determine the
spacetime structure and the tidal coupling through low-frequency GW observations.

Information about the central object can also come from the
transition to plunge at the end of the inspiral. In a black-hole
system, we expect GW emission to cut off sharply as the orbit
reaches the innermost stable orbit and then plunges rapidly through
the horizon. If there is no horizon in the system, the orbit may
instead enter a phase where it passes into and out of the
material of the central object. This was explored for boson-star
models in~\cite{kesden05}: Kesden  et al.\ found that persistent radiation after
the apparent innermost orbit could be a clear signature of the
presence of a supermassive, horizonless central object in the spacetime. This
analysis did not treat gravitational radiation or the interaction of
the inspiraling body with the central object accurately, but it does
illustrate a possible way to identify non--black-hole central objects.
Something similar might happen if the spacetime were to contain a naked
singularity rather than a black hole~\cite{gairbumpy}: in
principle, the nature of the emitted waveform after ``plunge'' would
encode information about the exact nature of the central object. However,
this has not yet been investigated. Naked-singularity spacetimes may
have very--high-redshift surfaces rather than horizons: these spacetimes
would be observationally indistinguishable from black holes,
unless the inspiraling object happened to move inside the high-redshift surface and then
emerged, and the two epochs of radiation could be connected observationally.

Another example of an object that can be arbitrarily close to a Schwarzschild black hole in the exterior but lack a horizon is a gravastar~\cite{2009PhRvD..80l4047P}. These are constructed by matching a de~Sitter spacetime interior onto a Schwarzschild exterior through a thin shell of matter, whose radius can be made arbitrarily close to the Schwarzschild horizon. It was shown in~\cite{2009PhRvD..80l4047P} that the oscillation modes of such a gravastar have quasinormal frequencies that are completely different from those of a Schwarzschild black hole. Therefore the absence of a horizon would be apparent if ringdown radiation was observed from such a system. In addition, the tidal perturbations that arise during the inspiral of a compact object into a gravastar during an EMRI~\cite{2010PhRvD..81h4011P} can resonantly excite polar oscillations of the gravastar as the orbital frequency passes through certain values over the course of the inspiral. The excitation of these modes generates peaks in the GW-emission spectrum at frequencies that are characteristic of the gravastar, and can also show signatures of the microscopic surface of the gravastar. This process would be apparent both in the amplitude of the detected GWs and in the rate of inspiral inferred from the gravitational-waveform phase, since the rate of inspiral will change significantly in the vicinity of each resonance due to the additional energy radiated in the excited quasi-normal modes. Although the gravastar model used in~\cite{2009PhRvD..80l4047P,2010PhRvD..81h4011P} may not be physically relevant, this work illustrates the more general fact that if the horizon of a black hole is replaced by some kind of membrane, then the modes of that membrane will inevitably be excited during an inspiral and these modes will typically be different to those of a black hole.

\subsubsection{Astrophysical perturbations: the influence of matter}
A change in the inspiral trajectory need not be caused only by
differences in the central object, but might arise due to the presence
of material in the spacetime, close to the black hole but external to
the event horizon. Such material could influence the inspiral
trajectory, and hence the emitted GWs, in two distinct
ways: the gravitational field of the matter could modify the
multipole moments of the spacetime and hence the orbit as discussed
above; if the orbital path intersected the material, it would
cause sufficient hydrodynamic drag on the object to alter the orbit.

The influence of the gravitational field of external material was
considered in~\cite{emrimatter}: Barausse et al.\ constructed a model spacetime
that included both a black hole and an external torus of material
very close to the central black hole. They examined the
properties of orbits in two systems: one with a torus of
comparable mass and spin to the central black hole (spacetime ``A''),
and one with a torus of low mass, but much greater angular momentum
than the central black hole (spacetime ``B''). Their comparisons were
based on computing equatorial geodesics and then ``kludge'' gravitational
waveforms in the spacetime and in a corresponding Kerr spacetime, and
then evaluating the waveform overlap. Orbits were identified by
matching the radial and azimuthal frequencies between the orbits in
the two spacetimes in two ways: altering the orbital
parameters, while setting the mass and spin to be the same in the
corresponding Kerr spacetime; or altering the mass and spin, while
keeping the orbital parameters the same.

This approach identified a
confusion problem: over much of the parameter space the overlaps
were very high, particularly when the second approach was adopted. Overlaps
were lower in spacetime ``B'' than in spacetime ``A,'' and overlaps
for ``internal'' orbits between the black hole and the torus were
particularly low. This work suggested that it would not be possible to
distinguish between such a spacetime and a pure Kerr spacetime in low-frequency GW EMRI observations. However, it did not consider inspiraling
orbits. The need to constantly adjust the orbital parameters in order
to maintain equality of frequencies would lead to a difference in the
evolution of the orbit between the two spacetimes, which might break
the waveform degeneracies. The torus model was also not physical,
being much more compact than one would expect for AGN discs.

The effect of hydrodynamic drag on an EMRI was first considered in~\cite{2000ApJ...536..663N} and was found to be negligibly small for systems likely to be of interest to space-based gravitational-wave detectors. The problem was revisited in~\cite{emridrag}: Barausse and Rezzolla
considered a spacetime containing a Kerr black hole surrounded
by a non--self-gravitating torus with constant specific angular
momentum. The hydrodynamic drag consists of a short-range part that
arises from accretion, and a long-range part that arises from the
gravitational interaction of the body with the density perturbations
it causes in the disc. The accretion onto the small object was
modeled as Bondi--Hoyle--Lyttleton accretion~\epubtkFootnote{Bondi--Hoyle--Lyttleton accretion is the process that occurs when an object is moving through a stationary distribution of material. The material is gravitationally focused behind the object and counter-rotating streams collide, which dissipates their angular momentum. The material then falls onto the back side of the object~\cite{1944MNRAS.104..273B,bondi52}.\label{footnote:bondi}} and the
long-range force using  collisional dynamical-friction results from
the literature~\cite{kim07,EB2007}. The effect of the hydrodynamical
drag on the orbital evolution was computed for geodesic orbits and
compared to the orbital evolution from radiation reaction for a
variety of torus models, varying in mass and outer radius.

The
conclusion was that, for realistic outer radii for the torus,
$r_{\mathrm{out}} > 10^4 M$, the effect of hydrodynamic drag on the
orbital radius and eccentricity was always small compared to
radiation reaction. However, the relative importance increases
further from the central black hole. The hydrodynamic drag has a
greater relative effect on the orbital inclination, and tends to
cause orbits to become more prograde, which is opposite to the effect
of radiation reaction. For $r_{\mathrm{out}}=10^5M$, the hydrodynamic drag
becomes significant at $r\approx 35\,M$, but this radius is smaller for
a more compact torus. Thus, this effect will only be important for
LISA if we observe a system with a very compact accretion torus, or
for systems of low central mass. For the latter, the GWs are detectable from orbits at larger radii where hydrodynamic
drag can be important. However, the SNR of such events will be low,
so we are unlikely to see many of them~\cite{2009CQGra..26i4034G}.
Thus, although it seems that this effect is also marginal, this conclusion is
based on considering geodesic orbits, and the possible secular build
up of a drag signature over the inspiral has yet to be examined,

The influence of an accretion disc on the evolution of an EMRI embedded within it was explored in~\cite{2011PhRvL.107q1103Y,2011PhRvD..84b4032K}. One of the channels that has been suggested to produce EMRIs is the formation of stars in a disc, followed by the capture of the compact stellar remnants left after the evolution of those stars~\cite{ASReview,PauLRR}. The migration of such an EMRI through the accretion disc could potentially leave a measurable imprint on the GW signal. The literature distinguishes between two types of migration. In Type~I migration, which generally occurs for lower-mass objects, the disc persists in the vicinity of the object throughout the inspiral. The object excites density waves in the disc, which exert a torque on the object. In general, the torque from material exterior to the orbit is greater than that from material interior to the orbit, which causes the object to lose angular momentum and spiral inward on a timescale that is short relative to the lifetime of the disc. In Type~II migration, a gap opens in the disc in the vicinity of the inspiraling object. Material enters the gap on the disc's accretion timescale, driving the object and gap inwards on that timescale.

Yunes, Kocsis et al.~\cite{2011PhRvL.107q1103Y,2011PhRvD..84b4032K} considered migration in geometrically-thin and radiatively-efficient discs, in which thermal energy is radiated on a much shorter timescale than the timescale over which the material moves inward, so the disc can remain thin. Such discs can be described by the Shakura--Sunyaev $\alpha$-disc model~\cite{1973AA....24..337S}, in which the viscous stress in the disc is proportional to the total pressure at each point in the disc. These discs are known to be unstable to linear perturbations~\cite{1974ApJ...187L...1L,1976MNRAS.175..613S,1977AA....59..111B,1978ApJ...221..652P}. The alternative $\beta$-disc model, where stress is proportional to the gas pressure only, is stable to perturbations~\cite{1981ApJ...247...19S}. Both disc models were considered for EMRI migration. 
Yunes, Kocsis et al.\ showed that, over a year of observation, Type~I migration could lead to $\sim 0.01/10$ radian dephasings in an EMRI signal for both $\alpha$-discs and $\beta$-discs, while Type~II migration could lead to dephasings of $\sim 10/10^3$ radians. The effects are larger for $\beta$ discs, since these can support  higher surface density. For more massive central black holes $\sim 10^6\,M_{\odot}$ the dominant contribution is from Bondi--Hoyle accretion (see Footnote~\ref{footnote:bondi} for a description), while for less massive black holes of $\sim 10^5\,M_{\odot}$ the dominant contribution is from the migration. These dephasings were computed for the same system parameters (apart from maximizing over time and phase shifts), so they do not account for possible parameter correlations, but the authors argue that the migration dephasing decouples from GW parameters as the effect becomes weaker with decreasing orbital radius, while relativistic effects become stronger. The relative number of EMRIs that will be produced in discs rather than from other channels is not well understood. However, it will be straightforward to identify such EMRIs, which will be circular and equatorial, to look for and constrain effects of this type. 

Another important question that has not yet been addressed is how to
distinguish the effect of external material from a difference in the
structure of the central object. If the orbit
does not intersect the material, such identification would come from
the variation in the effect over the inspiral -- if the change in
the multipole structure comes from material, then at some stage the
object would pass inside the matter, and the qualitative effect on
the inspiral would be different from that of a central object with an
unusual multipole structure. If the orbit does intersect the
material, then the spacetime-mapping analysis described above no
longer applies, since the Geroch--Hansen multipole decomposition~\cite{geroch70,hansen74} applies to vacuum spacetimes only. If this
decomposition could be generalized to nonvacuum axisymmetric
spacetimes, then low-frequency GW observation could potentially recover not
only the spacetime metric but also the structure of the
external material. It would then be possible to verify that this
matter obeys the various energy conditions (see Section~\ref{gravtheory:bhstruct}). This is an
open area of research at present.

An independent indicator of the presence of material in the spacetime
would come from the observation of an electromagnetic counterpart to an
EMRI event. For instance, if an inspiraling black hole was moving
through an accretion disc, there might be emission from the material
that was accreted onto the inspiraling object or from shocks formed
in the disc. Again, this has not been explored, although it is
likely, given the poor sky-position determination of EMRI events in GW observations~\cite{AK},
that it would not be possible to conclusively identify such a weak
electromagnetic signature in coincidence with a GW event.

The presence of exotic matter outside a black hole, in the form of a cloud of axions, was discussed in~\cite{2011PhRvD..83d4026A}. The presence of large numbers of light axions would be one consequence of extra dimensions in string theory, so the detection of an axion cloud would provide strong evidence in support of their existence. The axion cloud would modify the motion of an inspiraling black hole in a similar way to regular matter, although estimates for the precision of measurements possible with future GW detectors have not yet been carried out.  The passage of an EMRI through the cloud could also lead to its disruption, which may rule out the cloud as a possible explanation for any observed deviations, but further theoretical work is required to properly quantify these processes. 

The existence of axion clouds can also have other observable GW signatures. The axions in the cloud exist in different quantum energy states. If multiple states with the same orbital momenta $l$ and magnetic moments $m$ but different principal quantum numbers $n$ are occupied, transitions between these states can generate GWs with characteristic strain
\begin{equation}
h \sim 10^{-22} \alpha^2 \left(\epsilon_0 \epsilon_1\right)^{1/2} \left(\frac{10\mathrm{\ Mpc}}{r} \right) \left(\frac{M_{\mathrm{BH}}}{2\,M_{\odot}}\right)
\end{equation}
for a black hole of mass $M_{\mathrm{BH}}$ at a distance $r$. The pre-factor $\alpha^2 \left(\epsilon_0 \epsilon_1\right)^{1/2}$ depends on the axion masses and coupling. The axions can also undergo annihilations, which generate GWs with very similar characteristic strains
\begin{equation}
h \sim 10^{-22} \alpha^7 \epsilon \left(\frac{10\mathrm{\ Mpc}}{r} \right) \left(\frac{M_{\mathrm{BH}}}{2\,M_{\odot}}\right).
\end{equation}
In both cases the frequency depends on the black-hole mass in the usual way $f \sim 1/M_{\mathrm{BH}}$, with typical values for $2\,M_{\odot}$ black holes of $100 \, \alpha^3\mathrm{\ Hz}$ and $30\, \alpha \mathrm{\ kHz}$ respectively. Both eLISA and LIGO could place interesting constraints on the axion parameter space through (non)detections of these events~\cite{2011PhRvD..83d4026A}. Finally, the self-interactions in the axion cloud could eventually lead to the collapse of the cloud in a ``bosenova'' explosion~\cite{2012IJMPS...7...84K,2012PThPh.128..153Y}, which would generate GWs with strain
\begin{equation}
h \sim 10^{-17} \left(\frac{\epsilon}{10^{-4}}\right) \left(\frac{100\mathrm{\ Mpc}}{r} \right) \left(\frac{M_{\mathrm{BH}}}{10^{8}\,M_{\odot}}\right)
\end{equation}
and frequency $\sim c^3/GM_{\mathrm{BH}}$. These could also be an interesting source for low-frequency GW detectors. Recent calculations~\cite{2012PThPh.128..153Y} suggest that a bosenova from the Milky Way black hole at Saggitarius A* would be marginally detectable by LISA. The bosenova explosion comes about due to a super-radiant interaction between the cloud of particles and the central black hole, which extracts rotational energy from the black hole and transfers it to the cloud of particles. Recent results on these super-radiant instabilities can be found in~\cite{2013PhRvD..87d3513W}.

In~\cite{2013arXiv1302.2646M} the observable signatures of the presence of a cloud of bosons outside a massive compact object was considered. It was shown that the motion of a particle through the cloud would be dominated by boson accretion rather than by gravitational radiation reaction. During this accretion-dominated phase, the frequency and amplitude of the gravitational-wave emission is nearly constant in the late stages of inspiral. The authors also considered inspirals exterior to the boson cloud, and found that resonances could occur when the orbital frequency matched the characteristic frequency associated with the characteristic mass of the bosonic particles. These resonances could lead to significant, detectable deviations in the phase of the emitted GWs.

\subsubsection{Astrophysical perturbations: distant objects}
Perturbations to the EMRI trajectory could also arise from the
gravitational influence of distant objects, such as other stars or a
second MBH. At present, it is thought to be very
unlikely that a second star or compact object would be present in the
spacetime sufficiently close to the EMRI to leave a measurable
imprint on the trajectory~\cite{ASReview}, although detailed
calculations for this scenario have not been carried out. However, if
the MBH that was the host of the EMRI was in a binary
with a second MBH, this could perturb the trajectory
by a detectable amount~\cite{2011PhRvD..83d4030Y}. The primary observable
effect is a Doppler shift of the GW signal, which
arises due to the acceleration of the center of mass of the EMRI
system relative to the observer. It was estimated that, for typical
EMRI systems, the presence of a second black hole within a few tenths
of a parsec would lead to a measurable imprint in the signal. The
frequency scaling of the Doppler effect differs from the scaling of
the post-Newtonian terms in the unaccelerated waveform, which
suggests that this effect will be distinguishable in GW
observations~\cite{2011PhRvD..83d4030Y}. The magnitude of the leading-order
Doppler effect scales as $M_{\mathrm{sec}}/r^2$, where $M_{\mathrm{sec}}$ is
the mass of the perturbing black hole, and $r$ is its distance. If
the second object is within a few hundredths of a parsec, higher-order time derivatives could also be measured from the GW
observations. These scale differently with $M_{\mathrm{sec}}$ and $r$,
which would in principle allow the mass and distance of the perturber
to be measured from the EMRI data~\cite{2011PhRvD..83d4030Y}.

The probability that a second black hole would be within a tenth of a
parsec of a system containing an EMRI is difficult to assess. At
redshifts $z < 1$, at any given time a few percent of Milky Way-like galaxies
will be involved in a merger, which suggests an upper limit of a few
percent of EMRIs that could have perturbing companions. However, there is
uncertainty as to how long the black-hole binary will spend at radii
of a few tenths of a parsec following a merger, and it is plausible
that the presence of a second black hole would increase the EMRI rate
by perturbing stars onto orbits that pass close to the other black
hole, introducing an observational bias in favor of these
systems~\cite{2011PhRvD..83d4030Y}. Given these uncertainties, the
possibility of a distant perturber will have to be accounted for in
the analysis, and, if it is observed in some systems, LISA-like detectors
could indirectly inform us of the processes that drive MBH mergers.

The gravitational influence of a second stellar mass black hole on the
evolution of an EMRI were considered in~\cite{2012ApJ...744L..20A}. It
was shown that a second black hole with a semi-major axis of $\lesssim
10^{-5}\mathrm{\ pc}$ could influence the orbital parameters of an
EMRI ongoing in the same galaxy and already in the low-frequency GW
band. This influence is chaotic, leading to an unpredictable evolution of
the orbital parameters. It was estimated that in 1\% of EMRIs a second
black hole could be within the required distance. However, the
timescale of the chaotic motion was $\sim10^{9}\mathrm{\ s}$, which is
much longer than a typical low-frequency observation, and the system
considered in~\cite{2012ApJ...744L..20A} had an orbital period of
$6\times10^{3}\mathrm{\ s}$, which would probably not yet be detectable by LISA-like detectors. On the timescale of a GW mission, the effect would probably manifest itself as a linear drift in the orbital parameters, and therefore it is unlikely that this effect would prevent the detection of an EMRI signal. A full analysis of the effect on parameter estimation has not yet been carried out.

\subsubsection{Properties of the phase space of orbits}

\paragraph*{Loss of the third integral.}
\label{intloss}

It was demonstrated by Carter~\cite{carter68} that the Kerr metric
has a complete set of integrals -- in addition to the energy and
angular momentum that arise as conserved quantities in any stationary
and axisymmetric spacetime, geodesics in the Kerr metric conserve the
Carter constant, $Q$. This is the analog of the third isolating
integral found in some classical axisymmetric systems, and has been
shown to arise due to the existence of a Killing tensor of the
spacetime~\cite{walker70}. In the Schwarzschild limit, $Q$ is the sum
of the squared angular momentum components in the two equatorial
directions. The Kerr metric is one of a very special class of metrics
that have this property. In fact, Carter~\cite{carter68} demonstrated
that it was the only axisymmetric metric not containing a
gravitomagnetic monopole for which both the Hamilton--Jacobi and
Schr\"{o}dinger equations were separable. The special nature of the
Kerr metric was emphasized by Will~\cite{2009PhRvL.102f1101W} who demonstrated
that, in Newtonian gravity, motion about a body with an arbitrary set
of multipole moments $M_l$ possesses a third integral of the motion
only if the multipole moments obey the conditions
\begin{equation}
M_{2l+1} = 0, \qquad M_{2l} = m (M_2/m)^l \quad \mathrm{for\,every\,}l,
\end{equation}
which are precisely the conditions satisfied by the mass moments of a
Kerr black hole, Eq.~(\ref{nohair}). 

The separability of the Kerr metric aids the analysis of inspirals in
that spacetime, but it also suggests another potential observable that
would show a deviation from the Kerr metric in a GW
observation. The specialness of the third integral in the Kerr
spacetime suggests that if a spacetime differed from the Kerr
solution, even by a small amount, the third integral might vanish,
which would potentially lead to the existence of chaotic orbits. If
such orbits were observed it would be a clear ``smoking gun'' for a
deviation from the Kerr metric. The existence of chaotic orbits in
various spacetimes has been explored by several authors. The standard
approach is via construction of a Poincar\'{e} map: a geodesic is
computed in cylindrical coordinates $(\rho, z, \phi, \theta)$, and
the values of $\rho$ and $\dot{\rho}$ recorded every time the
particle intersects a specified plane $z=$constant. If the resulting
plot of all these points on a ($\rho, \dot{\rho}$) plane yields a
closed curve, then a third integral exists, otherwise the orbit is
chaotic. This is illustrated in Figure~\ref{ergfig}.

\epubtkImage{Pmapout-Pmapin.png}{%
\begin{figure}
\centerline{
\includegraphics[angle=0,keepaspectratio=true,width=3in]{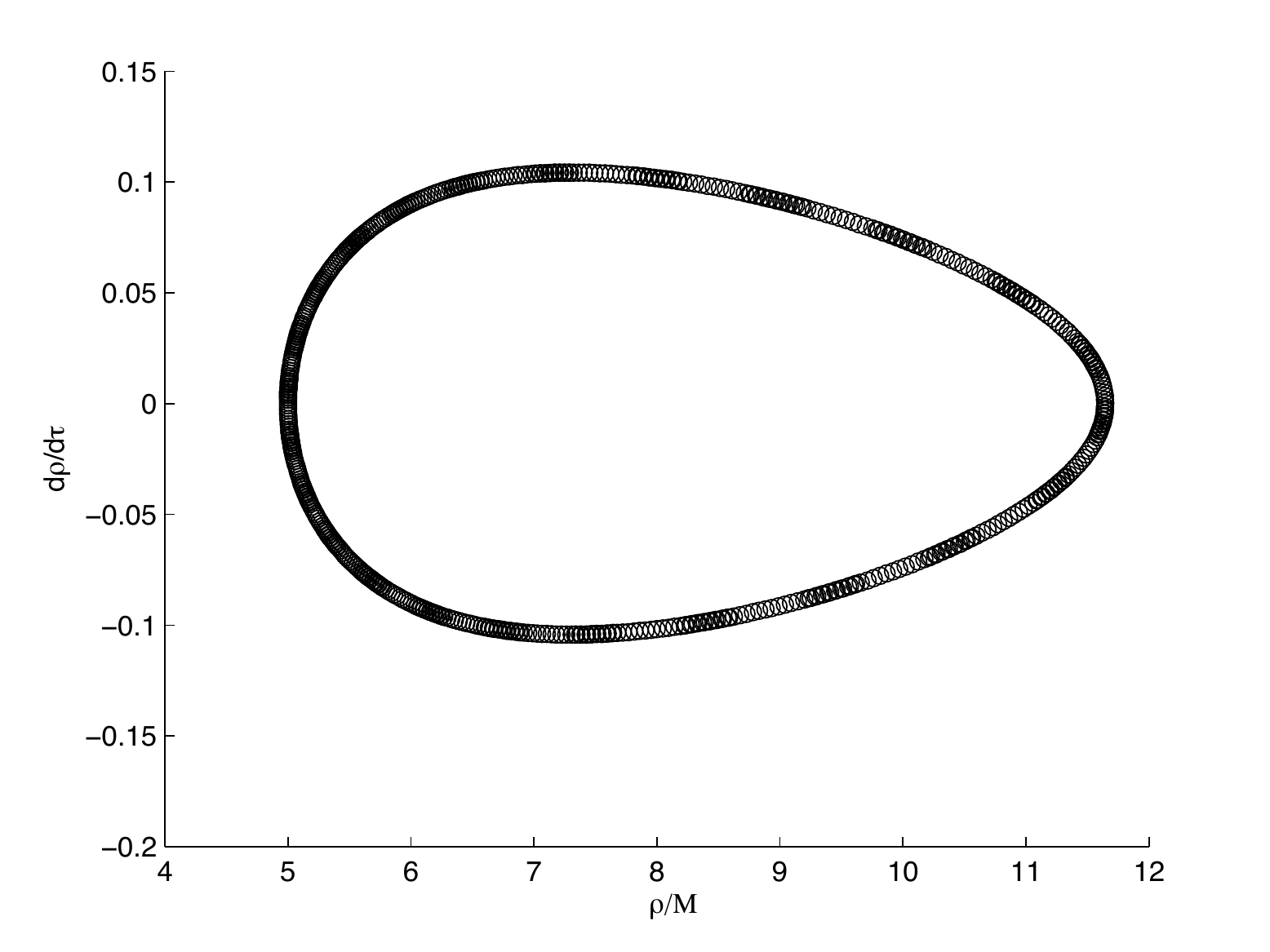}
\includegraphics[angle=0,keepaspectratio=true,width=3in]{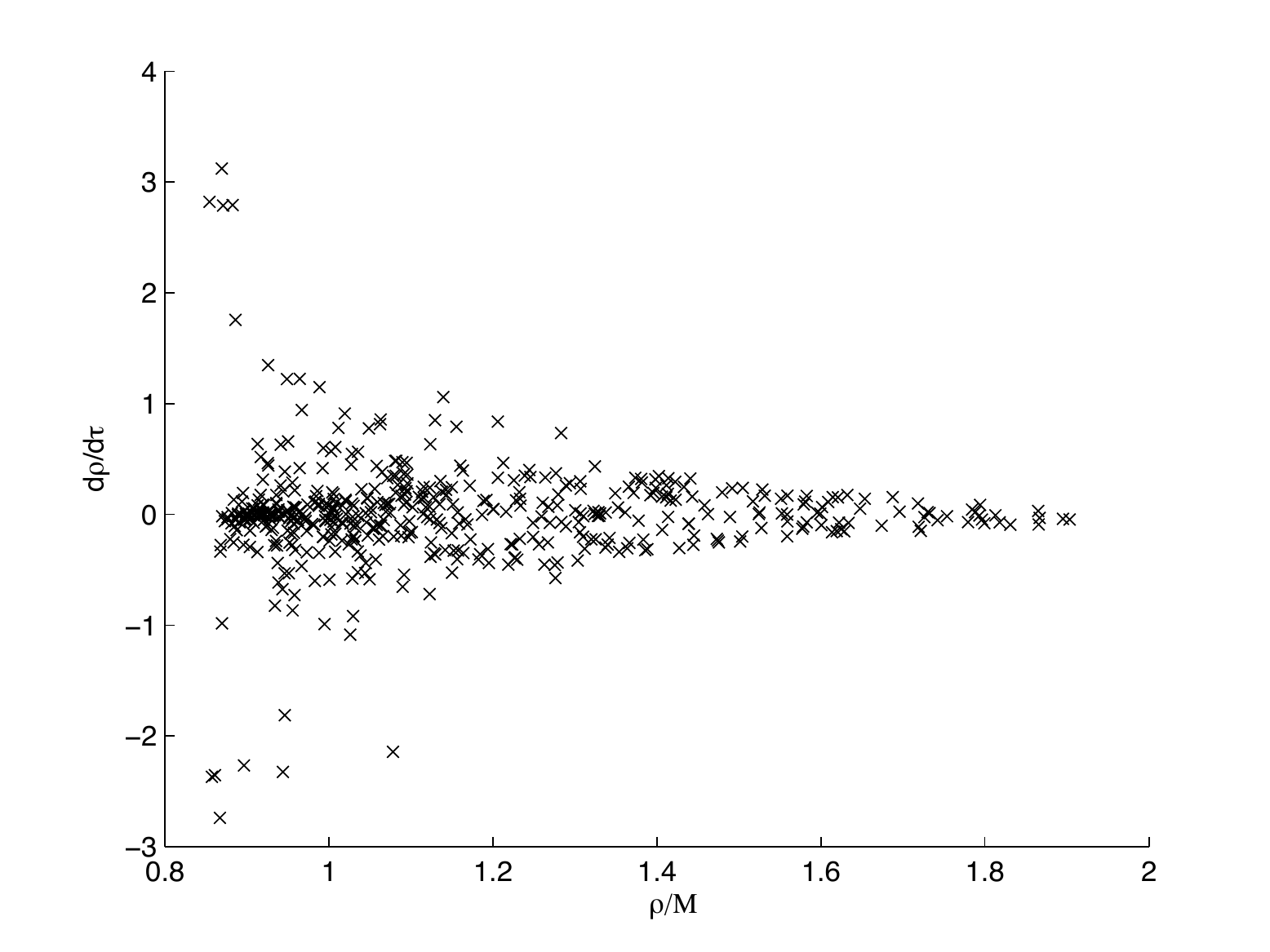}
}
\caption{Poincar\'{e} map for a regular orbit (left panel) and a
chaotic orbit (right panel) in the Manko--Novikov spacetime. 
Image reproduced by permission from~\cite{gairbumpy}, copyright by
APS.}
\label{ergfig}
\end{figure}}

Chaotic motion has been found by Sota et al. for orbits in the
Zipoy--Voorhees--Weyl and Curzon spacetimes~\cite{sota96}; by Letelier
and Viera for orbits around a Schwarzschild black hole perturbed by
GWs~\cite{letelier97}; by Gu\'{e}ron and Letelier for
orbits in a black-hole spacetime with a dipolar halo~\cite{gueron01},
and in prolate Erez--Rosen bumpy spacetimes~\cite{gueron02}; and by
Dubeibe et al. for some oblate spacetimes that are deformed
generalizations of the Tomimatsu--Sato spacetime~\cite{dubeibe07}.
None of these examples represented systems that were small deviations
from the Kerr metric. The only investigation to date of chaotic
orbits in the context of LISA was by Gair et al.~\cite{gairbumpy},
who explored geodesic motion in a family of spacetimes due to Manko
and Novikov~\cite{mankonov} that had arbitrary multipole moments, but
which included Kerr as a special case. Gair et al.~\cite{gairbumpy}
considered orbits in a family of spacetimes parameterized by a single
``excess quadrupole moment'' parameter, $q$, such that $q=0$
represented the Kerr solution. They found that, while the majority of
orbits in these spacetimes possessed an apparent third integral,
chaotic orbits existed very close to the central object for
arbitrarily small \emph{oblate} deformations of the Kerr solution. As
the spacetime was deformed away from Kerr, a second allowed region
for bound geodesic motion was found to appear close to the central
black hole, in addition to the allowed region present in the Kerr
metric. Chaotic orbits were found only in this additional bound
region. Gair et al.~\cite{gairbumpy} concluded that this chaotic
region would probably be inaccessible to an object that was initially
captured at a large distance from the central object. This analysis was revisited in~\cite{geraktransres} but the conclusions in that paper were the same. The only difference was that the authors in~\cite{geraktransres} identified a region of stable motion within the inner region that contains chaotic orbits. The chaotic orbit shown in the right hand panel of Figure~\ref{ergfig} appears to pass in and out of the region of stability which should not happen, so this might be a numerical artifact. However, the existence of chaotic orbits and the probable inaccessibility of these orbits to inspirals was confirmed by~\cite{geraktransres}.

Brink~\cite{brink08b} also explored integrability in arbitrary
stationary, axisymmetric and vacuum spacetimes, concentrating on
regions where the Poincar\'{e} maps indicated the presence of an
effective third integral. Brink hypothesized that some spacetimes
might admit an integral of the motion that was quartic in the
momentum, in contrast to the Carter constant, which is quadratic. This hypothesis is as yet unproven. There is also a
potential conflict with the example given in~\cite{gairbumpy}. The
Kolmogorov, Arnold, and Moser (KAM) theorem indicates that when a
Hamiltonian system with a complete set of integrals is weakly
perturbed, the phase-space motion will either be confined to the
neighborhoods of the invariant tori, or the motion will be
chaotic~\cite{tabor89}. Thus, if there is a region of the spacetime
where chaotic motion exists, there cannot be another region where a
third invariant exists. A mathematical demonstration that the orbits
can possess an approximate invariant, while technically being
chaotic, is lacking at present, although this does appear to be the
case from the numerical calculations~\cite{gairbumpy}. It is also not
entirely clear that the perturbation can be regarded as ``small''
everywhere, since the change in mutipole moments necessarily changes
the horizon structure and so the perturbation is infinitely large at
certain points.

A discussion of the reason for the existence of chaotic geodesics in
some spacetimes was given by Sota et al.~\cite{sota96}. They
suggested that it would arise either from a change in sign of the
eigenvalues of the Weyl tensor at a point, which would lead to a
local ``instability,'' or from the existence of homoclinic orbits.
The latter explanation applied only to non-reflection-symmetric
spacetimes, while most spacetimes of astrophysical interest should be
reflection symmetric. The Weyl-tensor analysis has not been carried
out for the Manko--Novikov family of spacetimes~\cite{gairbumpy}. One
other proposed explanation was the existence of a region of closed
timelike curves, which was found to touch the region in which chaotic
motion was identified~\cite{gairbumpy}.

In conclusion, it seems unlikely that chaotic geodesics will be found
in nature, but if they were identified we would know immediately that
the spacetime was not Kerr. However, detecting chaotic motion from a
GW observation is challenging. One possibility would
be to observe the transition from regular to chaotic motion in a
time-frequency analysis: the regular motion would be characterized
by a few well-distinguished peaks in a Fourier transform of the
signal, while chaotic motion would show a much broader band
structure~\cite{gairbumpy}. However, it is not clear that it would be
possible to distinguish the chaotic phase from detector noise, and
hence there would be no way to identify an inspiral that ``ends'' by
entering a chaotic phase as opposed to one which ends at plunge into
a black hole. If an orbit passed into a chaotic phase and then back
into a regular phase we might see a signal turn ``on'' and ``off''
repeatedly. However, the chaotic motion would randomize the phase at
the start of the regular motion, so to detect such a signal we would
need each regular phase to be long enough that they were individually
resolvable by matched filtering. This would require extreme fine
tuning of the system parameters~\cite{gairbumpy}.

\paragraph*{Persistent resonances.}
Eccentric and inclined EMRIs will generically
pass through transient resonances at which the radial and $\theta$
frequencies become commensurate. For EMRIs in GR, these resonances
will be isolated, and the transition through resonance will proceed on the usual radiation-reaction timescale but with a temporary modification in the energy flux on
resonance~\cite{2012PhRvL.109g1102F}. However, according to the
Poincar\'{e}--Birkhoff theorem, when an integrable system is perturbed
it causes the appearance of a Birkhoff chain of islands whenever the
frequencies of the system are at resonance. Therefore, in a perturbed
Kerr spacetime, another observable consequence would be that the EMRI
frequencies could remain on resonance for many more cycles, providing
another ``smoking gun'' for a deviation from
Kerr~\cite{aposres,geraktransres}. Detection of a persistent
resonance in a matched-filtering search will require a modification
of the search pipeline, but it should be considerably more
straightforward than detection of chaos, as the signal will be
coherent and could therefore be identified using time-frequency
methods, or a phenomenological waveform model. However, this has not
yet been studied in any detail.

As mentioned in Section~\ref{sub.quadrupole}, in massive scalar-tensor theories, a different type of persistent resonance can occur, in which a super-radiant scalar flux balances the GW flux~\cite{2011PhRvL.107x1101C,2012PhRvD..85j2003Y}. Such resonances can last a significant fraction of a Hubble time and so observing a single EMRI offers significant constraining power on the space of massive scalar-tensor theories. This is not a test of black-hole structure, since the resonance is between the scalar and gravitational fluxes, rather than in the geodesics of the central object, but the observable effects are similar so we mention it here.

\subsubsection{Black holes in alternative theories}
\label{sec:BHinalt}

\paragraph*{Kerr black holes.}
The Ryan mapping approach uses observations of precession frequencies
as functions of the orbital frequency to extract the spacetime metric
from GWs. However, even if the metric is found to be Kerr, this is not enough to verify GR. It was pointed
out by Psaltis et al.~\cite{psaltis08} that the Kerr metric is a
solution to the field equations for several alternative theories of
gravity. Essentially, since the Kerr metric has vanishing Ricci
tensor, $R_{\mu\nu}=0$, any theory in which the vacuum field
equations depend only on $R_{\mu\nu}$ will also admit Kerr as a
solution. Allowing for a nonzero cosmological constant, $\Lambda$, a
black-hole solution in GR satisfies 
\begin{equation}
R_{\mu\nu}=\frac{R}{4}g_{\mu\nu}, \qquad R=4\Lambda, \qquad \mbox{and}
\qquad R_{,\mu}=0,
\end{equation}
in which $R$ denotes the Ricci scalar and $g_{\mu\nu}$ is the
spacetime metric~\cite{psaltis08}. Psaltis et al.\ discussed four
different alternative theories, already described in Section~\ref{standardmodel}

\begin{itemize}
\item \textbf{\textit{f(R)} gravity in the metric formalism.} If $f(R)$ is expanded as a Taylor series $a_0 + R + a_2 R^2 + \cdots$, there are three possible cases. (i) If $a_0=0$ the Kerr solution, which corresponds to $R=0$, is always a solution to the equations of motion. (ii) If the Taylor series
terminates at $a_2$, all constant curvature solutions of GR (with any $\Lambda$) remain exact solutions of the $f(R)$ theory. (iii)
If $a_0 \neq 0$ and the series does not terminate at $a_2$, constant-curvature solutions of GR will be solutions of the $f(R)$ theory with
different values of the curvature. The difference in
curvature will be small, however.

\item \textbf{\textit{f(R)} gravity in the Palatini formalism.} In this case, any constant-curvature solution of GR is also a solution to these equations, with the same Christoffel symbols. This is
unsurprising, as it is known that Palatini $f(R)$ gravity reduces to GR in vacuum~\cite{EBpsaltisreply}.

\item \textbf{General quadratic gravity.} For any black-hole solution in GR, the tensors
$K_{\kappa\lambda}$ and $L_{\kappa\lambda}$ that appear in the field equations (\ref{quadraticFEs}) both vanish, and the field equations reduce to those of GR. Hence all black-hole solutions of GR are solutions in this theory.

\item \textbf{Vector-tensor gravity.} In this case, we find once again that
all constant-curvature solutions of GR are solutions to the
equations, but with a shifted value of the curvature that depends on
the vector field strength $R = 16\Lambda/(4+(4\omega + 3\eta)K^2)$.
\end{itemize}

The action for general quadratic gravity introduced by Stein and Yunes~\cite{SYBHinalt} also admits the Kerr solution, but only in the non-dynamical version of the theory in which the functions $f_i(\theta)$ are constants. In that case the field
equations are once again satisfied by spacetimes with $R_{ab}=0$, and
so the vacuum solutions of GR are solutions to the field equations in
these theories as well. In the \textit{dynamical} version of the theory,
the Riemann tensor enters the field equations explicitly and so
$R_{ab}=0$ is no longer sufficient to satisfy them. We will discuss
those black-hole solutions in the following subsection.

Although it is true that all of these theories admit the Kerr metric
as a solution, this does not mean that we have no way to distinguish
between them via GW observations. This was not
discussed in~\cite{psaltis08}, but is argued in a comment on that
paper by Barausse and Sotitirou~\cite{EBpsaltisreply}. First, the
uniqueness theorems of
GR~\cite{penrose65,hawkpen70,israel67,carter71,rob75} do not
necessarily apply in these alternative theories. In other words, just
because the Kerr metric is a solution does not mean that we would expect
it to form as a result of gravitational collapse. This is an equally
important consideration as to what we would expect to see in the
universe, although this argument can be sidestepped by suitable
fine tuning. In $f(R)$ gravity, the metric of a spherically-symmetric body is not the Schwarzschild metric, but has a Yukawa correction~\cite{2009MPLA...24..659C}. The constraints that LISA could place on such a deviation from the Kerr solution were investigated in~\cite{BerryGair}. Expanding  $f(R)$ to quadratic order ($R+a_2 R^2/2$), it was found that EMRI observations could place a bound $|a_2| \lesssim 10^{17} \mathrm{\ m}^2$, about an order of magnitude better than the bound from observations of planetary precession in the solar system, $|a_2| < 1.2\times10^{18} \mathrm{\ m}^2$. However, the bound from the E\"{o}t-Wash laboratory experiments is many orders of magnitudes better, $|a_2| < 2\times10^{-9} \mathrm{\ m}^2$~\cite{2009PhRvL.102n1301C,BerryGair}.

The GW constraints will be obtained in a very different curvature regime and could be of interest if something like the ``chameleon mechanism'' is invoked. The chameleon mechanism was introduced to allow $f(R)$ models to explain cosmological acceleration without violating laboratory and solar-system constraints~\cite{2004PhRvL..93q1104K}. It is a nonlinear effect that could arise when the curvature is very different from the background value, e.g., in the vicinity of matter. If the matter density is high the scalar degree of freedom in $f(R)$ gravity acquires an effective mass, which means that the effective coupling to matter becomes much smaller than the bare coupling, which is relevant on cosmological scales. Therefore the bare coupling could be much higher than inferred from laboratory constraints, allowing the theories to explain cosmological acceleration (see~\cite{2010LRR....13....3D} for a full description of the mechanism and complete references). In a similar way, the effective coupling in the vicinity of a compact object could in principle be different from that in the laboratory and so the weak constraints from gravitational-wave observations are still interesting because they probe a different curvature scale.

Subsequent to the publication of~\cite{BerryGair}, it has been shown that the end state of gravitational collapse in $f(R)$ theories (and scalar-tensor theories) is not the point-mass limit of an extended body, but is in fact the Kerr solution~\cite{2012PhRvL.108h1103S}. Therefore the results in~\cite{BerryGair} do not apply to black holes in $f(R)$ gravity, but would apply for an exotic horizonless compact object if one existed. The constraints that low-frequency GW could place on this metric can also be considered to be constraints on a strawman Yukawa-like deviation from the Kerr solution, without reference to a specific theory in which such corrections arise.

Another 
consideration is that the 
spacetime metric is not the only
GW observable. GWs are generated by
perturbations of the spacetime and hence depend on the full dynamical
sector of the theory. Therefore the response to perturbations will be
different if the field equations are different, even if the unperturbed metric is
the same
. In~\cite{EBpsaltisreply} it was demonstrated that, for
metric $f(R)$ gravity, linearizing about the Minkowski spacetime gives rise
to massive graviton modes in addition to the standard
transverse-traceless modes of GR. These cannot be zeroed out by a
gauge transformation. To date, no one has considered the generation
of such massive modes in a binary system nor the observational
consequences. However, if these modes are generated, there would be
two natural ways to find them: a direct GW detection (the
modes have different polarization states and propagation velocities),
and an indirect detection, as the binary system would inspiral faster than
expected due to loss of additional energy in the extra modes. It is
not clear whether the latter will be distinguishable from tidal
interaction effects, and quantitative estimates of the power of tests
that would be possible with LISA-like detectors have not yet been done.

A second example of an alternative theory in which a spacetime metric
is the same as a GR solution but behaves differently perturbatively is
dynamical Chern--Simons (CS) gravity. The nonrotating--black-hole
solution in that theory is Schwarzschild, as in GR, but it was shown
in~\cite{2011PhRvD..83j4048P} that there are 7PN and 6PN corrections
respectively to the scalar and gravitational energy flux radiated to
infinity by a circular EMRI. This could lead to a 1-cycle difference
in waveform phasing over a year of inspiral if the CS coupling
parameter $|\zeta| \gtrsim 0.1$. In~\cite{2012PhRvD..85f4022Y}
post-Newtonian results were obtained for the emission from spinning
black-hole binaries in a general quadratic-gravity theory. One model
within the general class considered corresponds to dynamical CS
gravity and for that case the results were found to be consistent with
the perturbative calculations in~\cite{2011PhRvD..83j4048P}.

In interpreting GW observations, it will be necessary to verify
both the static and dynamic aspects of the theory. The Ryan mapping
algorithm is the natural way to start, but if a metric is found to be
consistent with Kerr it will then be necessary to verify that the
observed GWs and energy loss are also in agreement
before concluding that we are observing a system consistent with a
Kerr black hole in GR.

\paragraph*{Non-Kerr black-hole solutions.}\label{nonKerrBH}
Certain alternative theories of gravity do not admit the Kerr metric
as a solution: these include the dynamical version of general
quadratic gravity described by Eq.~(\ref{genquadgrav}). In general,
it is difficult to solve the field equations in alternative theories,
so few analytic black-hole solutions are known outside of GR.
Some solutions are known, but only under certain approximations.
Solutions for slow rotation are known in dynamical CS gravity, which is a special case of Eq.~(\ref{genquadgrav}) with $\alpha_1=\alpha_2=\alpha_3=0$, to leading order in spin~\cite{YPCSBH} and now to quadratic order~\cite{2012PhRvD..86d4037Y}. Slow-rotation solutions are also known in Einstein-Dilaton-Gauss--Bonnet
(EDGB) gravity~\cite{EDGBsol}, in which gravity is coupled to
dilaton and axion fields. In the weak-coupling limit, spherically
symmetric and stationary solutions to the general quadratic-gravity
theory described by Eq.~(\ref{genquadgrav}) take the same form as the
solutions to EDGB gravity and were derived in~\cite{SYBHinalt}. Solutions to this same class of theories are also known for arbitrary coupling, but only under the slow rotation approximation, and these were given in~\cite{2011PhRvD..84h7501P}. 

In dynamical CS gravity, it was shown that the metric
describing a slowly rotating black hole at linear order in the spin~\cite{YPCSBH} differs from the
slow-rotation limit of the Kerr metric beginning at the fourth multipole,
$l=4$~\cite{SYCSemri}. It was also shown that the equations describing perturbations of this metric in dynamical CS gravity coincide with those of GR at leading order in spin and coupling parameter, with corrections due to the sourcing of metric perturbations by the scalar field entering at higher order. Therefore it was argued that the deviations from GR will manifest themselves primarily through differences in the geodesic orbits in the spacetime.  Therefore this is completely analogous to
spacetime mapping within GR, as described in Section~\ref{testnohair}, so the prospects for detection of such deviations with LISA-like detectors are comparable. The energy-momentum tensor of the GWs was also shown to take the same form in terms of the metric perturbation as it does in GR, so the leading-order correction to the GW energy flux is determined by these modifications to the conservative dynamics. 
Energy balance also applies in this context and can be used to relate the adiabatic evolution of an orbit to the flux of energy and angular momentum at infinity. However, energy and angular momentum are also carried to infinity by the scalar field, and these corrections to the evolution were not estimated in~\cite{SYCSemri}.
Considering non-inspiraling geodesic orbits, Sopuerta and Yunes~\cite{SYCSemri} estimated that IMRI or EMRI events observed by LISA would be able to put constraints on the coupling parameter $\xi$ of
dynamical CS gravity of $\xi^{\frac{1}{4}} \lesssim 10^2\mathrm{\ km}$. A more complete study that included inspiral and used a Fisher matrix analysis to account for parameter correlations is described in~\cite{2012PhRvD..86d4010C,2012JPhCS.363a2019C}. There it was found that LISA observations of EMRIs could place a bound $\xi^{\frac{1}{4}} < 10^4\mathrm{\ km}$. This is somewhat weaker than the bound estimated for IMRIs in~\cite{SYCSemri}, but better than the estimate for EMRIs in that paper, so IMRI systems (involving a $\sim10\,M_{\odot}$ object inspiraling into a $\sim10^3\,M_{\odot}$ object) may be able to place even more stringent constraints than originally estimated, if they were observed. The EMRI bound $\xi^{\frac{1}{4}} < 10^4\mathrm{\ km}$ is four orders of magnitude better than the best solar-system bound, which is based on data from Gravity Probe B and LAGEOS satellites~\cite{GPBxiBound,LAGEOSxibound}.

In~\cite{YPCSBH} the bound from current binary-pulsar observations was estimated to be four orders of magnitude better than that in the solar system and hence comparable to the expected GW result. However, this bound was based on an upper limit on the rate of precession. It was argued in~\cite{AliHamCSBound} that an upper bound could not place a constraint on the CS deviation since the sign of the CS contribution is opposite to that in GR, and so a bound could only be placed if a lower bound on the precession lying below the GR value could be found. Therefore the solar-system bound is the best current constraint, which GW observations will improve by several orders of magnitude. The Chern--Simons black-hole metric has recently been derived to quadratic order in the spin~\cite{2012PhRvD..86d4037Y}, so the results discussed here can now be extended to this higher-order metric. This second-order solution had several interesting properties, in particular the Petrov type~\epubtkFootnote{The Petrov type of a metric describes the algebraic properties of the Weyl tensor, the vacuum part of the Riemann tensor. The Petrov classification is based on the multiplicities of the null eigenvectors (principal null directions) of the Weyl tensor. A metric of Petrov type D is algebraically special, having two repeated (double) principal null vectors, while metrics of type I have no algebraic symmetry, possessing four simple principal null directions.} is type I, whereas it was type D in the linear-spin case; furthermore, there is no second-order Killing tensor, which means orbits do not have a third Carter-constant--like integral of the motion. This could have important observable consequences for GWs from EMRIs, as discussed in Section~\ref{intloss}. The analogous calculation to~\cite{2012PhRvD..86d4010C}, the evolution of an EMRI in the strong-field region of the metric, has not yet been carried out for the second-order CS metric. However, the GW emission from quasi-circular binaries in this theory has been determined without any restriction on the spin magnitude or coupling, but in the post-Newtonian limit~\cite{2012PhRvL.109y1105Y}, i.e., with a restriction on the magnitude of the velocity. This waveform model was constructed using energy balance and expressions for the gravitational-wave energy flux emitted in general quadratic-gravity theories that were derived in~\cite{2012PhRvD..85f4022Y}. The model was used to estimate possible GW constraints on CS gravity. These were given in Section~\ref{sub.quadrupole} and are two orders of magnitude better than the EMRI constraints estimated using the linear-spin metric~\cite{2012PhRvD..86d4010C}. These constraints were derived in a very different mass-ratio regime (a binary with mass ratio of 1:2), but the fact that they are so much better is probably not surprising since the quadratic in spin corrections to the metric enter at a lower post-Newtonian order than the linear-spin corrections. Therefore it is likely that the bounds that EMRIs could place on CS gravity are much stronger than quoted here. This possibility and possible qualitative signatures of the loss of the third integral in the higher-order CS metrics should be further investigated.

In the small-coupling limit, it can be shown that the energy-momentum
tensor of GWs in the general quadratic-gravity theory
[Eq.~(\ref{genquadgrav})] also follows the same quadrupole
formula as in GR~\cite{SYTinalt}. Using this result, a post-Newtonian expression for the energy flux emitted from a quasi-circular binary in these theories was obtained in~\cite{2012PhRvD..85f4022Y}. Corrections to the energy flux come from changes to the conservative dynamics, i.e., to the orbits that test particles follow in the background metric, from energy lost in scalar field radiation and from the contribution to the GW energy-momentum integral from metric perturbations sourced by the scalar field. The authors found that, for these theories, the scalar dipole radiation dominates the correction to the energy flux and enters at a post-Newtonian order of ``--1'', i.e., at a power of $v/c$ one order before the GW energy flux of GR. There are also $0$PN and $2$PN corrections to the flux from the scalar-metric interaction and the modification to the conservative dynamics. 

The observability of these waveform differences with space-based gravitational detectors has, so far, only been directly estimated for the specific case of Einstein-Dilaton-Gauss--Bonnet gravity~\cite{2012PhRvD..86h1504Y}. In that case, the constraints derivable with space-based GW detectors were compared to those that can be obtained from observations of low-mass X-ray binaries. Some of these are observed to have orbital decay rates that are larger than predicted by radiation reaction in GR. Assuming this excess orbital-decay rate arises from additional scalar radiation emitted from the binary, it is possible to obtain a bound six times stronger than that derivable from solar-system experiments (using the binary A0620-00) of $\sqrt{|\alpha|} < 1.9\times 10^5\mathrm{\ cm}$. eLISA would be able to place a bound that is slightly stronger (a factor of two) than this, while combining multiple DECIGO observations would yield a bound three orders of magnitude better~\cite{2012PhRvD..86h1504Y}. 

For other theories in this general quadratic class, GW bounds have not yet been directly assessed. However, the ``post-Einsteinian'' parameters have been calculated for circular-equatorial inspirals in this class of theories. The parameters characterizing the modification to the GR waveform in the ppE framework are defined through the expression
\begin{equation}
\tilde{h} = |\tilde{h}_{\mathrm{GR}}| \left(1 + \alpha \eta^c u^a\right) \exp\left[\mathrm{i} \Psi^{\mathrm{GR}}_{\mathrm{GW}} \left(1 + \beta \eta^d u^b\right)\right],
\end{equation}
in which $\eta = m_1 m_2/ (m_1+m_2)^2$ is the mass ratio of a binary with component masses $m_1$ and $m_2$, $u = \pi {\cal M} f$, ${\cal M} = \eta^{3/5} (m_1+m_2)$ is the chirp mass and $\tilde{h}_{\mathrm{GW}}$/$\Psi^{\mathrm{GR}}_{\mathrm{GW}}$ are the general-relativistic waveform and waveform phase respectively. Under the weak-coupling approximation, the metric of non-spinning black holes in these general quadratic-gravity theories depends only on a parameter $\zeta \propto \alpha_3^2$. For mergers of such black holes, the ppE parameters were found to be $\alpha=(5/6)\zeta$, $\beta=(50/3)\zeta$, $a=4/3=b$, $c=-4/5=d$~\cite{SYBHinalt}. More recently, the ppE corrections to the waveform phase (which are parametrized by $b$ and  $\beta$) were found without making a weak-coupling approximation and for binaries with spinning components. The exponent $b$ depends on whether the black holes are spinning and on which coefficients of the theory are allowed to be non-zero. The authors considered the cases of odd parity (i.e., including the Pontryagin term $\alpha_4 \neq 0$) and even parity (i.e., including the other terms quadratic in the curvature, $\alpha_1, \alpha_2, \alpha_3 \neq 0$) separately. The exponent of the ppE phase correction was found to be $b=-7/3$ for spinning black holes in even-parity theories, $b=-1/3$ for spinning black holes in odd-parity theories, and $b=3$ for non-spinning black holes~\cite{2012PhRvD..85f4022Y}. The authors also obtained equations, which we do not present here, relating the ppE amplitude, $\beta$, to the coupling constants of the theory. If a GW observation is used to place bounds on these ppE parameters, these expressions can be used to translate the bound into a constraint on this set of alternative theories of gravity.

As discussed in Section~\ref{sec:ppE}, the authors of~\cite{2011PhRvD..83j4027V} considered more general possible forms for black-hole solutions without specifying a particular alternative theory. By imposing the existence of a Carter-like third integral of the motion, asymptotic flatness, and the ``peeling theorem''\epubtkFootnote{``Peeling'' refers to the fact that different parts of the metric (specifically the Weyl tensor) fall off at different rates with distance. The dominant monopole component is effectively the mass of the central object, while the next-order component, the current dipole, is effectively the spin or angular momentum of the central black hole. The ``peeling'' constraints imposed in~\cite{2011PhRvD..83j4027V} ensured that the mass and spin were not changed by the perturbation. Such changes are already contained within the Kerr metric family and do not represent modifications to GR.}, and by setting as many metric components to zero as possible while still recovering all known modified gravity spacetime metrics, they obtained a family of generic modified black-hole solutions.  It was subsequently realized that all solutions admitting a Carter constant had been found in~\cite{GeneralCarterConst}, but~\cite{2011PhRvD..83j4027V} identifies the physically-realistic subset of these general solutions. This approach is suited to the construction of eccentric, inclined inspirals, and in~\cite{GairYunesGenericEMRI} the GWs for generic EMRIs occurring in these metrics were constructed by extending the ``analytic-kludge'' framework~\cite{AK}. As yet, these waveforms have not been used to assess the ability of a space-based detector to constrain such deviations from the Kerr solution, but work in this area is ongoing.

\subsubsection{Interpretation of observations}
The previous subsections~\ref{testnohair}--\ref{sec:BHinalt} have identified some of the possible causes
of deviations in the structure of the central object from the Kerr
metric, as well as some observational consequences of these deviations.
Nevertheless, interpreting GW observations correctly is a nontrivial challenge. Our working assumption is that
the massive compact objects that occupy the centers of most
galaxies are indeed Kerr black holes. Therefore it is reasonable to
design an approach to spacetime mapping that looks for inspirals into
Kerr black holes and quantifies any deviations from such inspirals
that may be present. One such approach to spacetime mapping was
described in~\cite{2008RSPTA.366.4365G}. The starting point is to assume
that GR is correct, and that the source's spacetime is vacuum and axisymmetric. The
multipole moments can be extracted from the precession frequencies
via Ryan's algorithm, and then the expected GW and
inspiral rate for such a spacetime in GR can be computed.

If the
observations are consistent there is no evidence to contradict the
initial assumptions, but we can still do several tests, by asking the following questions: (i) Does
the ``no-hair'' property hold for the multipoles? If there is a
deviation, the spacetime must either contain closed-timelike curves,
or lack a horizon. (ii) Does the emission cut off at plunge or does
the radiation persist? (iii) Is the tidal interaction consistent with
a Kerr black hole? If the radiation still cuts off at a plunge and
the multipole structure is not consistent with the Kerr metric, it
might indicate the presence of a naked singularity, which would be a
violation of the cosmic-censorship hypothesis.

If the observations
are not consistent with GR, the first assumption to relax would be
that the spacetime was vacuum. In principle, it might then be
possible to deduce both the multipole moments of the spacetime and the
energy-momentum tensor of the matter distribution from the
observations. The GW could then be computed for
such a spacetime in GR, and checked for consistency with the
observations. If the observations are consistent, similar questions can
be asked: (i) Does the emission cut off at plunge or does the
radiation persist? (ii) Is the tidal interaction consistent with a
Kerr black hole? (iii) Does the matter distribution obey the strong,
null, and weak energy conditions, or have we identified an exotic
matter distribution? Only if the observations are inconsistent will
there be evidence of a failure in GR itself.

The main complication to this approach is that the detection of EMRIs
will rely on matched filtering using waveform templates. If a system
differs significantly from a Kerr inspiral, then it may not be picked
out of the data stream. An alternative approach was discussed by Brink~\cite{brink08a}.  She suggested that the
spacetime-mapping problem could be thought of as analogous to inverse-scattering problems in quantum mechanics, where the potential to be
determined is the Ernst potential~\cite{ernst68} (see discussion in Section~\ref{testnohair}), but the technical
details of such a method have not yet been worked out. Brink also
suggested an approach with a similar philosophy to what we discuss above: assuming that the instantaneous geodesic is triperiodic (i.e., that it has a
complete set of integrals), there is a good chance it will have a high
overlap with a Kerr geodesic. Thus, each ``snapshot'' along the inspiral defines a point in the Kerr geodesic space, and so matched
filtering with Kerr snapshots can pick out an inspiral trajectory.
The inspiral trajectory in the Kerr spacetime can be computed, and we
can check to see if the observed inspiral deviates from the predicted
one. This approach has two drawbacks: first, it relies on the
GWs from a geodesic in an arbitrary spacetime having
a high overlap with those from at least one Kerr geodesic; second, the
interpretation of deviations in the trajectory relies on being able to
compute GW emission in alternative metrics in GR, or in alternative theories of gravity. However, these will be problems common
to all spacetime-mapping algorithms. One approach to addressing the first of these problems is to search for EMRIs using a template-free approach. This could be done by searching for individual harmonics of the EMRI signal separately. One method would be to search for tracks of excess power in the time-frequency domain~\cite{tfGairJones,tfGairWen,tfWenGair}. However, excess-power algorithms lose information on the phase of the waveform, which is the most important observable for carrying out tests of GR. More recently, a method that attempts to match the phase of individual harmonics using a Taylor expansion with a small number of coefficients was described in~\cite{2012PhRvD..86j4050W}, which built on ideas used to search for GR EMRIs in the Mock LISA Data Challenges~\cite{mcmcBGP}. While the method appears promising, it must be kept in mind that so far these approaches have only been used on highly simplified datasets containing a single EMRI source. Issues related to confusion between the harmonics of the multiple sources expected in real data have not yet been properly explored. Nonetheless, these techniques warrant further investigation.

In summary, it is clear that low-frequency GW detectors will be able to check that an EMRI is
consistent with Kerr, which can be interpreted as a statement on how
much the spacetime could differ from the Kerr
metric~\cite{ASReview,2007PhRvD..75d2003B} and still be consistent with our search
templates. What is more difficult is to say exactly what LISA will
tell us if we observe something that \emph{does} differ from the Kerr
metric. More work is still needed in this area.

\subsubsection{Extreme-mass-ratio bursts}

Another class of low-frequency GW sources that are closely connected
to EMRIs are Extreme-Mass-Ratio Bursts EMRBs~\cite{rubbo06}. These are the
precursors to EMRIs: in the standard EMRI formation channel,
compact objects are perturbed onto orbits that pass very close to the
central black hole, emitting bursts of GWs as they do, which can lead to capture if sufficient orbital energy and angular momentum are lost to GWs.
Furthermore, just after capture, EMRI compact objects proceed on very eccentric
orbits that radiate in bursts near pericenter rather than continuously.
Similar bursts will be generated by ``failed'' EMRIs -- objects that
only encounter the central black hole a few times before being
perturbed onto an unbound or plunging orbit.

For a sufficiently nearby system (in the Virgo cluster or closer~\cite{rubbo06}),
these bursts could be individually detected by LISA-like detectors. The low
SNR of such events compared to EMRIs is partially
outweighed by the significantly larger number of compact objects
undergoing ``flybys'' at a given time, so the detection rate
could be as high as 18 yr$^{-1}$, of which 15 yr$^{-1}$ would originate
in the Milky Way~\cite{rubbo06}. These rate estimates employed
several simplifying assumptions. A more careful analysis predicted
only $\sim1$ event per year~\cite{hopman07}, and a proper analysis of
our ability to identify such bursts in the presence of signal
confusion has not been carried out. Therefore it is unclear at
present whether any of these events would be identified in the data of LISA-like detectors.

It has been suggested~\cite{yunes08} that EMRBs could be used to test
black-hole structure in a similar way to EMRIs. This assertion was
based on comparing approximate burst waveforms coming from spinning
and nonspinning central black holes with the same system parameters.
In~\cite{2013MNRAS.429..589B}, an analysis of parameter estimation for EMRBs accounting for parameter correlations was performed, which concluded that a space-based detector like LISA could measure many of the parameters of the system to moderate precision, including the mass and spin, provided that the periapse of the burst orbit was within $\sim 10\,GM/c^2$. This analysis did not consider sky position or eccentricity to be free parameters, taking the former to be the center of the Milky Way, and the latter to be maximal, and it treated the distance and compact-object mass as a single amplitude parameter $m/D$. The small object mass cannot be separately determined because there is no significant orbital evolution over the duration of one burst. The analysis did not consider constraints on GR through measurements of other multipoles, but it was found that the mass and spin could both be measured to high accuracy for orbits that pass sufficiently close to the black hole. Therefore it may also be possible to place some kind of constraint on GR deviations using these systems. However, the event rate for such relativistic systems is very low~\cite{hopman07} so we will have to be lucky to see a useful event.

Useful burst events might also be seen from nearby galaxies other than our own~\cite{2013MNRAS.433.3572B}, which are thought to have black holes in the appropriate mass range. Other unknown black holes in the local universe could also provide sources, but for such extra-galactic systems, the sky position will be unknown and so a single burst will not provide enough information to allow the determination of all the system parameters. If an EMRB was observed to ``repeat'' several times over the lifetime of a GW mission, then more information could be derived about the system. However, there are serious practical issues with associating multiple, well-separated bursts with the same
source. The inspiral orbit might also
be perturbed between pericenter passages by encounters with other
stars. The prospects for doing useful tests of black-hole structure
with EMRBs therefore seem rather bleak. Moreover, any
tests that can be carried out will inevitably be much weaker than
those based on EMRIs, as an EMRB event will generate several waveform cycles rather than the several hundred thousand from an EMRI.

\subsection{Tests of black-hole structure using ringdown radiation:
black-hole spectroscopy}
\label{sec:ringdowntests}

When a black hole is perturbed, it rapidly settles back down to a
stationary Kerr state via the radiation of exponentially damped
sinusoidal GWs known as quasi-normal modes (QNMs).
For a Kerr black hole in GW, the QNM frequencies and damping
times are uniquely determined by the mass and spin of the black hole.
Observations of two or more QNMs from the
same system will therefore test that the end-state of the merger is a Kerr
black hole, if they provide consistent estimates for the mass and
spin of the central object. QNMs will be excited during the merger of supermassive black holes
and will dominate the late-stage radiation, so LISA-like detectors should observe
QNM radiation with relatively high SNRs~\cite{flanhughes,dreyer04,berti06}. In fact, the SNR from the
QNMs alone may be comparable to the total SNR in the
inspiral~\cite{flanhughes}. This is illustrated in Figure~\ref{fig.ringdownInspiral}, which shows the SNR with which LISA would detect the inspiral and ringdown for equal-mass, nonspinning MBH binaries as a function of black-hole mass. The ringdown dominates the SNR for redshifted masses greater than $3\times10^6\,M_{\odot}$. 

\epubtkImage{smbhSNRs.png}{%
\begin{figure}
\centerline{\includegraphics[width=4in,keepaspectratio=true]{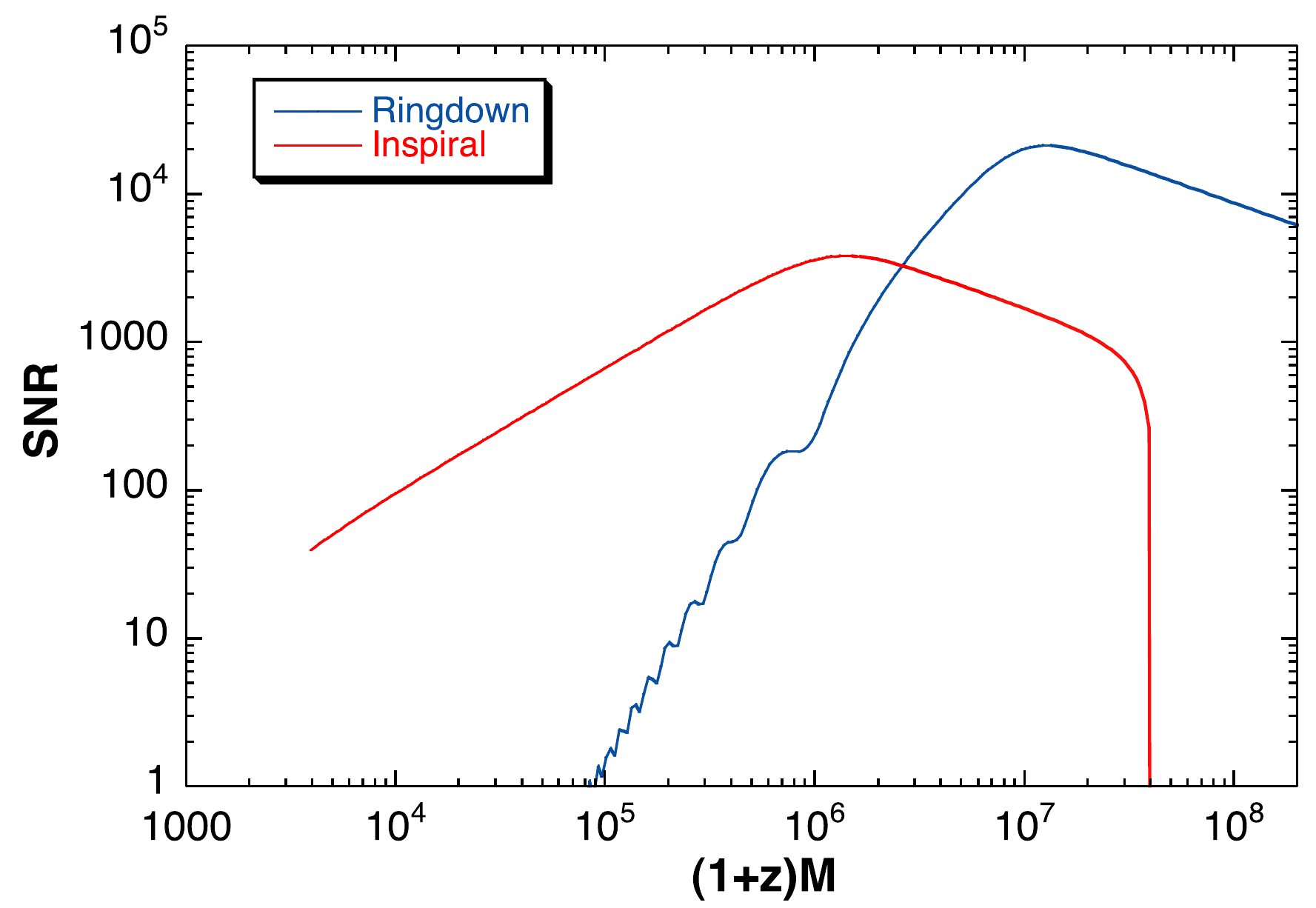}}
   \caption{Comparative SNRs, as a function of redshifted black-hole
   mass $(1+z) M$, for the last year of inspiral of an equal-mass MBH
   binary and for the ringdown after the merger of the system.  The
   method used to generate this figure follows that
   of~\cite{flanhughes}, updated to use a modern LISA sensitivity
   curve~\cite{SCG} with a low-frequency cutoff of $f = 1 \times
   10^{-4}\mathrm{\ Hz}$. The redshift is set to $z = 1$, at which the
   luminosity distance is $D_{L} = 6.6\mathrm{\ Gpc}$ using WMAP
   7-year parameters~\cite{WMAP7}.}
\label{fig.ringdownInspiral}
\end{figure}}

Reviews of the theory of quasi-normal modes can be found in Kokkotas
and Schmidt~\cite{QNMKokkotas}, in Nollert~\cite{QNMNollert} and in Berti, Cardoso and Starinets~\cite{2009CQGra..26p3001B},
but we will briefly summarize the relevant results here. At late times,
the gravitational radiation from a perturbed black hole can be
written as a sum of exponentially-damped sinusoids. It is known that
QNMs do not form a complete set in a mathematical sense, but
numerical results confirm that the late-time behavior is well
described by such an expansion~\cite{berti06}. The angular dependence
can be decomposed into a sum of spheroidal harmonics $S_{lm}$ of spin weight
2, which are labeled in analogy to standard spherical
harmonics. For each $(l,m)$, there are infinitely many resonant QNMs,
which can be labeled in order of decreasing damping time by a third
parameter $n$. The waveform expansion takes the form
\begin{equation}
h_++\mathrm{i}h_{\times} = \frac{M}{r} \sum_{lmn} {\cal A}_{lmn} \mathrm{e}^{\mathrm{i}(\omega_{lmn} t+\phi_{lmn})} \, \mathrm{e}^{-t/\tau_{lmn}}
S_{lmn} \,,
\end{equation}
where $\omega_{lmn}$, $\tau_{lmn}$, ${\cal A}_{lmn}$, and $\phi_{lmn}$
are the frequency, damping time, amplitude, and initial phase of the
mode, and $S_{lmn}$ is a complex number obtained by evaluating the
spheroidal harmonics for the particular orientation and mode
frequency. The frequency and damping time,  $\omega_{lmn}$ and
$\tau_{lmn}$, are determined solely by the mass and frequency of the
central black hole and are the key observables. The damping time can
also be considered as the reciprocal of the imaginary part of a
complex mode frequency, The amplitude and phase depend on the exact
nature of the perturbation that excited the QNMs, so they are less
useful. One complication is that the mode numbers that are excited [i.e., the $(l,m,n)$ parameters] are not known \emph{a priori}. In the first
analysis of SNR from QNMs~\cite{flanhughes}, it was assumed that only
the $l=m=2$, $n=0$ was excited, which simplifies analysis of the
observations. However, this simplifying assumption is not needed in
general.

The use of QNMs as a probe of black-hole structure was first proposed
by Dreyer et al.~\cite{dreyer04}, who coined the phrase ``black-hole
spectroscopy'' for this technique. A measurement of $\omega_{lmn}$
and $\tau_{lmn}$ for a single QNM will give several discrete
solutions for the black-hole mass and spin, corresponding to
different choices for the mode numbers. A measurement of a second QNM
will give another discrete set of values, and in general these will
not be the same as the first set except for one combination, which are
the true values~\cite{dreyer04}. This not only determines the black-hole parameters, but provides a test that the central object is a
Kerr black hole, since the relationship between black-hole parameters
and frequencies depends on that assumption. If the object was something else, we would find no consistent
combination of mass and spin parameters to arise from the analysis.
Dreyer et al.~\cite{dreyer04} considered a frequentist approach to
analyzing the QNM problem. The observable is a set of $N$ complex QNM
frequencies. We denote a set of choices for the $N$ modes to which
the frequencies correspond by ${\cal Q}$. Each ${\cal Q}$ defines a
two-dimensional curve in the $2N+2$ dimensional space, parameterized
by $(a,M)$. The probability distribution for the observed
frequencies, $\mathbf{\omega}$, may be denoted by
$P(\mathbf{\omega}|a,M,{\cal Q})$ and can be used to define a
confidence interval by determining the $p_0$ such that
\begin{equation}
\int_{\{\mathbf{\omega}:P(\mathbf{\omega}|a,M,{\cal Q})>p_0\}} \, P(\mathbf{\omega}|a,M,{\cal Q}) \, \mathrm{d}\omega^{2N} .
\end{equation}
Dreyer et al.~\cite{dreyer04} deemed that an observation was
inconsistent with GR if the actual observed frequencies lay
outside such confidence intervals for all possible choices of $a$,
$M$ and ${\cal Q}$. Using a toy model for a non-black-hole source,
and assuming that two QNMs were observed, drawn from four possible
excited modes, they estimated that, with a false alarm probability of
1\%, the false dismissal rate would be 60\% if the
SNR in the weakest mode was 10, falling to 10\%
at an SNR of 100. Assuming a reasonable amount of energy deposition
into the QNMs, they estimated that this would be satisfied by all
LISA MBH merger sources, so the prospects for such tests are good.

Berti et al.~\cite{berti06} took a more detailed look at QNMs in
general. They computed SNRs that improved on those in~\cite{flanhughes} by
using an up-to-date detector noise curve and including black-hole spin. For a source of mass $10^5\,M_{\odot} < M < 10^7\,M_{\odot}$
at a distance of 3~Gpc, they found SNRs between 10 and $2\times10^4$ for an
excitation energy of 3\% of the rest mass, and between 1 and
3000 for an excitation energy of 0.1\%. There is little
dependence of the SNR on the central black-hole spin. Putting the
source at higher $z$ shifts the sensitive range of masses in the
usual way, but even at redshift $z=10$ the maximum SNR was found to
be several hundred. Berti et al.~\cite{berti06} also looked at
parameter-estimation accuracy from observations of one QNM (with
known mode numbers) and found that LISA observations would
measure the mass and spin to fractional accuracies in the range
$10^{-2}\mbox{\,--\,}10^{-5}$ for sources at $D=3\mathrm{\ Gpc}$ and for the masses as quoted
above. 

Berti et al.~\cite{berti06} also performed a Fisher-matrix analysis of a
two-mode problem, which accounted for all correlations between the
modes. They considered both ``pseudo-orthonormal'' modes with
different angular dependence ($l' \neq l$ or $m' \neq m$) and
``overtones'' -- that is, modes with the same angular dependence ($l'=l$
and $m'=m$). In both cases, the accuracy of determination of the mass
and spin was comparable to the results quoted for a single mode,
although the parameter space was simplified significantly to just
five parameters: $a$, $M$, two mode amplitudes ${\cal A}_1$,
${\cal A}_2$, and a common initial phase $\phi$. Berti et al.\ define two distinct QNMs to have resolvable frequencies if the difference in these is larger than the error in both (and similarly for the damping times). By recasting the two-mode model in
terms of the two mode frequencies and damping times rather than $a$
and $M$, resolvability can be examined by looking at the Fisher-matrix errors. Berti et al.~\cite{berti06} considered two cases
(the resolvability of \emph{either} frequencies or damping times, and 
the resolvability of \emph{both}), and they computed the
source SNR needed in each case. When the two modes are overtones, the
SNR required to resolve ``either'' was $\sim$~100, the exact value depending
on the black-hole spin and the ($l$,$m$) numbers of the mode, but the
SNR required to resolve ``both'' was $\sim$~1000. When the
two modes are ``pseudo-orthonormal,'' the SNR range to resolve
``either'' is $\sim$ a few if the modes differ in $l$, and
$\sim$ a few tens if the modes have the same $l$ but differ in $m$ and the
central black-hole spin is $a>0.4$. There is a mode degeneracy, which means
the SNR required blows up when $l=l'$ and $a \rightarrow 0$. The SNRs
required to resolve ``both'' modes are in the range 100\,--\,10000.

Berti et al.~\cite{berti06} conclude that even under pessimistic
assumptions about the amount of energy radiated in ringdown
radiation, it should be possible to resolve one QNM and either the
damping time or frequency of a second QNM, provided that the first
overtone radiates $\sim10^{-2}$ of the total ringdown energy. This
will provide enough for a test of the no-hair property of the central
object. A stronger test would come from detecting frequencies and
damping times for both QNMs, but this would require ringdown SNRs
$\sim$~1000, which is rather unlikely. The main uncertainty in this
analysis was in the excitation coefficients for the various modes, but numerical relativity simulations have now provided some information on the excitation of ringdown modes in a merger. Berti et al.\ revisited the QNM problem in~\cite{Berti2007}, using numerical relativity results presented in~\cite{2007PhRvD..76f4034B}. This paper revised downward the SNR required to detect either the frequency or damping time of a second mode, which is sufficient for a test of the no-hair theorem, to $\rho_{\mathrm{crit}} \approx 30$ for any binary with mass ratio greater than $\sim$~1.25. A significant fraction of LISA events should satisfy this criterion (see for example~\cite{2013BrJPh.tmp...27B}).


In~\cite{Kamaretsos2011}, the authors used numerical-relativity simulations of nonspinning black-hole binaries, with a variety of mass-ratios ranging from 1:1 to 1:11, to compute the amplitude to which several different ringdown modes were excited, and hence an estimate of the SNR with which LISA would be able to observe them. For all the mass ratios considered, the $(3,3)$ mode was found to radiate between 2\% and 20\% of the energy of the $(2,2)$ mode, and for mass ratios more unequal than 1:2 the $(4,4)$ also radiated more than 1\% of the energy of the $(2,2)$ mode. For sources at a redshift of $z=1$, they estimated a total ringdown SNR between 1000 and 20000 for black-hole masses in the range $10^{6}\mbox{\,--\,}10^{8}\,M_{\odot}$ and mass ratios from 1:1 to 1:25, with individual SNRs in the $(2,2)$, $(2,1)$, and $(3,3)$ modes between several hundred and 10000. The mode SNRs do not add in quadrature, but these SNRs more than meet the requirements of~\cite{berti06} for LISA to be able to carry out black-hole spectroscopy. The SNRs realizable with eLISA will be a few factors lower and the systems will tend to have lower masses, for which the intrinsic SNRs are also smaller. Therefore constraints from eLISA will be weaker and it is possible that eLISA will not be able to resolve sufficient separate modes, but this has not yet been explored.

In~\cite{2012PhRvD..85l4056G}, Gossan et al.\ applied the results of~\cite{Kamaretsos2011} to explore the practicality of using QNM radiation to test relativity. The authors considered mergers with mass ratios of 1:2 and constructed a waveform model comprised of four ringdown modes. They used Bayesian methods and approached the problem of testing relativity in two ways: (i) determining the parameters of each ringdown mode separately and checking for consistency; (ii) comparing the Bayesian evidence of the GR model to a non-GR model constructed by allowing for deviations of the mode parameters from the GR predictions. The analysis was carried out for eLISA and for the Einstein Telescope, the proposed third-generation ground-based detector. Gossan et al.\ showed that method (i) could reveal inconsistencies of 1\% in the QNM frequency for events of mass $10^{8}\,M_{\odot}$ at a distance of 50~Gpc and inconsistencies of 10\% for systems of mass $10^{6}\,M_{\odot}$ at 6~Gpc. Better constraints will come from systems at smaller distances and will therefore be observed with higher SNR. Using method (ii), in the case of a signal from a $10^{6}\,M_{\odot}$ black hole, 2\%, 6\%, and 10\% deviations in the frequency of the dominant mode would be identifiable at distances of approximately 1~Gpc, 5~Gpc, and 8~Gpc, respectively. Deviations of 2\%, 6\% and 10\% in the damping time of the dominant mode would only be detectable at distances of 200~Mpc, 700~Mpc, and 1.2~Gpc, respectively. For a more massive system, of $10^{8}\,M_{\odot}$, a 2\% deviation in the frequency/damping time of the dominant mode would be detectable out to 35/25~Gpc, but deviations of 5\% or more would be detectable at greater than 50~Gpc. Such massive systems are very rare, however, and the choice of a 1:2 mass ratio means that the QNM radiation is stronger than would be expected for typical eLISA events, which are likely to have larger mass ratios. Therefore this study was limited in extent, but these preliminary results suggest that QNMs could place interesting constraints on GR modifications for the strongest signals detected by eLISA.

Space-based QNM observations could be used directly to put constraints on
alternative theories of gravity, but this will require a calculation
of the QNMs in those alternative theories. It was shown in~\cite{YSCSqnm}
that the equations governing black-hole perturbations in dynamical
CS gravity were different from those in GR, with the
consequence that the QNM spectrum would also be different. The
authors did not compute QNMs, but QNM frequencies for non-spinning black holes in dynamical CS gravity were computed in~\cite{2010PhRvD..81l4021M}. The QNM spectrum was found to be different, due to the coupling of the black hole oscillations to the scalar field, but the detectability of these deviations by a gravitational-wave detector was not estimated. Therefore, it is
not yet clear at what level ringdown radiation will be able to constrain the
CS coupling parameter.

If a LISA-like observatory observes a MBH inspiral and does not
detect QNM ringing from the merger remnant where radiation would be
expected, this might indicate a violation of the cosmic-censorship
hypothesis. QNM ringing arises as a result of GWs
becoming trapped near the horizon of the black hole. If a horizon was
absent, for instance if a super-extremal ($a/M > 1$) Kerr metric was
formed, we would not expect to observe the QNM radiation. This would
be evidence for the existence of a naked singularity, and a
counterexample to cosmic censorship, although not an indication of a problem
with GR.

\subsection{Prospects from gravitational-wave and other observations}
Although there are still many open questions, current research clearly
indicates that low-frequency GW detectors will be able to make strong statements
about the structure of the massive compact objects in the centers of
galaxies by observing EMRIs and by detecting ringdown radiation following supermassive black-hole mergers. At the very least, it will be possible
to test that an observation is consistent with an inspiral into a
Kerr black hole, and to state quantitatively how so large a deviation from
Kerr could have been masked by instrumental noise. If a system
differed significantly from Kerr, we would identify the deviation and should
be able to quantify its nature. How well we would
be able to distinguish between different types of deviation (e.g.,
external matter vs.\ a different multipole structure of the central
object) is still not totally clear, but there are ongoing efforts in
this direction.

Another potential issue relates to modeling EMRIs. The tests of
black-hole structure outlined in this section will rely on the
detection of small differences between the observed GWs and the GWs
expected in GR. This will of course rely on having reliable EMRI
waveforms within GR. These waveforms can be computed using black-hole
perturbation theory, by evaluating the gravitational self-force, but
despite extensive work in this area the calculation of the self-force
is not yet complete even at first order in the mass ratio. The
first-order gravitational self-force has been computed for
circular~\cite{2007PhRvD..75f4021B} and generic orbits in the
Schwarzschild spacetime~\cite{2010PhRvD..81h4021B} and the scalar
self-force is known for circular equatorial orbits in the Kerr
spacetime~\cite{2011PhRvD..83l4038W}. Recently, the first
self-force--driven inspirals have been computed, under an adiabatic
approximation for the gravitational
self-force~\cite{2012PhRvD..85f1501W}, and self-consistently for the
scalar self-force~\cite{2012PhRvL.108s1102D}.%
\epubtkFootnote{The difference between the ``adiabatic'' and
``self-consistent'' evolutions is in the treatment of the past history
of the particle, which is what determines the self-force. In the
``adiabatic'' approximation the past history is assumed to be
geodesic, so the self-force is computed for objects on geodesics and
then the final trajectory is evolved through a sequence of
``osculating'' geodesics using this geodesic self-force. The
``self-consistent'' approach uses the actual trajectory to describe
the past history of the particle. The two approaches differ at second
order in the mass ratio, which is the same order as other corrections
that have been ignored in both approaches.}
EMRI models will require knowledge of the gravitational self-force for generic orbits in Kerr, which is still not available at first order. It has also been recognized that the radiative part of the second-order self-force will also be required to derive EMRI waveform phase to sufficient accuracy~\cite{2009PhRvD..79h4021H}. A formalism for the evaluation of the second-order self-force has been developed~\cite{2012PhRvL.109e1101P}, but not yet implemented. Complete reviews of the current status of the self-force program can be found in~\cite{2009CQGra..26u3001B,poissonLR}. Additional complications in EMRI modeling arise from the need to accurately follow the transition of the inspiral through transient resonances~\cite{2012PhRvL.109g1102F}. Although many of these issues will have been resolved by the time data from a space-based gravitational-wave interferometer is available, they must be borne in mind in the interpretation of future results on tests of GR. 

Low-frequency GW detectors will provide information on black-hole structure to a much
higher precision than can be achieved by other techniques. It has
been suggested that observations of IMRIs with Advanced LIGO will have some power to do similar
tests~\cite{imriprl}. This data will be available ten or more years
prior to any space-based experiment. However, IMRI-based tests will be much weaker, since
many fewer cycles of the GWs will be detected in a
single observation due to the larger mass ratio ($\eta \sim
10^{-2}\mbox{\,--\,}10^{-1}$); furthermore, event rates are highly
uncertain~\cite{imrirate} and accurate modeling of IMRIs is far behind EMRI modeling~\cite{2011PhRvD..83d4020H,2011PhRvD..83d4021H}. An IMRI observation would be able to
detect a deviation from Kerr of fractional order unity in the
quadrupole moment of a source~\cite{imriprl}, compared to the
$\sim10^{-3}$ fractional accuracies achievable with LISA
EMRIs~\cite{2007PhRvD..75d2003B}.

Observations in the electromagnetic spectrum can also probe black-hole structure. Psaltis~\cite{psaltisLR} provides a thorough review
of tests of strong-field gravity using electromagnetic observations.
The most relevant tests of black-hole structure are as follows.

\begin{itemize}
\item \textbf{Horizon detection.} X-ray interferometers (e.g., the Black
Hole Imager) will soon have the angular resolution to directly image
the horizon of extragalactic black holes at distances of
$\sim$~1~Mpc~\cite{psaltisLR}. This is already almost possible with
sub-mm/infrared observations. The accretion flow for the Milky Way
black hole, Sgr~A*, can already be imaged directly down to its
innermost edge. However, interpretation of the observations is complicated by the need to simultaneously constrain the
disc properties. Observations at multiple wavelengths
will be required to fit out the disc and directly image the shadow of
the black hole on the disc. 

Evidence for the existence of horizons also comes from observations
of quiescent X-ray binaries. An X-ray binary typically comprises a star filling its Roche lobe and transferring matter to a compact-object companion. That compact object can either be a neutron star or a black hole. If the object is a neutron star then most of the gravitational potential energy of the accreting matter has to be radiated away, but for black holes a significant amount of this energy can be advected through the black-hole horizon and is therefore not radiated to infinity~\cite{psaltisLR}. Among systems in the same state of mass transfer, those containing neutron stars are expected to be systematically more luminous than those containing black holes, as has been observed in our galaxy.

Electromagnetic observations will really indicate the existence of a
high-redshift surface in the system, and not necessarily an actual
horizon. If a system contained a naked singularity with a high-redshift surface, but not a true event horizon, this would not be
evident from the electromagnetic observations alone. By probing the
multipole structure and verifying consistency with the no-hair
theorem, LISA-like detectors will go much further.

\item \textbf{ISCO determination.} The highest-temperature emission from a
disc comes from the innermost stable circular orbit (ISCO), and the flux at
that temperature is proportional to the ISCO radius squared. This
allows the determination of the spin of the black hole. If the inferred
spin was found to exceed one, this might indicate failure in the
black-hole model. Such spin determinations have typical errors
of $\sim$~10\%, much greater than LISA's expected errors of
$\sim$~0.01\%~\cite{AK}.

Indirect inference of the location of the inner edge of the accretion disc could also come from the interferometric observations mentioned above in the context of horizon detection. In~\cite{2012Sci...338..355D} the authors use radio interferometry to resolve the base of the jet in the galaxy M87, finding it to be $5.5 \pm 0.4$ Schwarzschild radii. The jet-base radius is interpreted as being greater than or equal to the radius of the innermost edge, so this system provides evidence for prograde accretion onto a spinning black hole. Observations of jets in other nearby galaxies may follow in the future, but the distance to which these structures could be resolved is quite small.

\item \textbf{Accretion-disc mapping.} Particles in the accretion disc
around a black hole move on circular geodesics of the metric. If the
orbits in the disc could be mapped, this would allow spacetime
mapping along the lines of Ryan's theorem~\cite{ryan95}. Two possible
probes of accretion-disc structure have been identified: quasi-periodic oscillation (QPO) pairs and iron K$\alpha$ lines. QPOs
have been used to constrain possible values of black-hole masses and
spins, but there are uncertainties in the radius at which they
originate, and in the resonance that gives rise to the pairs of lines.
Iron lines show broadening due to differential gravitational redshift
and Doppler shift at different points in the disc, but again their
interpretation depends on unknown details of the disc geometry. In
principle, the time variability of an iron line after a single flare
event could constrain both the geometry and map the disc. Future
observations with instruments such as IXO/ATHENA 
will have the time-resolution to attempt this~\cite{psaltisLR}.
\end{itemize} 

Electromagnetic observations are generally hampered by a lack of
knowledge of the physics of the material that is generating the
radiation. GW systems are, by contrast, very
``clean'' since the same theory describes the spacetime and the
radiation generation. While there is some hope that future
electromagnetic observations will perform crude spacetime mapping,
LISA EMRI observations will improve any previous constraints by
orders of magnitude.



\section{Discussion}\label{sec.discuss}

Although the bibliography of this review lists more than 500 references,
fundamental tests of gravity remain the least explored sector of GW
astronomy.  It is also one where new discoveries may be hardest to
come by -- even if they would be momentous if they do.
Space-based gravitational-wave detectors will improve certain existing constraints on alternative theories; even more important, they will provide novel and unique opportunities to precisely characterize the poorly explored nonlinear, dynamical sector of gravitation, as well as the properties of gravitational radiation fields. 
These measurements will provide a definite confirmation that Einstein's theory, born from insight and beauty, applies in the most extreme regimes; or they may expose new phenomena, leading to new models for gravitational physics.

The research that we have reviewed spans a broad assortment of GW observations, which probe many disparate aspects of gravitational phenomenology. Some of these investigations have yet to be refined to the same level of formal rigor as other, more astrophysical applications of GW astronomy. Nevertheless, they paint an exciting picture of the expected fundamental-physics payoff of space-based detectors. We expect that this will remain an active and intriguing research area for many years: thus, it is appropriate that this review should be \emph{living}, and we welcome the suggestions of our readers in improving it and keeping it up to date.

\newpage


\section{Acknowledgements}
\label{sec.acknowledgements}

We thank Masaki Ando, Stanislav Babak, Christopher Berry, Emanuele Berti, Vitor Cardoso, Naoki Seto, Kent Yagi, and the Living Reviews referees for useful comments on the manuscript; we thank Christopher Moore for providing the image in Figure~\ref{fig.discoverySpace}. JG's work is supported by the Royal Society. MV's work was carried out at the Jet Propulsion Laboratory under contract with the National Aeronautics and Space Administration, with support from the LISA project and PCOS program. JB's work was partly supported by NASA grant 11-ATP11-046.

\newpage





\bibliography{refs}

\end{document}